\documentclass[superscriptaddress,twocolumn,showpacs,a4paper,longbibliography,amssymb,amsmath,nobibnotes,aps,prd,
showkeys,nofootinbib,notitlepage,floatfix]{revtex4-2}
\pdfoutput=1

\usepackage[dvipsnames,table]{xcolor}
\usepackage{verbatim}
\usepackage[T1]{fontenc}
\usepackage[utf8]{inputenc}
\usepackage[american]{babel}
\usepackage{epsfig}
\usepackage{booktabs}
\usepackage{multirow}
\usepackage{dcolumn}
\usepackage{amsmath}
\usepackage{mathtools}
\usepackage{amsfonts}
\usepackage{amssymb}
\usepackage{ulem}
\usepackage{epstopdf}
\usepackage{bm}
\usepackage{siunitx}
\usepackage{braket}
\usepackage{enumitem}
\usepackage{soul}
\usepackage{placeins}
\usepackage{transparent}
\usepackage{pifont}
\usepackage{float}
\usepackage[caption=false]{subfig}

\definecolor{navyblue}{rgb}{0.0, 0.0, 0.5}
\definecolor{royalblue}{rgb}{0.25, 0.41, 0.88}
\definecolor{cadmiumgreen}{rgb}{0.0, 0.42, 0.24}
\definecolor{blue-violet}{rgb}{0.54, 0.17, 0.89}
\definecolor{darkviolet}{rgb}{0.58, 0.0, 0.83}
\definecolor{orange(colorwheel)}{rgb}{1.0, 0.5, 0.0}

\usepackage{microtype}
\usepackage{xurl}
\usepackage{hyperref}
\hypersetup{colorlinks,linkcolor={blue},citecolor={red},urlcolor={cyan}}

\usepackage{colortbl}

\newcommand{\cmbdesi}{CMB+\allowbreak DESI}
\newcommand{\cmbdesipantheon}{CMB+\allowbreak DESI+\allowbreak Pantheon\allowbreak Plus}
\newcommand{\cmbdesiunion}{CMB+\allowbreak DESI+\allowbreak Union3}
\newcommand{\cmbdesidovekie}{CMB+\allowbreak DESI+\allowbreak DES-\allowbreak Dovekie}

\begin{document}

\title{Do DESI-DR2 BAO data imply a coupling of dark matter and dark energy?}

\author{Weiqiang Yang}
\email{d11102004@163.com}
\affiliation{Department of Physics, Liaoning Normal University, Dalian, 116029, People's Republic of China}

\author{Sibo Zhang}
\email{sbzhang02@163.com}
\affiliation{Department of Physics, Liaoning Normal University, Dalian, 116029, People's Republic of China}

\author{Dinorah Barbosa}
\email{dinorahbarbosaft@gmail.com}
\affiliation{Instituto de Fisica, Universidade Federal Fluminense, 24210-346 Niteroi, RJ, Brazil}

\author{Marco Antonio Alvarez}
\email{marcocardoso@id.uff.br}
\affiliation{Instituto de Fisica, Universidade Federal Fluminense, 24210-346 Niteroi, RJ, Brazil}

\author{Supriya Pan}
\email{supriya.maths@presiuniv.ac.in}
\affiliation{Department of Mathematics, Presidency University, 86/1 College Street, Kolkata 700073, India}
\affiliation{Institute of Systems Science,
Durban University of Technology, Durban 4000, Republic of South Africa}

\author{Leila Graef}
\email{leilagraef@id.uff.br}
\affiliation{Instituto de Fisica, Universidade Federal Fluminense, 24210-346 Niteroi, RJ, Brazil}

\begin{abstract}
We revisit an interacting dark matter (DM) -- dark energy (DE) model characterized by the interaction function $Q = \Gamma \rho_x$, where $\Gamma$ is a constant coupling parameter and $\rho_x$ is the energy density of DE. This type of interaction is independent of the Hubble rate or other external parameters, but depends only on the fundamental properties of DE, such as its equation of state (EoS), $w_x$. We pay special attention to $w_x$ and study three distinct interacting scenarios distinguished by the nature of $w_x$, i.e. $w_x =-1$ (when DE corresponds to the vacuum energy), $w_{x} < -1$ (when DE has a phantom behavior), and $w_x > -1$ (quintessential DE). We constrain all of them using the most recent cosmological datasets, including CMB from Planck 2018, BAO from DESI DR2, and three compilations of SNIa (PantheonPlus, Union3, and DES-Dovekie). Our analyses reveal that evidence of interaction is supported in scenarios with $w_x =-1$ and $w_x < -1$ when all three datasets are combined, but from the Bayesian evidence analysis, $\Lambda$CDM remains favored over these interacting scenarios. Regarding the $S_8$ parameter, when $w_x > -1$, this interacting scenario leads to mildly lower estimates across all datasets.
\end{abstract}
\maketitle

\begingroup
\small
\textcopyright\ 2026 American Physical Society. This is the accepted manuscript of the following article: W. Yang, S. Zhang, D. Barbosa, M. A. Alvarez, S. Pan, and L. Graef, ``Do DESI-DR2 BAO data imply a coupling of dark matter and dark energy?'', \textit{Phys. Rev. D} \textbf{114}, 063549 (2026). The final published version is available at \url{https://doi.org/10.1103/6kdf-vzq4}. This manuscript version is posted with permission for non-commercial scholarly use.
\par
\endgroup

\section{Introduction}
Modern cosmology, grounded in Einstein’s General Relativity, reflects the interdependence between the constituents of the universe and its dynamical evolution. While the properties of baryons and radiation can be independently constrained by observational data, dark matter (DM) and dark energy (DE) are accessible only through their gravitational effects.
In the standard cosmological model, $\Lambda$CDM, DM is treated as non-relativistic (“cold”) matter and DE as a cosmological constant $\Lambda$. Despite its predictive success, $\Lambda$CDM faces conceptual difficulties and emerging observational challenges   \cite{DiValentino:2021izs,Perivolaropoulos:2021jda, Anchordoqui:2021gji, Abdalla:2022yfr}. Besides the well-known cosmological constant problem~\cite{Weinberg:1988cp} and cosmic coincidence problem~\cite{Zlatev:1998tr}, the $H_0$ tension \cite{DiValentino:2021izs,Perivolaropoulos:2021jda,Schoneberg:2021qvd, DiValentino:2020vnx, H0LiCOW:2019pvv} has been an intriguing and persistent issue. Hence, alternative cosmological models often attempt to avoid or at least alleviate the aforementioned difficulties while also aiming to accurately describe the cosmological observations associated with the dark sector. A cosmological model that seeks to go beyond the standard framework should ideally not only address these open problems but also explore alternative dynamical behaviors within the dark sector.

Recently, the quest for understanding dynamical dark energy scenarios became unprecedentedly interesting. In March 2025, the Dark Energy Spectroscopic Instrument (DESI)  Data Release 2 (DR2) results were released~\cite{DESI:2025zgx}, marking the start of the Stage IV DE era. These results, when combined with other external probes, provide indications of a dynamic DE component. These signs were already present in the DR1 release \cite{DESI:2024mwx}. However, the statistical improvement and extended redshift range of DESI DR2 yielded a level of statistical significance comparable to or greater than that reported with DESI DR1 in favor of a dynamical dark energy. Dynamical dark energy models parameterized by $w_0$ and $w_a$ were shown to be preferred over $\Lambda$CDM at 3.1$\sigma$ for the combination of DESI BAO and CMB data and over 2.8$\sigma$ to 4.2$\sigma$ when also including SNe \cite{DESI:2025zgx}.
An extended analysis was also performed, including more general dark energy parametrization \cite{DESI:2025fii,Giare:2024gpk} showing that "the evidence for dynamical dark energy, particularly at low redshift ($z \lesssim 0.3$), is robust and stable under different modeling choices" \cite{DESI:2025fii}. Also, a diverse set of model-independent approaches aiming at reconstructing the dark energy properties using DESI BAO data in combination with CMB data and different Supernovae compilations, support an evolving dark energy behavior \cite{DESI:2024aqx, Jiang:2024xnu, Berti:2025phi, Gonzalez-Fuentes:2025lei}.

After the new results, models with a modified dark sector, initially proposed to handle the inconsistencies of the standard model ~\cite{Copeland:2006wr,Bamba:2012cp}, have attracted even more attention.
In particular, coupled dark energy models,  are among the models that stand out in the literature~\cite{Amendola:1999er, Cai:2004dk,Huey:2004qv,Barrow:2006hia,Valiviita:2008iv,Boehmer:2008av,Quartin:2008px,Quercellini:2008vh,Cataldo:2008tx,Kim:2008pi,MohseniSadjadi:2008na,Jesus:2008xi,Borges:2008ii,Chen:2008pz,Caldera-Cabral:2008yyo,Chen:2008ft,Gavela:2009cy,Jackson:2009mz,Costa:2009mv,Chimento:2009hj,Cataldo:2010kc,Gavela:2010tm,Baldi:2010vv,Lip:2010dr,Tarrant:2011qe,Solano:2011ie,Cotsakis:2012sh,vandeBruck:2013yxa,Shojai:2013gsa,Bolotin:2013jpa,Yang:2014gza,Yang:2014okp,Faraoni:2014vra,Yang:2014hea,vandeBruck:2015ida,Li:2015vla,Feng:2016djj,Yang:2016evp,Kumar:2016zpg, Wang:2016lxa,vandeBruck:2016jgg,Pourtsidou:2016ico,vandeBruck:2016hpz,DiValentino:2017iww,Mifsud:2017fsy, VanDeBruck:2017mua,Paliathanasis:2019hbi,Yang:2019uzo,Mifsud:2019fut,Barrow:2019jlm,Li:2019ajo,DiValentino:2019jae,Sa:2021eft,Zhang:2021yof,Potting:2021bje,DiValentino:2020kpf,Paliathanasis:2021egx,Chatzidakis:2022mpf,Yang:2022csz,Nunes:2022bhn,Zhai:2023yny,Teixeira:2023zjt,Wang:2024vmw,Giare:2024ytc,Giare:2024smz,Li:2024qso,Ghedini:2024mdu,Halder:2024aan,Zhai:2025hfi,Li:2025owk,Liu:2025pxy,Pan:2025qwy,vanderWesthuizen:2025iam,Li:2025ula,Liu:2025vda,Guedezounme:2025wav,Wu:2025vrl,Figueruelo:2026eis,Li:2026xaz,Paliathanasis:2026ymi,Wang:2026wrk,Paliathanasis:2026acn,Zhai:2026uwr}, since they are not only capable of softening the coincidence problem~\cite{Cai:2004dk, Huey:2004qv, delCampo:2008jx}, but in some cases, they also provide a reduction of the $H_0$ tension~\cite{Kumar:2016zpg, DiValentino:2017iww, Yang:2018euj, Pan:2019gop}.
Furthermore, interactions in the dark sector can also naturally occur in running vacuum models \cite{Sola:2015wwa, Sola:2016jky, SolaPeracaula:2017esw, Sola:2017znb, SolaPeracaula:2023swx, deCruzPerez:2025dni}. Moreover, from the field theory perspective, it is natural to consider an interaction between DM and DE if they are fundamental fields.

In the present article, we consider coupled scenarios in which DE interacts with DM through $Q = \Gamma\rho_x$, where $\Gamma$ is a constant, known as the coupling parameter of the interaction, and $\rho_x$ is the energy density of DE.\footnote{In this work, we restrict ourselves to cases where the interaction occurs within the dark sector, while baryons, radiation and neutrinos remain unaffected.}
This interaction does not depend on external factors, such as the curvature or expansion rate of the universe, rather on intrinsic properties of the dark sector components. This compelling feature distinguishes this class of models from interacting scenarios in which external parameters are involved.
Interaction functions such as the one above have been studied
in Ref.~\cite{Clemson:2011an,Shafieloo:2016bpk}.
In order to assess the role of the equation of state (EoS) of the DE in our coupled models, we considered three distinct scenarios: (i) an interacting vacuum ($w_x = -1$) scenario ({\bf IVS}); (ii) a phantom ($w_x < -1$)  interacting DE ({\bf IDEphan}); and (iii) a quintessential ($w_x > -1$) IDE ({\bf IDEquin}). Since these models can encompass distinct underlying mechanisms, the sound speed is not uniquely determined. Therefore, we also tested different values for the speed of sound to analyze whether our results remain  consistent under modifications in the value of $c_s^2$, or if they are noticeably sensitive to this quantity. We constrained these models using the most recent cosmological probes, including the cosmic microwave background from Planck 2018, baryon acoustic oscillations from DESI DR2, and three different compilations of Type Ia Supernovae.

The remainder of this paper is organized as follows. In Section \ref{sec-2} we present the coupled dark energy models considered and derive their background and perturbation equations. In Section \ref{sec-data} we describe the datasets and methodology used. In Section \ref{sec-results} we report our results.
Finally, Section \ref{sec-summary} summarizes our conclusions and outlines prospects for future work. Additional plots and tables are presented in Appendix~\ref{sec-appendix}.

\section{Coupled DE: Set-up}
\label{sec-2}

We consider a scenario in which  DM and DE interact with each other, whereas the remaining components (baryons, radiation, and neutrinos) are conserved separately.\footnote{The sum of neutrino masses, $\sum m_{\nu}$, and the number of neutrino species, $N_{\rm eff}$, are fixed to $0.06$ eV and $3.044$. }
We assume that the matter sector is minimally coupled to gravity, as in the standard case.

We begin by introducing the background equations for the models considered in this work.

\subsection{Background Equations}

In the interacting scenarios analyzed here, the universe is well approximated by the Friedmann-Lema\^{i}tre-Robertson-Walker (FLRW) geometry, as usual, with the line element assuming the form

\begin{align}
    ds^2 = -dt^2 + a^2(t) \left[\frac{dr^2}{1-Kr^2} + r^2 (d\theta^2 + \sin^2 \theta d \phi^2) \right].
\end{align}
Above, $a(t)$ is the expansion scale factor of the universe, $(t, r, \theta, \phi)$ are the comoving coordinates, and $K$ corresponds to the spatial geometry of the universe.
Assuming a flat FLRW universe which corresponds to $K =0$, the Friedmann equations are,
\begin{eqnarray}\label{Friedmann-eqns}
\rho_{\rm tot} = \frac{3H^2}{\kappa^2},\;\;\;\;\;\; p_{\rm tot} = -\frac{2 \dot{H} + 3 H^2}{\kappa^2},
\end{eqnarray}
as usual.
Above, $\kappa^2 = 8 \pi G$ is Einstein's gravitational constant ($G$ is Newton's gravitational constant), and $\rho_{\rm tot}$ and $p_{\rm tot}$ are the sum of the energy density and pressure of the individual fluid present in the universe, respectively. Also, $H$ is the Hubble parameter and following (\ref{Friedmann-eqns}), it can be written as

\begingroup
\small
\begin{equation}
\frac{H(a)}{H_0} = \Bigg[\frac{\rho_{r0}\,a^{-4}}{\rho_{\rm cr,0}}
 + \frac{\rho_{b0}\,a^{-3}}{\rho_{\rm cr,0}}
 + \frac{\rho_{c}(a)}{\rho_{\rm cr,0}}
 + \frac{\rho_{x}(a)}{\rho_{\rm cr,0}}
 + \frac{\rho_{\nu}(a)}{\rho_{\rm cr,0}}\Bigg]^{\frac12}.
\end{equation}
\endgroup

Above, the indices ($b,r,c,x,\nu$) stand for baryons, radiation, CDM, DE and neutrinos, respectively. The subscript $0$ indicates quantities evaluated today ($a=1$) and $\rho_{\rm cr}$ refers to the critical density. Since the models considered in this work contemplate interactions between DE and DM, we separate above the density parameter of dark matter from the baryonic density parameter. The latter, as well as the radiation density parameter,  follows the standard evolution. We can also see that, unlike in the $\Lambda$CDM model, the  density parameter of the DE component evolves with the scale factor, implying also in a non-standard evolution of the DM.

The evolution of the DM and DE energy densities can be obtained from the conservation equations, which, in the presence of a coupling $Q (t)$ can be written respectively as

\begin{eqnarray}
&& \dot{\rho}_{x}+ 3H (1+w_x) \rho_x=Q (t),\label{cons-de}\\
&& \dot{\rho}_{c}+3H \rho_{c}=-Q (t),\label{cons-dm}
\end{eqnarray}
where $w_x$ is a constant barotropic EoS of DE. Note that for $Q (t) > 0$ the energy transfer occurs from DM to DE, while $Q (t) < 0$ indicates the reverse scenario (DE to DM).

To illustrate this possibility, we shift the interaction term $Q(t)$ to the left-hand side of each equation and rewrite the resulting expressions as follows,
\begin{eqnarray}\label{coupled-eqns}
\left\{\begin{array}{ccc}
\dot{\rho}_{x}+ 3H (1+w_x^{\rm eff}) \rho_x=0,\\[0.5em]
\dot{\rho}_{c}+3H (1+w_{c}^{\rm eff})\rho_{c}= 0.
\end{array}\right.
\end{eqnarray}
Above  $w_x^{\rm eff}$ and $w_{c}^{\rm eff}$ are defined as,

\begin{eqnarray}\label{weff}
w_x^{\rm eff} = w_x - \frac{Q(t)}{3H \rho_x}, \quad w_{c}^{\rm eff} = \frac{Q(t)}{3H \rho_c},
\end{eqnarray}
and they are respectively termed as the effective EoS parameters of DE and DM. Thus, one can see that a coupling between DE and DM is equivalent to a non-interacting scenario between DE and DM with "non-constant" (or dynamical) effective EoS parameters as shown in  (\ref{weff}). This has also been observed in several articles, for instance in Refs. ~\cite{Pan:2020bur,Yang:2025ume,You:2025uon}.

Allowing both interpretations of these models creates room for a wider range of new-physics mechanisms  that could account for the underlying dynamics. On one hand, a coupling within the dark sector is a natural expectation in the context of quantum field theory. Such interactions help alleviate the coincidence problem and incorporate the benefits of interacting dark energy models, which have been extensively explored in the literature. On the other hand, the notion of a dynamical dark energy with a dynamical EoS has garnered support from various theoretical frameworks. Instabilities associated with a cosmological constant, $\Lambda$, have been suggested for a long time ago in different contexts \cite{Mottola:1985qt,Mazur:1986et,Tsamis:1992sx, Tsamis:1994ca,Polyakov:2012uc,Obied:2018sgi,Brandenberger:2018fdd}. By considering  time-dependent DE equations of state coming  from Eq.(\ref{weff}),  we contemplate a broader range of parameterizations than those typically explored in the literature \cite{DESI:2025fii}, encompassing  variations of vacuum decay, phantom and quintessence behaviors. Furthermore, we extend the analysis commonly considered by also allowing a variation in the dark matter equation of state, as described in Eq. (\ref{weff}). The interplay between these two interpretations -- dark sector coupling and dynamical dark energy -- may offer a broader stage for model builders exploring the fundamental physics behind these phenomena.

The nature of the effective EoS parameters depends on the type of interaction function used. For the local interaction rates which do not involve $H$ or other external time-dependent quantities, as in the case of  this work, both the effective EoS parameters are time-dependent, irrespective of the choice of the interaction function. Moreover, if  $Q(t)$ exhibits  oscillating behavior, this characteristic will be reflected in  the effective EoS parameters as well. Alternatively, if the equation of state
is treated as a dynamical variable with oscillating nature~\cite{Escamilla:2024fzq,Alvarez:2025kma}, then this can be interpreted as an oscillating interaction in the dark sector.
Concerning  DM, a possible dynamical behavior, or alternatively the presence of a non-zero EoS of DM (abbreviated as "non-cold" DM following  Ref. \cite{Pan:2022qrr}),  has recently received significant attention in the cosmology community~\cite{Pan:2022qrr,Naidoo:2022rda,Yao:2023ybs,Sardar:2024gmo,Yang:2025ume,Kumar:2025etf,Chen:2025wwn,Giani:2025hhs,Abedin:2025dis,Li:2025eqh,Li:2025dwz,Wang:2025hlh,Yao:2025twv,Braglia:2025gdo,Poulot:2024sex}. Such a non-cold behavior of DM may indicate the presence of an interaction in the dark sector.
It can be further noticed that the sign of $w_{c}^{\rm eff}$ could be either positive or negative depending on the sign of $Q(t)$.
On the other hand,  the nature of effective DE could alter with the sign of $Q$ and one can expect more from this interacting mechanism:

\begin{enumerate}
\item  If $w_x$ mimics the cosmological constant (i.e. $w_x =-1$), then phantom behavior will happen for $Q > 0$ (transfer of energy from DM to DE), whereas, for $Q < 0$ (transfer of energy from DE to DM), $w_{x}^{\rm eff}$ could behave like a quintessential DE along with a dynamical DM with negative effective EoS parameter.

\item A phantom ($w_x < -1$) in this interacting framework may lead to an effective DE EoS which could be quintessential if $Q < 0$ or more phantom if $Q > 0$.

\item  A quintessential DE ($w_x> -1$) may lead to an effective DE EoS which could be phantom if $Q > 0$ (transfer of energy from DM to DE).
On the other hand, for $Q <0$ (transfer of energy from DE to DM), $w_x$ deviates significantly from $-1$ in the quintessential direction.

\item Mathematically, it is possible to realize a scenario with $w_{x}^{\rm eff} = 0$ which leads to $Q = 3 H w_x \rho_x$. In this case, $\rho_x $ evolves as $a^{-3}$, and consequently, since $w_c^{\rm eff} = w_x \rho_x/\rho_c < 0$ (as $w_x <0$), the roles of DE and DM are reversed. This possibility is interesting and could have many future implications. The strength of the coupling plays a critical role in this case because if the decay rate from DE to DM increases significantly, this scenario could be realized.
\end{enumerate}

Once the coupling function $Q (t)$ is given, using the conservation equations (\ref{cons-de}) and (\ref{cons-dm}), it is possible to obtain the energy density evolution  of the individual dark fluids either analytically or numerically. Consequently, using the Hubble equation in (\ref{Friedmann-eqns}), the evolution of the scale factor can be determined. The choice of the coupling expression in coupled DE–DM scenarios is a particularly delicate issue, as no fundamental theoretical framework currently exists to uniquely determine the interaction function in a general setting.
In this work, we consider the following model,~\cite{Clemson:2011an,Li:2019san,Yang:2020zuk}
\begin{eqnarray}\label{interaction-model}
    Q (t) = \Gamma \rho_x,
\end{eqnarray}
where $\Gamma$ is the coupling parameter, which has a constant value and dimension of the Hubble parameter. This particular form  represents a local interaction rate which depends only on the energy density of DE, i.e., the intrinsic nature of DE.  Following the direction of energy transfer,  $\Gamma >0$ indicates the transfer of energy from DM to DE, and $\Gamma <0$ denotes the opposite. Now, for the present interaction function, one can now see the explicit expressions of the effective EoS parameters as,
\begin{eqnarray}\label{eff-eos-dm-de}
 w_{x}^{\rm eff} = w_x - \Gamma/3H, \quad \quad w_{c}^{\rm eff} = (\Gamma/3H) \times \rho_x \rho_c^{-1}.
\end{eqnarray}

Before discussing the impact of the present interacting model at the perturbative level, we present a formalism discussing the challenges in obtaining analytical solutions for such local interaction rates. In order to do so, let us denote $\rho_t \equiv \rho_c +\rho_x$, and with this notation, the conservation equations (\ref{cons-de}) and (\ref{cons-dm}) lead to, $\dot{\rho}_t + 3 H (\rho_t + w_x \rho_x) =0$ and it allows us to express $\rho_x$ and $\rho_c$ as ~\cite{Pan:2016ngu,Yang:2018xlt}

\begin{eqnarray}
    \rho_x = - \frac{1}{w_x} \left(\frac{d\rho_t}{dN} +\rho_t\right),\\
    \rho_c  = \rho_t +  \frac{1}{w_x} \left(\frac{d\rho_t}{dN} +\rho_t\right)
\end{eqnarray}
where $N = 3 \ln a$. Now, using the above expression for $\rho_x$ in Eqn. (\ref{cons-de})
one arrives at the following equation,

\begin{eqnarray}\label{diff-eqn-gen-Q}
  \frac{d^2\rho_t}{dN^2} + (2+w_x) \frac{d\rho_t}{dN} + (1+w_x) \rho_t = -\frac{Qw_x}{3H},
\end{eqnarray}
which for the present interaction model (\ref{interaction-model}) assumes the form,

\begin{align}\label{diff-eqn-present-Q}
    \frac{d^2\rho_t}{dN^2} + \left(2+w_x - \frac{\Gamma}{3H (N)} \right) \frac{d\rho_t}{dN} \nonumber\\+ \left(1+w_x - \frac{\Gamma}{3H (N)} \right) \rho_t =0.
\end{align}
From the above equation, we can see that finding the analytical solutions for the dark components is complicated, and a numerical method is convenient for this purpose.
We can also see that if the coupling between DE and DM was described by $\widetilde{Q} \equiv (\Gamma\times H)\rho_{x}$, instead of Eq.(\ref{interaction-model}), then one could easily solve the above equation for constant $\Gamma$. However, in this case, the interaction function would correspond to a global one, which is not our focus for the present article (we refer to Refs. \cite{Pan:2016ngu,Yang:2018xlt} for more details).

We now turn to the perturbative level, where we introduce the remaining equations to be employed in testing the models against the observational data.

\subsection{Perturbed Equations}
\label{subsec:pert_eq}

The evolution of perturbations on the dark sector is fundamental to understanding if a model can accurately describe the observable universe. When dealing with coupled models, one has to be particularly careful to check for stability conditions on larger scales. It is well-known that interactions on the dark sector can introduce instabilities that arise at perturbative level ~\cite{Valiviita:2008iv, Jackson:2009mz, Yang:2018xlt}, even for a weak dark sector coupling \cite{Valiviita:2008iv}.
We consider the perturbed line element in the synchronous gauge, given by
\begin{eqnarray}
\label{perturbed-metric}
ds^2 = a^2(\tau) \left [-d\tau^2 + (\delta_{ij}+h_{ij}) dx^idx^j  \right],
\end{eqnarray}
where $\tau$ denotes the conformal time, $\delta_{ij}$ denotes the unperturbed metric tensor and $h_{ij}$ is the perturbed metric tensor.
Following \cite{Yang:2018xlt}, assuming zero anisotropic stress, and no momentum transfer in the DM rest frame, the linear perturbation equations in Fourier space in the synchronous gauge for the case $w_x \neq -1$, corresponding to the phantom (\textbf{IDEphan}) and quintessential (\textbf{IDEquin}) scenarios are described by:

    \begin{eqnarray}
        \delta _x^{\prime } =&&-(1+w_x)\left( \theta _x+\frac{h^{\prime }}{2}%
        \right) -3\mathcal{H}(c_{s, x}^{2}-w_x)\Bigg[ \delta _x \nonumber \\  &&+3\mathcal{H}%
        (1+w_x)\frac{\theta _x}{k^{2}}\Bigg] -3\mathcal{H}w_x^{\prime}\frac{\theta_x}{k^{2}}\nonumber \\
        &&+3\mathcal{H}\Gamma a(c_{s,x}^{2}-w_x)\frac{\theta _x}{k^{2}}, \label{IDE_deltax}\\
        \theta _x^{\prime } =&&-\mathcal{H}(1-3c_{s,x}^{2})\theta _x+\frac{%
            c_{s,x}^{2}}{(1+w_x)}k^{2}\delta _x \nonumber \\ && +\Gamma a \left[ \frac{%
            \theta _c-(1+c_{s,x}^{2})\theta _x}{1+w_x}\right], \label{IDE_thetax}\\
            \label{IDE_deltac}
        \delta _c^{\prime } =&&-\left( \theta _c+\frac{h^{\prime }}{2}%
        \right)+\Gamma a \frac{\rho _x}{\rho _c}\left( \delta _c-\delta_x\right), \\
            \label{IDE_thetac}
        \theta _c^{\prime } =&&-\mathcal{H}\theta_c.
    \end{eqnarray}
    \label{IDE_pert}

While, for the interacting vacuum scenario (\textbf{IVS}) $w_x = -1$, the DE has no contribution at the perturbative level and the equivalent equations for this case are \cite{Yang:2020zuk}:

    \begin{eqnarray}
            \label{IVS_deltac}
        &&\delta _c^{\prime } =-\left( \theta _c+\frac{h^{\prime }}{2}%
        \right)+\Gamma a \frac{\rho _x}{\rho _c}\delta _c, \\
           \label{IVS_thetac}
        &&\theta _c^{\prime } =-\mathcal{H}\theta_c.
    \end{eqnarray}
    \label{IVS_pert}

Here, $\delta_{i}=\delta\rho_i/\rho_i$ and $\theta_i \equiv i\,k^j v_{j(i)}$ are respectively the density contrast and the divergence of the fluid velocity for the i-th component $i = \{c,x\}$. The trace part of the metric perturbations is denoted by $h$, while $k$ characterizes the respective scalar mode. Prime denotes derivative with respect to the conformal time  $\tau$, and $\mathcal{H}\equiv a'/a$ is the conformal Hubble factor. Lastly, the propagation of pressure fluctuations, or rather, the comoving sound speed squared, in the DE reference frame is

\begin{equation}
    c_{s,x}^2 \equiv \Big(\frac{\delta p_x}{\delta \rho_x}\Big)_{ref}.
\end{equation}

\begin{figure*}[h]
    \centering
    \includegraphics[width=0.47\textwidth]{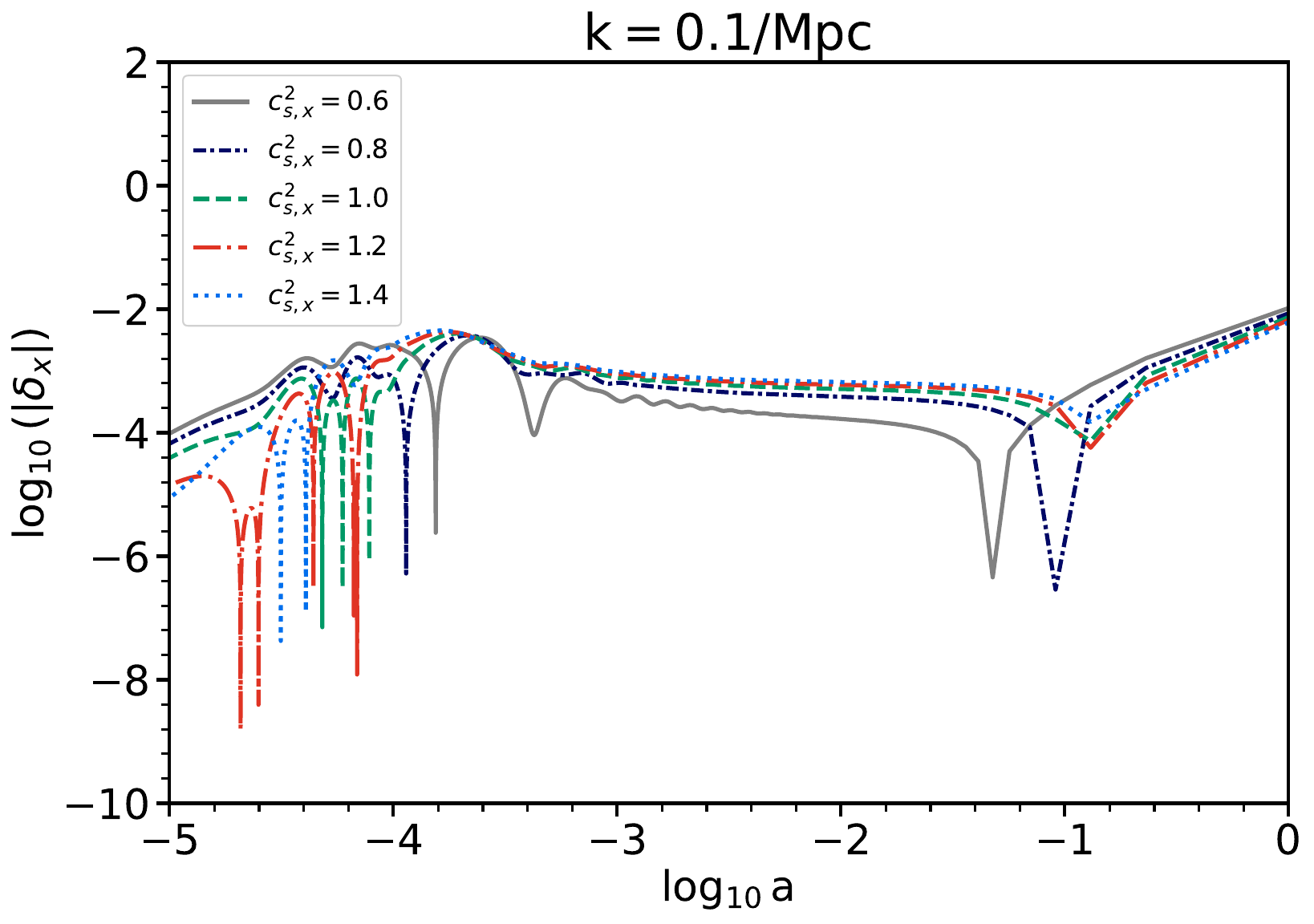}
    \includegraphics[width=0.47\textwidth]{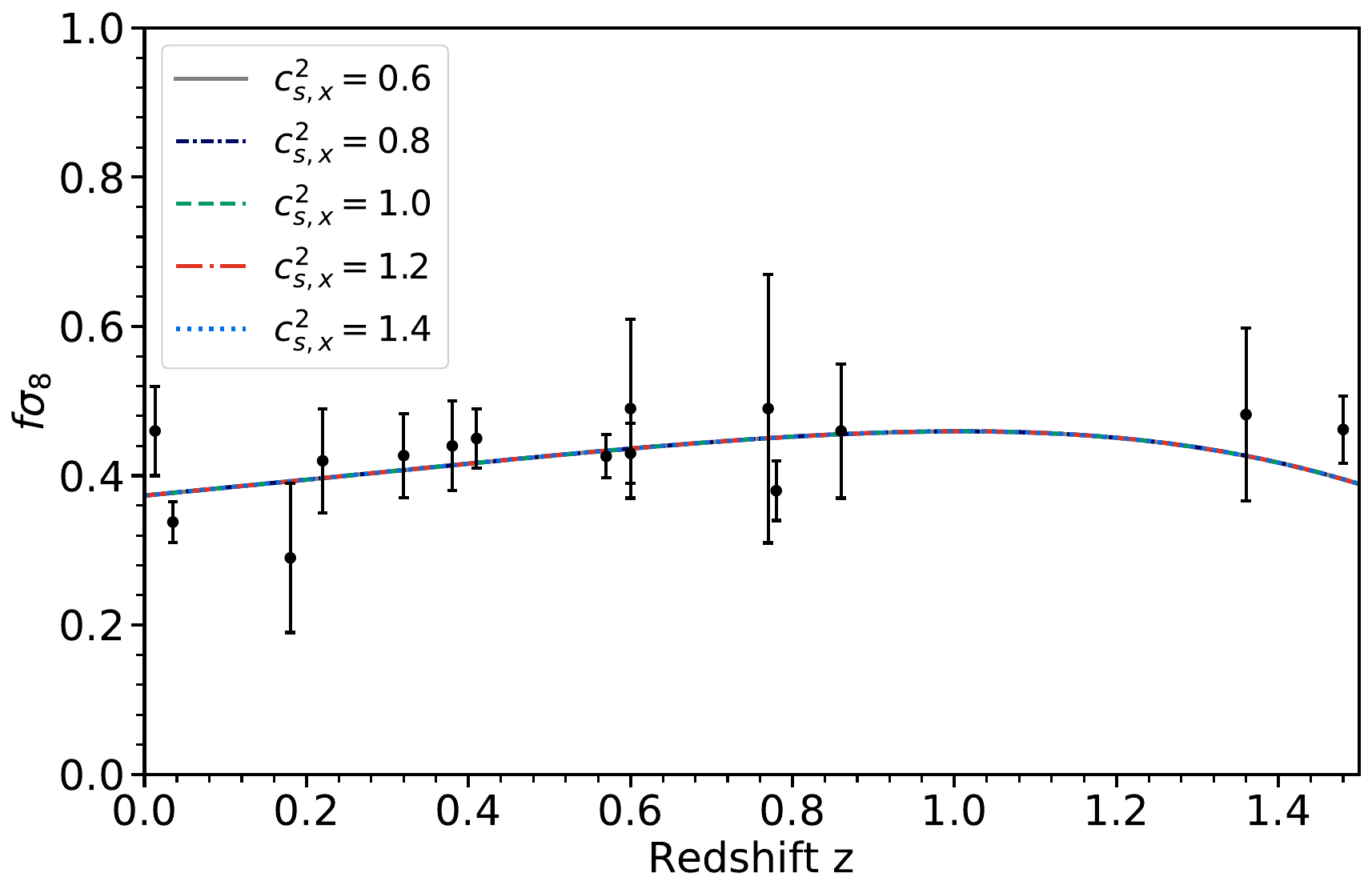}
    \caption{{\bf (IDEphan)} DE density contrast fluctuations $\delta_x$ at $k = 0.1/\mathrm{Mpc}$ (left) and $f\sigma_8$ curves (right) for variations of DE sound speed $c_{s,x}^2$. The $f\sigma_8$ measurements (black) can be found at Appendix \ref{sec-appendix}, in Table \ref{tab:fs8_values}. For constructing $\delta_x$, we set $w_x = -1.01$, $\Gamma/H_0 = -0.1$, $H_0=67.5$ km/s/Mpc, $\Omega_b h^2=0.022$ and $\Omega_c h^2=0.122$, and for the $f\sigma_8$ plot we fix $n_s =0.965$.   }
    \label{fig:idephan_deltax_fs8_cs2}
\end{figure*}

Early time instabilities that arise from pressure density fluctuations can be associated to a constant EoS parameter, $w_x$ \cite{Valiviita:2008iv}. We separate the {\bf IDEphan} and {\bf IDEquin} regimes, as well as treating {\bf IVS} as a special case, since perturbation equations in {\bf IVS} (Eqs. \ref{IVS_deltac}, \ref{IVS_thetac}) vanish for DE.
Likewise, such instabilities are also related to $c_{s,x}^2$. The sound horizon for DE can be described as
\begin{equation}
    r_{s}(z) = \int_{\infty}^{z}\frac{c_{s,x}(z')dz'}{H(z')(1 + z')}.
\end{equation}
A sound speed close to zero means the DE agglomerates similar to matter, while for $c_{s,x}^2 \simeq 1$ the pressure perturbations do not allow DE to cluster at sub-horizon scales. In a standard non-interactive vacuum scenario, the definition of the sound speed would lead to a negative value of  $c_{s,x}^2$, therefore we impose $c_{s,x}^2>0$ to avoid instabilities and nonphysical properties. Most of the time, it is convenient to set $c_{s,x}^2 = 1$, a value motivated by scalar field DE. While the perturbation equations in {\bf IVS} are independent of $c_{s,x}^2$, variations in the DE sound speed could directly impact the growth of structures for {\bf IDEphan} and {\bf IDEquin} scenarios. For instance, the growth rate $f \equiv \frac{d \, ln \, \delta_{m}}{d \, ln \, a}$ depends on the DM density contrast, which aside from the changes introduced by an interacting dark sector \cite{Caldera-Cabral:2009hoy}, it will also be modified in the presence of a clustering DE \cite{Mehrabi:2015hva,Yang:2026sound}.
As mentioned, for $c_{s,x}^2$ close to zero, DE perturbations may contribute to the gravitational potentials and can in principle affect structure-growth observables such as \(f\sigma_8\). However, in this work, we limit ourselves to the general case $c_{s,x}^2 \neq 1$. A careful analysis of the regime $c_{s,x}^2 \to 0$ will be left for future work.
In Fig. \ref{fig:idephan_deltax_fs8_cs2}, we present how the choice of $c_{s,x}^2$ could affect our model in the \textbf{IDEphan} case at $k = 0.1/\rm{Mpc}$.
We rescale our interaction parameter, defining $\Gamma/H_0$ as the dimensionless interacting parameter, and set $w_x = -1.01$, $\Gamma/H_0 = -0.1$ for this plot. On the left, we notice that higher values of $c_{s,x}^2$ aggravate instabilities in the early universe, causing them to happen at earlier times. On the right panel of Fig. \ref{fig:idephan_deltax_fs8_cs2}, we calculate the growth of structures $f\sigma_8$, where $\sigma_8$ is the amplitude of mass fluctuations in a $8\rm{Mpc/h}$ scale. We show that, in this case, the choice of $c_{s,x}^2$ for fixed $w_x$ and $\Gamma/H_0$ do not affect the growth rate. We compare these curves with $f\sigma_8$ measurements found in Table \ref{tab:fs8_values}, adapted from \cite{Toda:2024fgv}.

\begin{figure*}[h]
    \centering
    \includegraphics[width=0.47\textwidth]{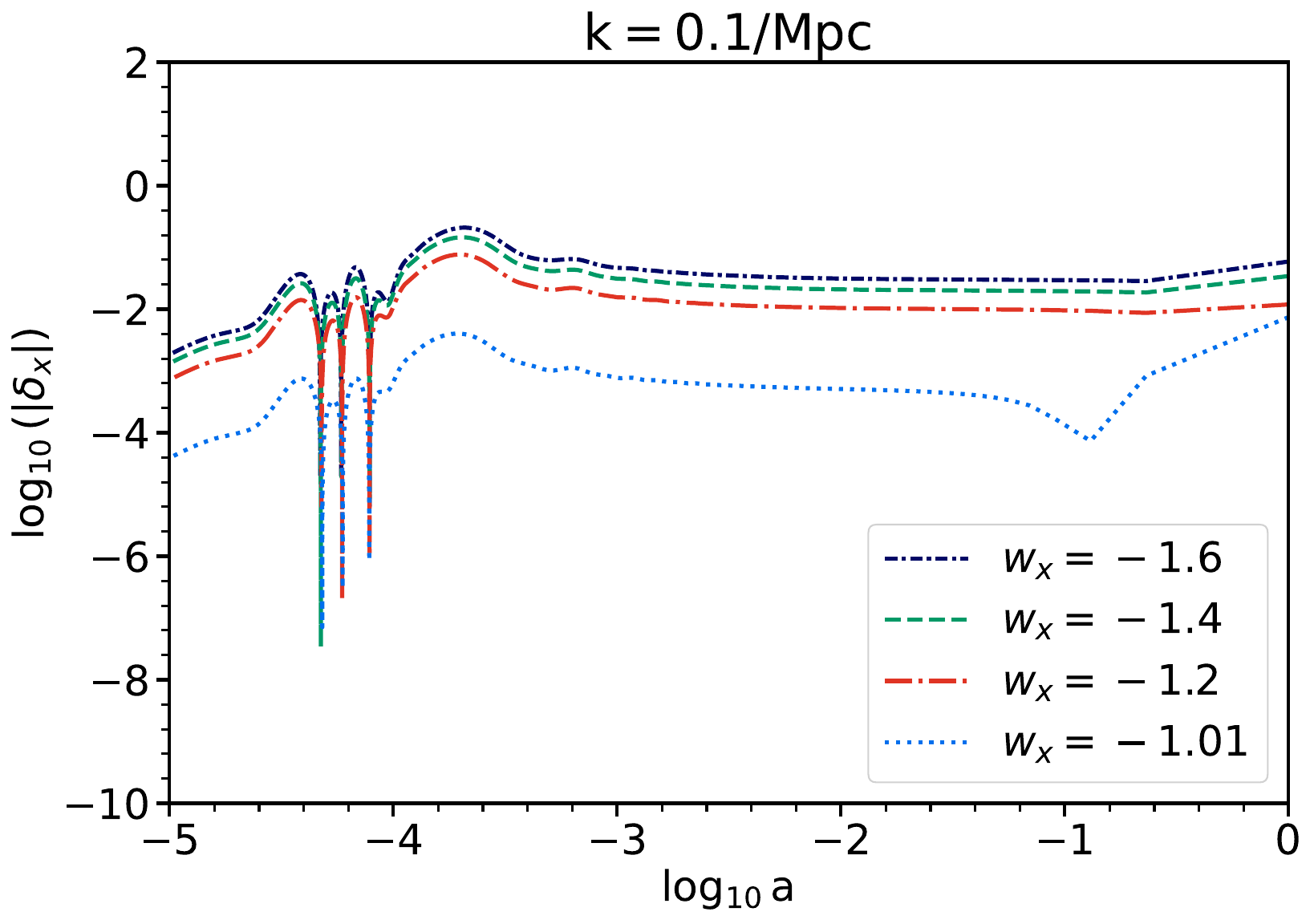}
    \includegraphics[width=0.47\textwidth]{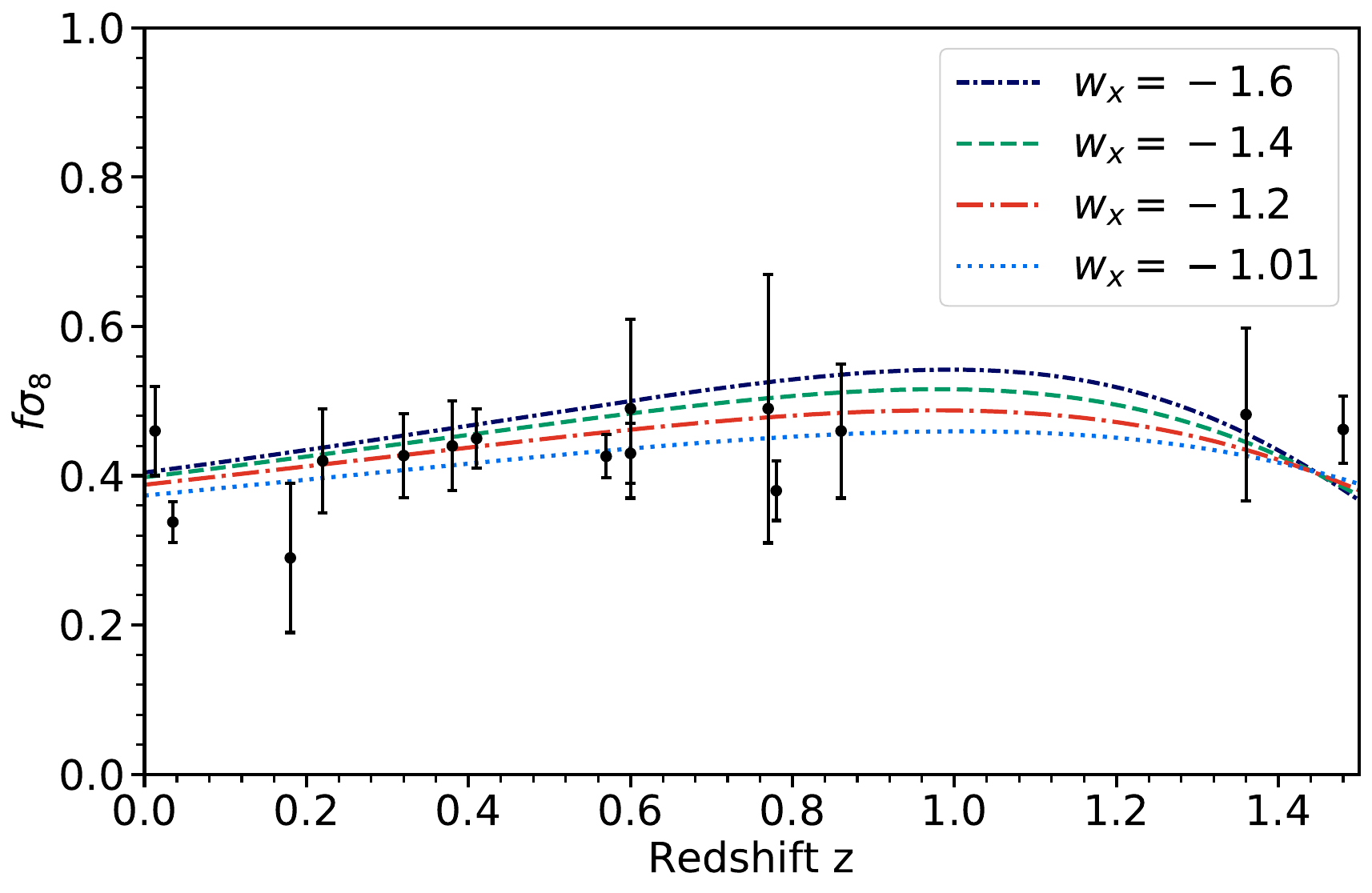}
    \caption{{\bf (IDEphan)} DE density contrast fluctuations $\delta_x$ at $k = 0.1/\mathrm{Mpc}$ (left) and $f\sigma_8$ curves (right) for variations of EoS parameter $w_{x}$. The $f\sigma_8$ measurements (black) can be found at Appendix \ref{sec-appendix}, in Table \ref{tab:fs8_values}. For constructing $\delta_x$, we set $\Gamma/H_0 = -0.1, c_{s,x}^2 = 1$, $H_0=67.5$ km/s/Mpc, $\Omega_b h^2=0.022$ and $\Omega_c h^2=0.122$, and for the $f\sigma_8$ plot we fix $n_s =0.965$.  }
    \label{fig:idephan_deltax_fs8_w0}
\end{figure*}

\begin{figure*}[h]
    \centering
    \includegraphics[width=0.47\textwidth]{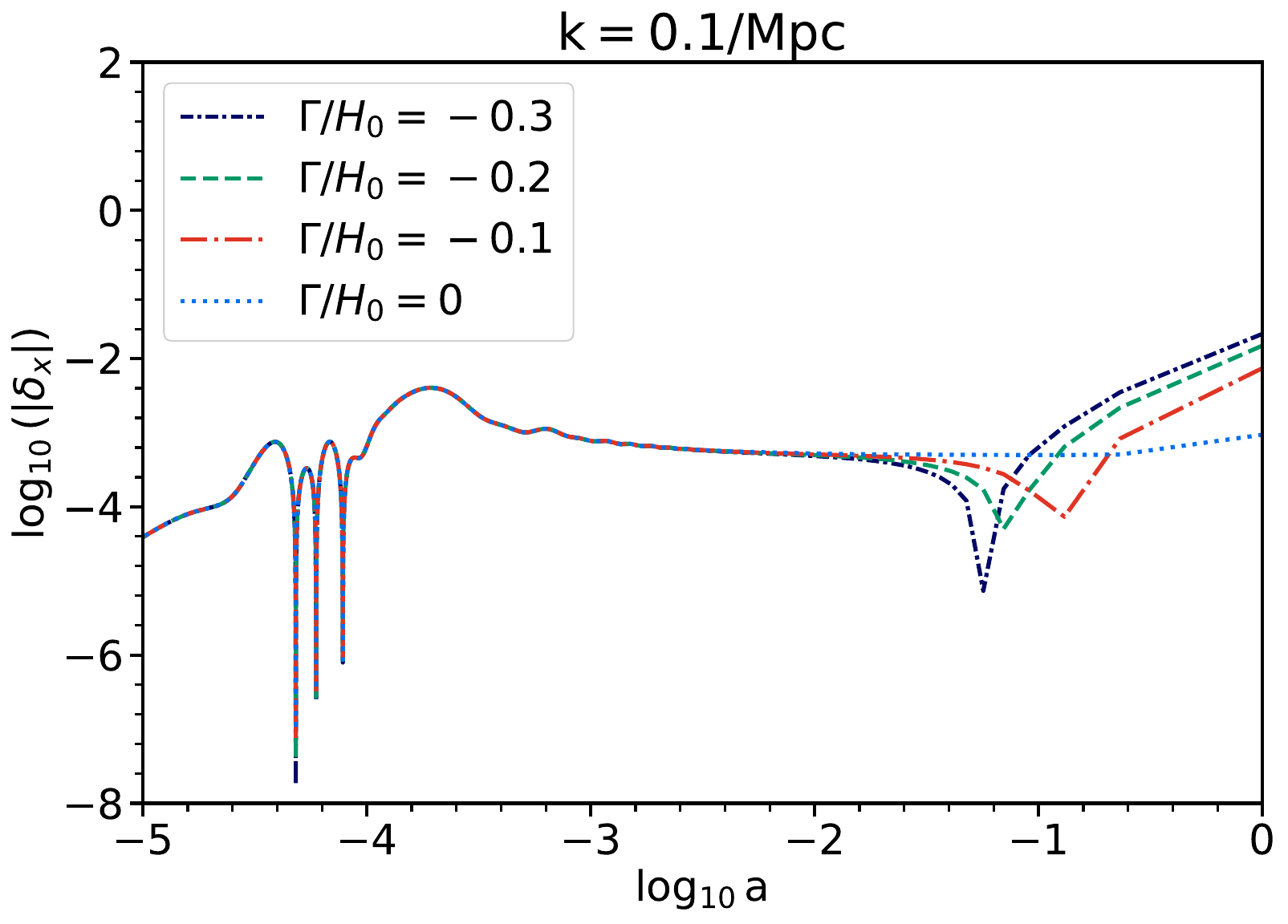}
    \includegraphics[width=0.47\textwidth]{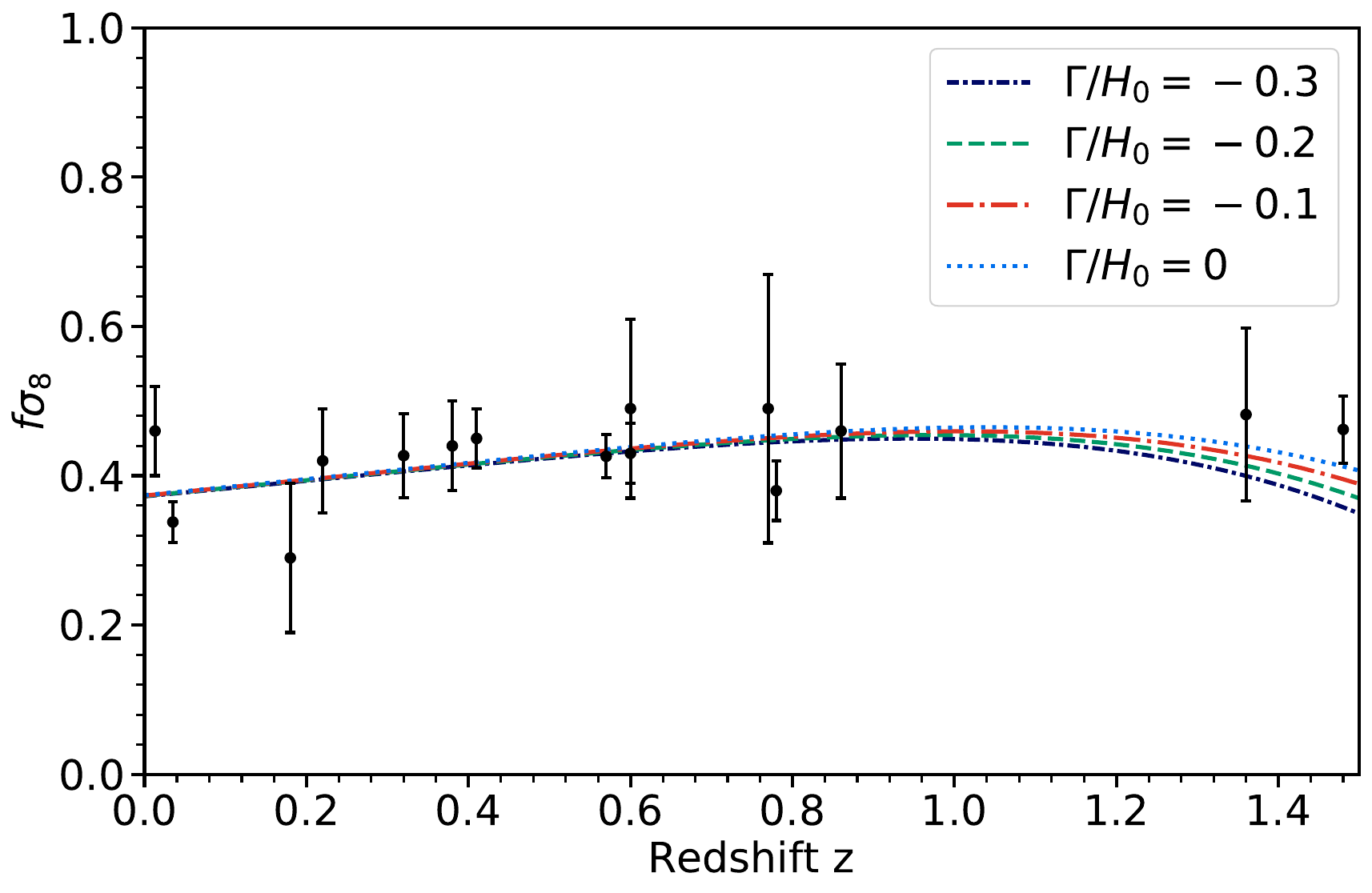}
    \caption{{\bf (IDEphan)} DE density contrast fluctuations $\delta_x$ at $k = 0.1/\mathrm{Mpc}$ (left) and $f\sigma_8$ curves (right) for variations of $\Gamma/H_0$. The $f\sigma_8$ measurements (black) can be found at Appendix \ref{sec-appendix}, in Table \ref{tab:fs8_values}. For constructing $\delta_x$, we set $w_x = -1.01, c_{s,x}^2 = 1$, $H_0=67.5$ km/s/Mpc, $\Omega_b h^2=0.022$ and $\Omega_c h^2=0.122$, and for the $f\sigma_8$ plot we fix $n_s =0.965$. }
    \label{fig:idephan_deltax_fs8_xi}
\end{figure*}

Additionally, we study variations for both $w_{x}$ and $\Gamma/H_0$ in Figures \ref{fig:idephan_deltax_fs8_w0} and \ref{fig:idephan_deltax_fs8_xi}. This time, in Fig. \ref{fig:idephan_deltax_fs8_w0}, we can observe that increasing $w_x$ values shift the curves downwards, and the evolution of $f\sigma_{8}$ changes significantly. As for $\Gamma/H_0$ (Fig. \ref{fig:idephan_deltax_fs8_xi}), we notice that $\Gamma/H_0$ values mainly affect the density contrast $\delta_x$ at later times, when DE becomes more relevant, while $f\sigma_8$ is sensitive to the interaction strength at high redshifts. We find somewhat similar results, now with $w_x$ and $\Gamma/H_0$ behaviors inverted, for the {\bf IDEquin} case in Fig. \ref{fig:idequin_deltax_fs8_cs2}. The {\bf IVS} scenario is displayed in Fig. \ref{fig:ivs_fs8_xi}, with positive $\Gamma/H_0$ values leading to higher $f\sigma_8$ and negative $\Gamma/H_0$ reaching lower values in comparison to $\Lambda$CDM ($\Gamma/H_0 = 0$). Additional {\bf IDEquin} results are shown in Figs. \ref{fig:idequin_deltax_fs8_w0} and \ref{fig:idequin_deltax_fs8_xi}.


Let us now discuss the spectra of CMB temperature anisotropies (CMB TT) and the matter power spectra of the interacting model considered in this study.

\subsection{CMB Spectrum}

\begin{figure}
    \centering
    \includegraphics[width=0.47\textwidth]{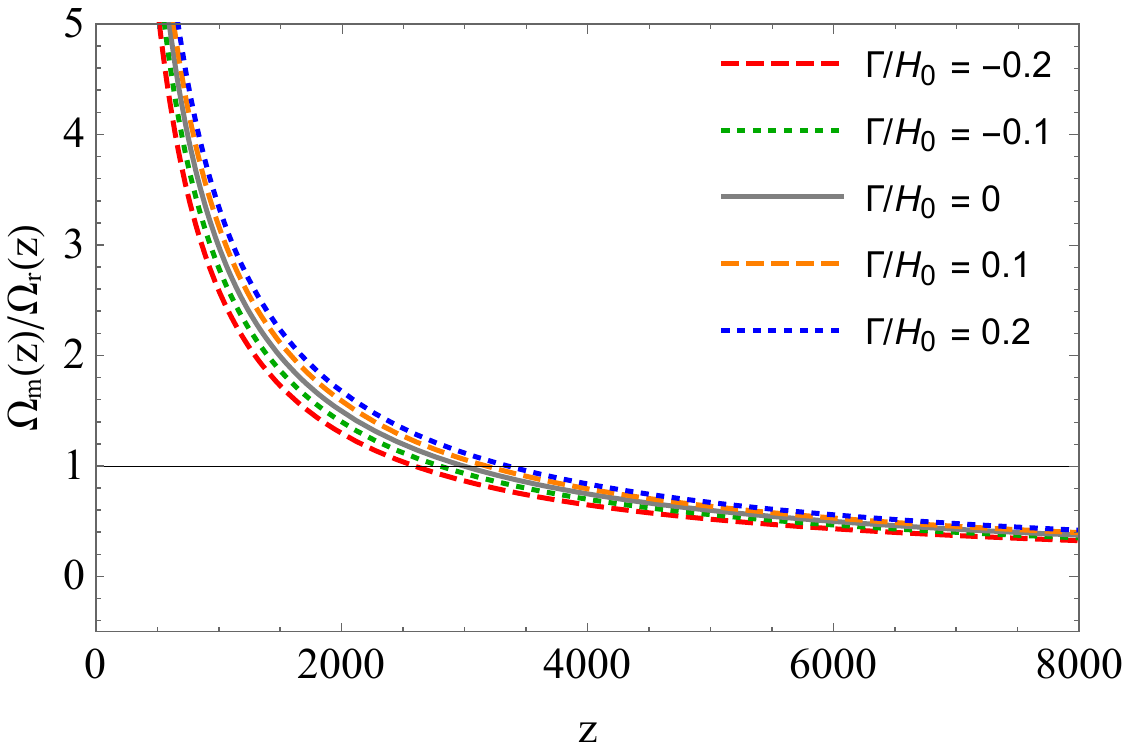}
    \caption{{\bf (IVS)}  Evolution of (total) matter over radiation, $\Omega_{m}(z)/\Omega_{r}(z)$ for variations of dimensionless interacting parameter $\Gamma/H_0$. The radiation-matter equality is indicated with a solid black line. $\Gamma/H_0 = 0$ corresponds to the $\Lambda$CDM model. While drawing the curves, we have fixed $H_0=67.5$ km/s/Mpc, $\Omega_b h^2=0.022$ and $\Omega_c h^2=0.122$.}
    \label{fig:rad_mat_eq_IVS}
\end{figure}

\begin{figure*}
    \centering
    \includegraphics[width=0.47\textwidth]{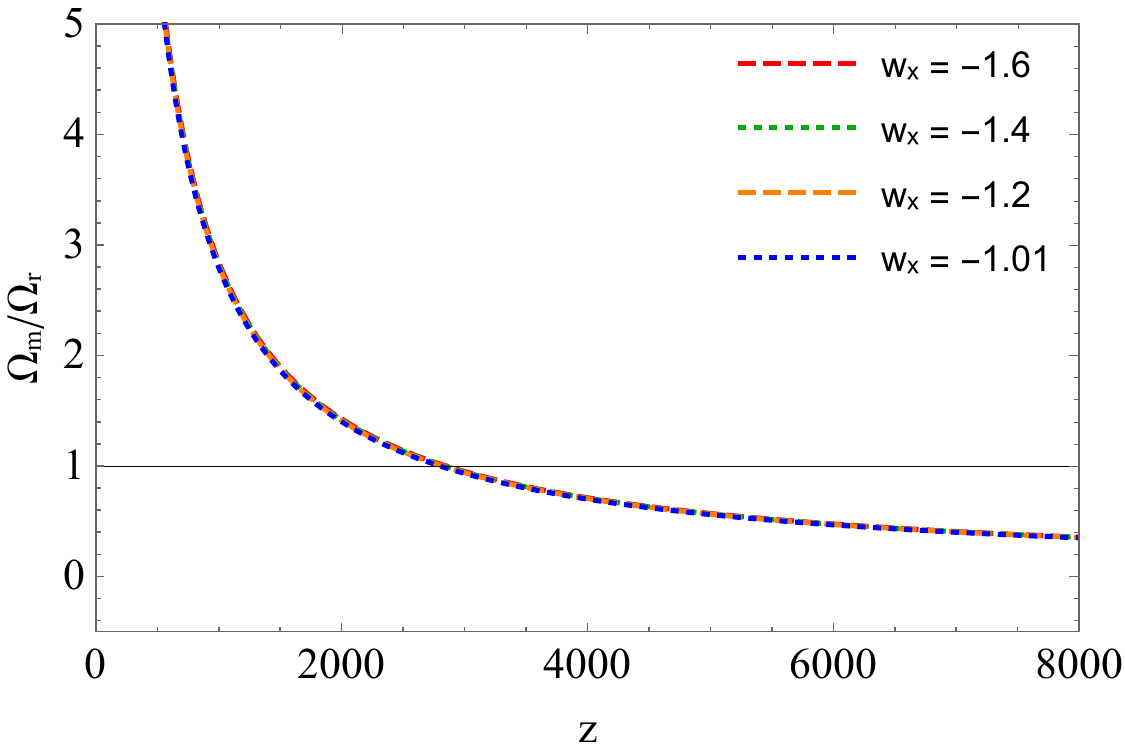}
    \includegraphics[width=0.47\textwidth]{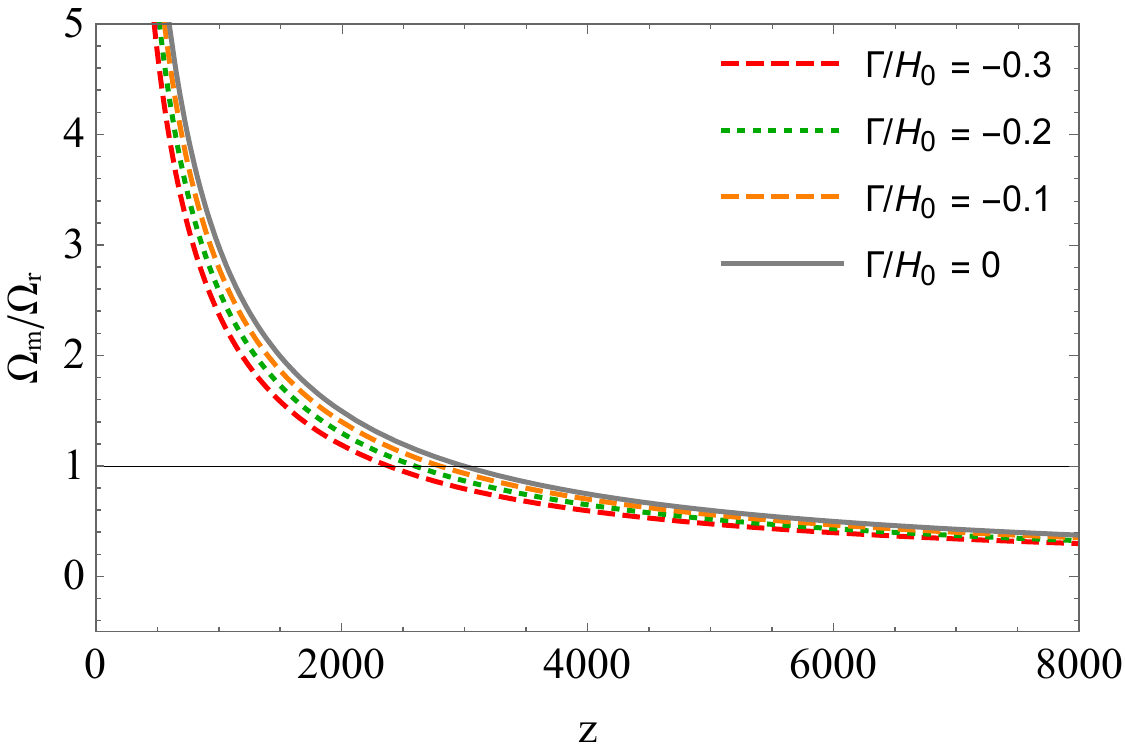}
    \includegraphics[width=0.47\textwidth]{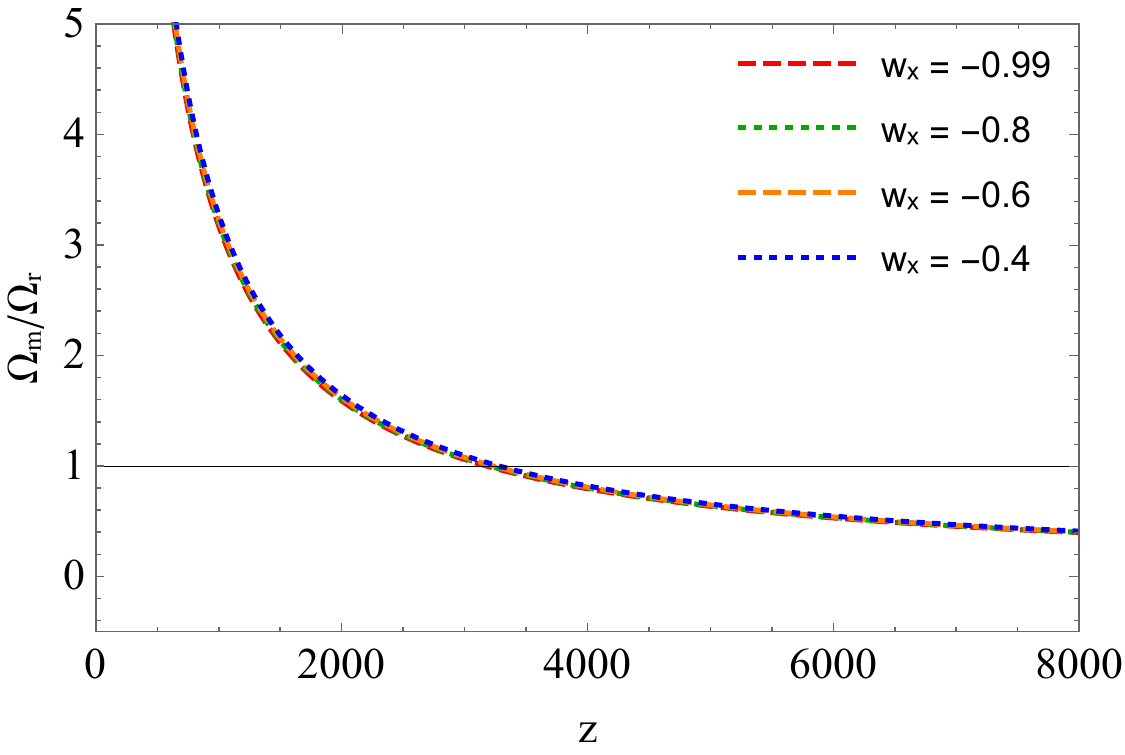}
    \includegraphics[width=0.47\textwidth]{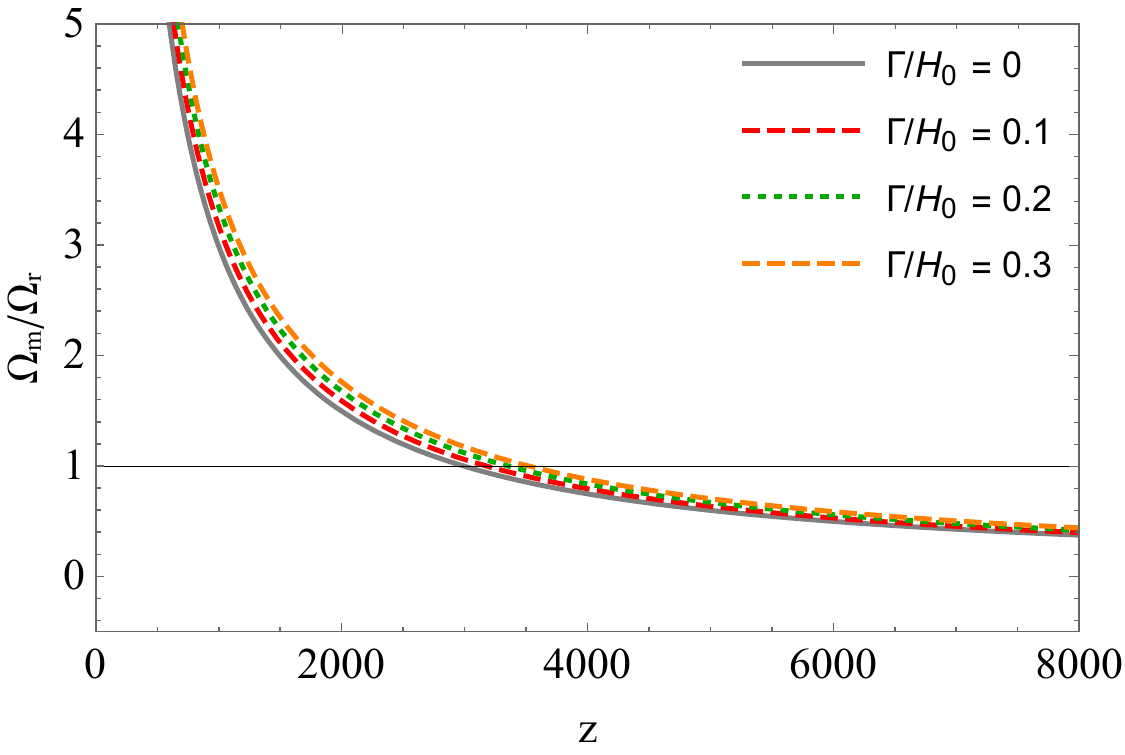}
    \caption{{\bf (IDEphan and IDEquin)} Evolution of (total) matter over radiation, $\Omega_{m} (z)/\Omega_{r} (z)$ for variations of EoS parameter $w_{x}$ and dimensionless interaction parameter $\Gamma/H_0$ in the {\bf IDEphan} (top) and {\bf IDEquin} (bottom) models. Higher absolute values of $\Gamma/H_0$ are equivalent to stronger coupling in both scenarios. The radiation-matter equality is indicated with a solid black line. $\Gamma/H_0 = 0$ corresponds to the $\Lambda$CDM model. While drawing the curves, we have fixed $H_0=67.5$ km/s/Mpc, $\Omega_b h^2=0.022$ and $\Omega_c h^2=0.122$.    }
    \label{fig:rad_mat_eq_IDEphan_IDEquin}
\end{figure*}

We now focus on the effects of the interaction on the CMB TT spectrum and the matter power spectrum. This analysis is important for understanding how the presence of an interaction between DE and DM could affect the large-scale structure of the universe. Before describing the effects on the CMB TT and matter power spectra,  we recall the sound horizon at the time of photon-baryon decoupling $r_{s}(z_{\star}) \equiv \int_{\infty}^{z_{\star}} c_{s}(z')dz'/[H(z')(1 + z')]$, with $z_{\star}$ the redshift at photon decoupling and $c_{s}$ is the sound speed of the photon-baryon fluid. An interacting dark sector changes the background evolution of the universe, altering the time at which matter-radiation equality happens. Since this epoch is related to the sound horizon $r_{s}$, we should expect IDE to impact the CMB power spectrum as well. To illustrate this effect, we plot the (total) matter-radiation ratio evolution for all three models considered in this work in Figures \ref{fig:rad_mat_eq_IVS} and \ref{fig:rad_mat_eq_IDEphan_IDEquin}. In the {\bf IVS} scenario (Fig. \ref{fig:rad_mat_eq_IVS}), variations in $\Gamma/H_0$ can slightly alter the matter-radiation equality. Because of the direction of energy flow between the dark components, quantified through the sign of $\Gamma/H_0$, their standard evolution is modified. As a consequence, the epoch of matter-radiation equality differs from that in the non-interacting scenario. The top and bottom of Fig. \ref{fig:rad_mat_eq_IDEphan_IDEquin} present, respectively, the {\bf IDEphan} and {\bf IDEquin} models. In these cases, we notice that the choice of $w_{x}$ (left) is not pivotal to determining matter-radiation equality, while increasing values of interaction strength (right) significantly shift the curves down (up), leading to later (earlier) matter-radiation equality for the
{\bf IDEphan} ({\bf IDEquin}) scenario.

\begin{figure*}
    \centering
    \includegraphics[width=0.47\textwidth]{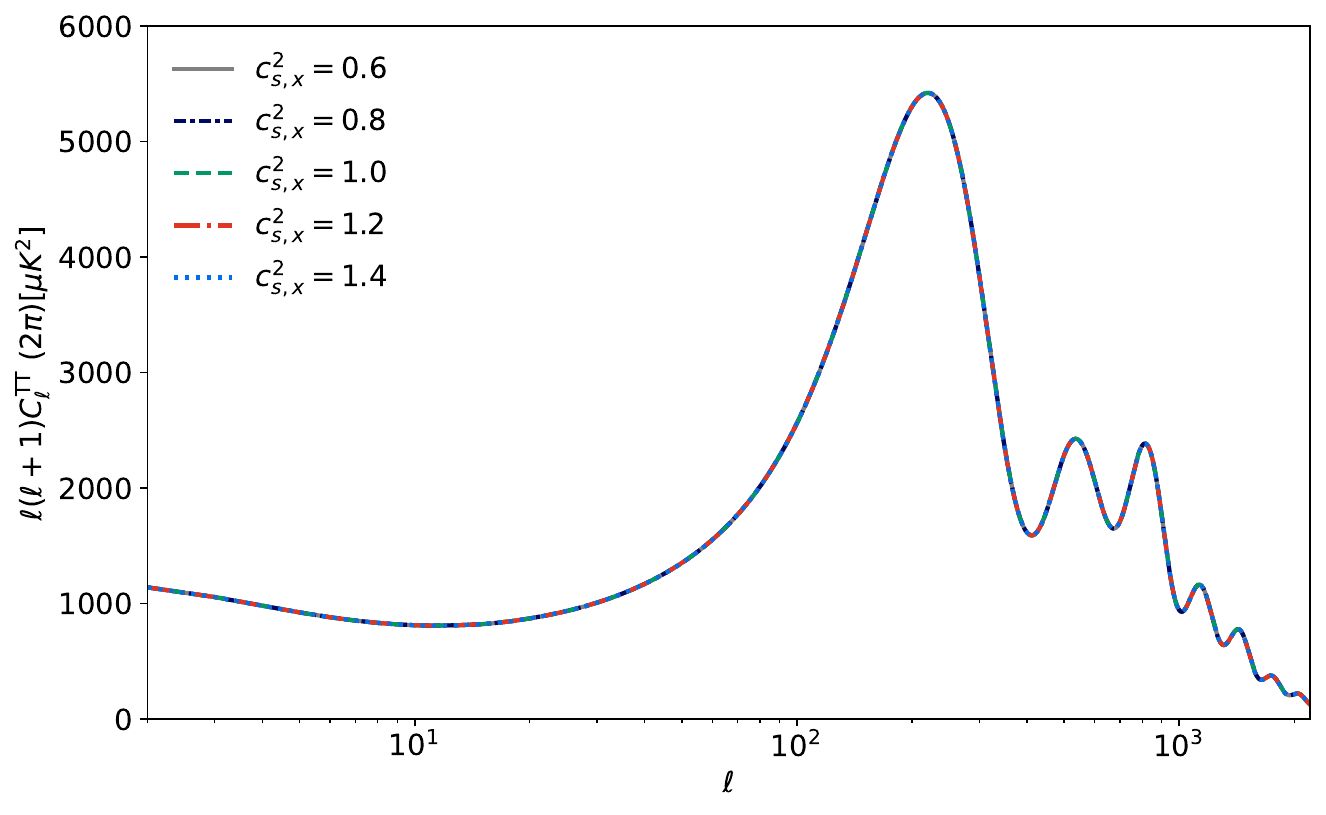}
    \includegraphics[width=0.47\textwidth]{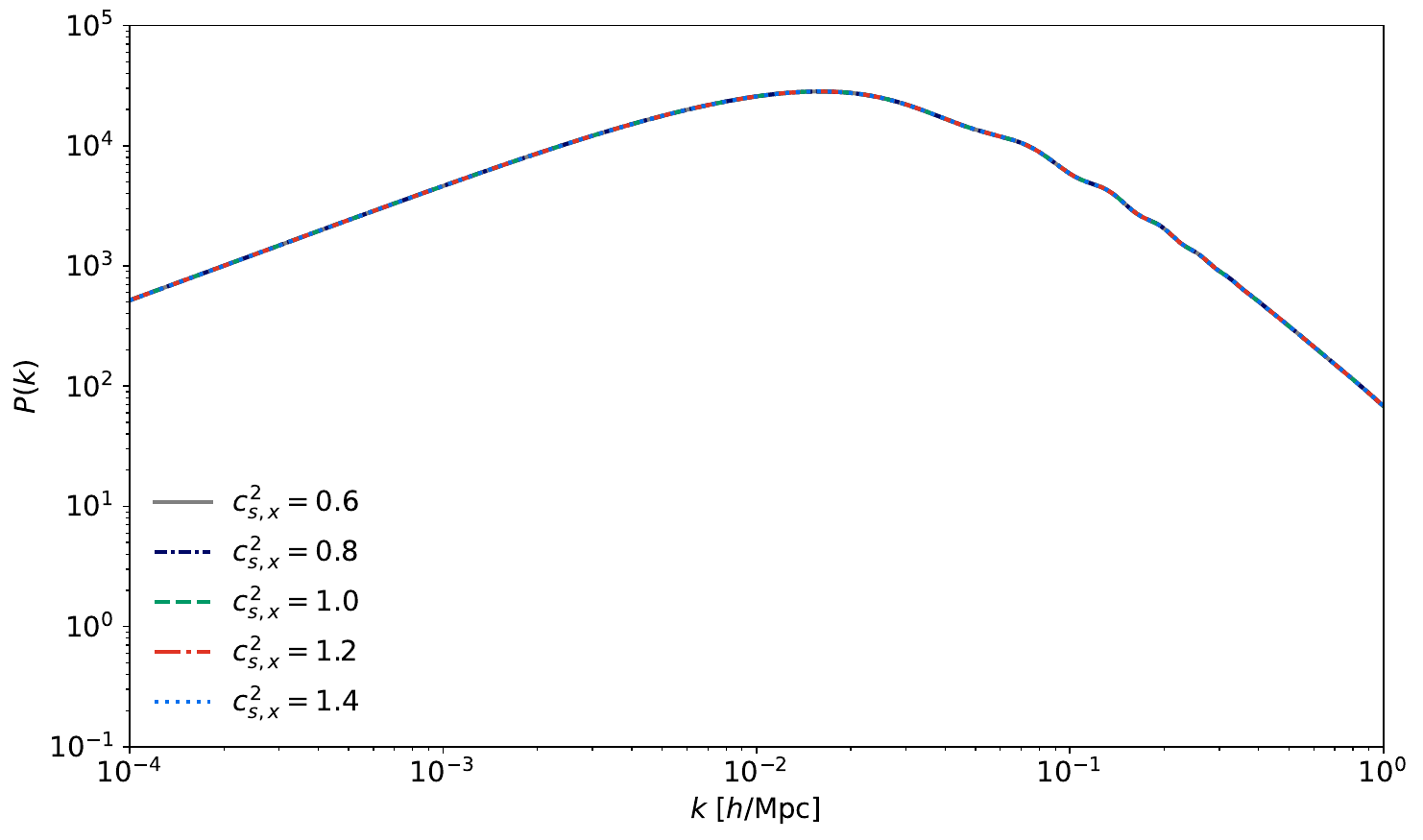}
    \caption{{\bf (IDEphan)} CMB TT spectrum (left) and matter power spectrum (right) for different values of $c_{s,x}^2$. We have fixed $\Gamma/H_0 =-0.1$ and $w_{x} = -1.01$. The mean values of the other parameters, e.g. $\Omega_bh^2$, $\Omega_ch^2$ and $H_0$ required to generate the plots are taken from the combined analysis \cmbdesipantheon (Table~\ref{tab:IDEphan}).  }
    \label{fig:idephan-cs2}
\end{figure*}

We begin discussing the CMB and matter power spectra by giving special attention to $c_{s,x}^2$, which is usually fixed to $c_{s,x}^2=1$ in perturbative analyses. In the present article, we seek an answer to how different values of $c_{s,x}^2$ could affect the observables.  As mentioned in Sec. \ref{subsec:pert_eq}, the {\bf IVS} case is insensitive to the DE sound speed, therefore, no changes in CMB and $P(k)$ are expected. Alternatively, we analyzed both spectra in the other two interacting scenarios for different values of $c_{s,x}^2$, namely, {\bf IDEphan} and {\bf IDEquin}. In Fig. \ref{fig:idephan-cs2}, we first plot CMB TT spectrum and matter power spectrum for varying $c_{s,x}^2$ in the {\bf IDEphan} case, where we use, as an example, a typical value of $\Gamma/H_0 = -0.1$. This figure evidences that even if we take different choices of $c_{s,x}^2$, CMB TT and matter power spectra are unaffected. Hence, constraints in $c_{s,x}^2$ are insensitive to the cosmic microwave background data, and setting $c_{s,x}^2$ to $1$ is justified in order to understand how the coupling affects other cosmological parameters. We find identical results for the {\bf IDEquin} scenario, which can be visualized in Fig. \ref{fig:idequin-cs2}, in Appendix \ref{sec-appendix}. Considering this, we fix $c_{s,x}^2 =1$ in the next analyses, including the statistical simulations. All the conclusions obtained in this work will remain robust regardless of the value of $c_{s,x}^2 =1$.
\begin{figure*}
    \centering
    \includegraphics[width=0.47\textwidth]{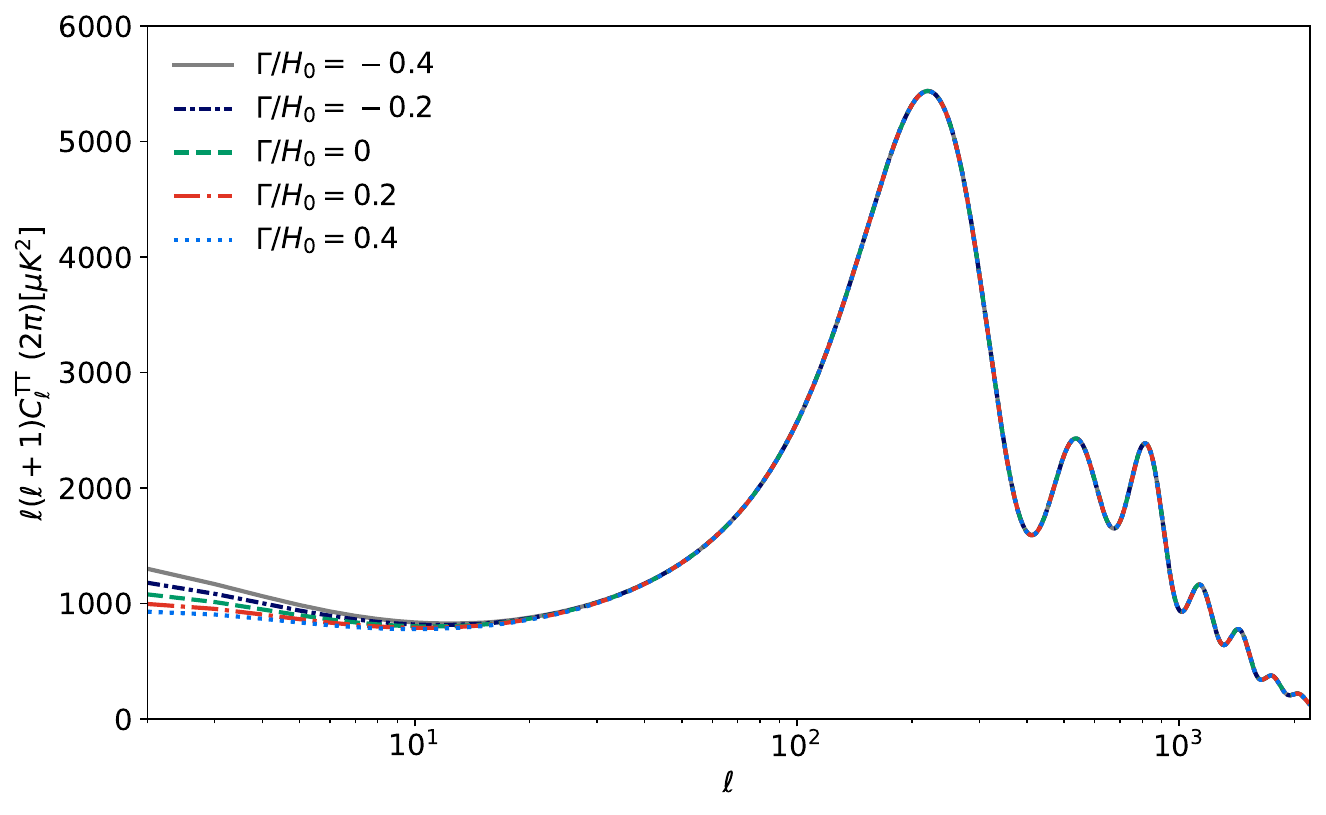}
    \includegraphics[width=0.47\textwidth]{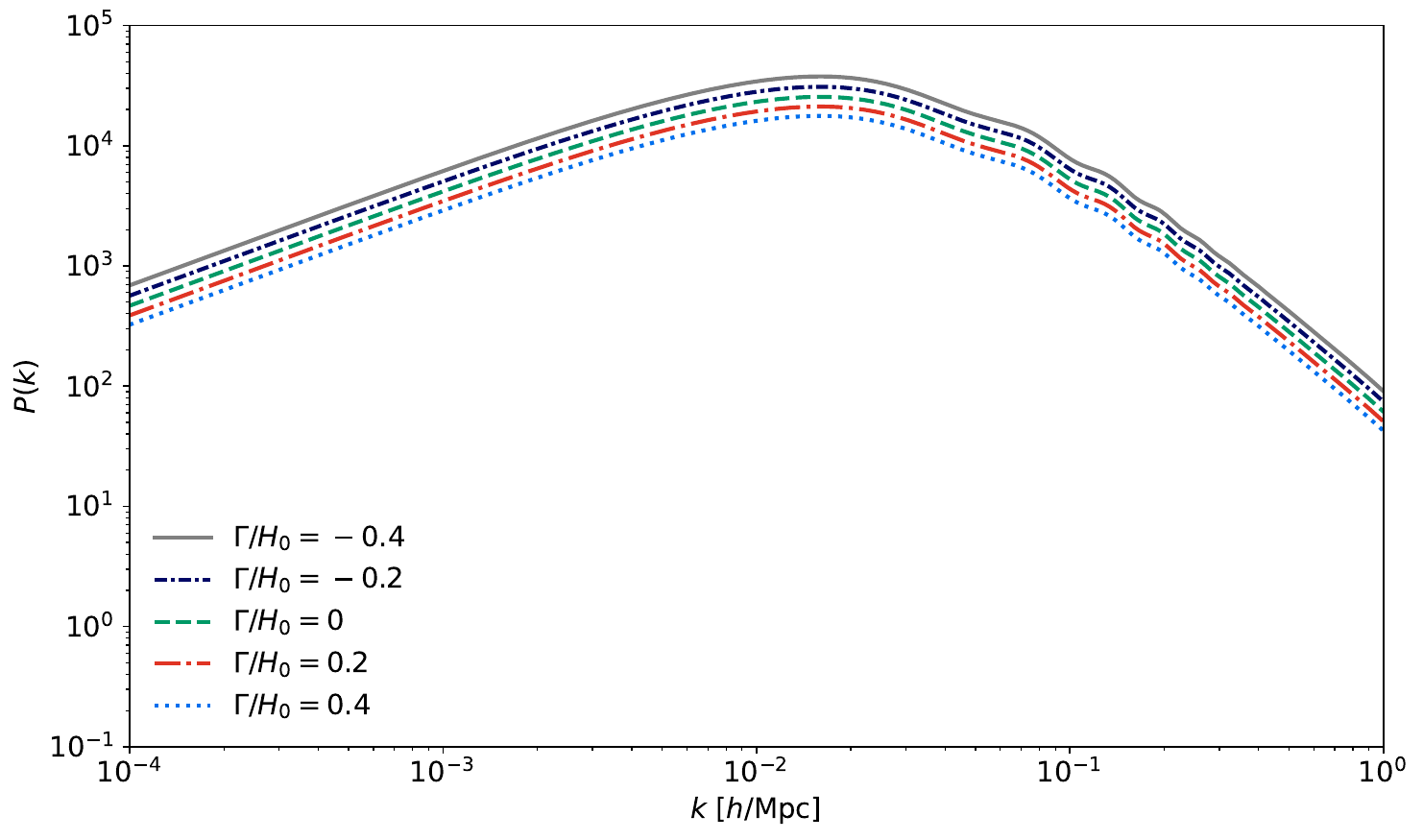}
    \caption{({\bf IVS}) The left graph corresponds to CMB TT spectrum and the right graph corresponds to matter power spectrum for different values of $\Gamma/H_0$ in the IVS. The mean values of the other parameters, e.g. $\Omega_bh^2$, $\Omega_ch^2$ and $H_0$ required to generate the plots are taken from the combined analysis \cmbdesipantheon (Table~\ref{table:IVS}).  }
    \label{fig:ivs-spectra}
\end{figure*}
\begin{figure*}
    \centering
    \includegraphics[width=0.47\textwidth]{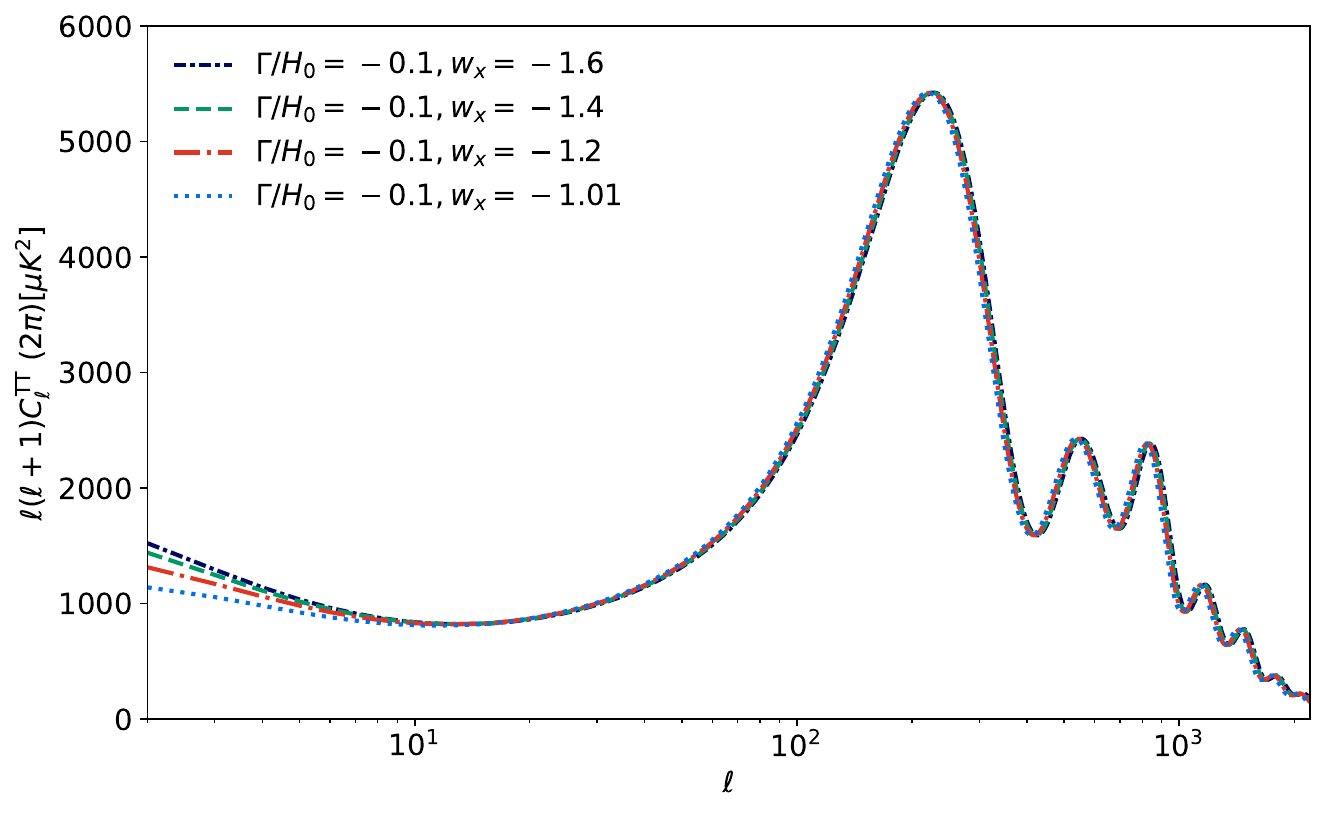}
    \includegraphics[width=0.47\textwidth]{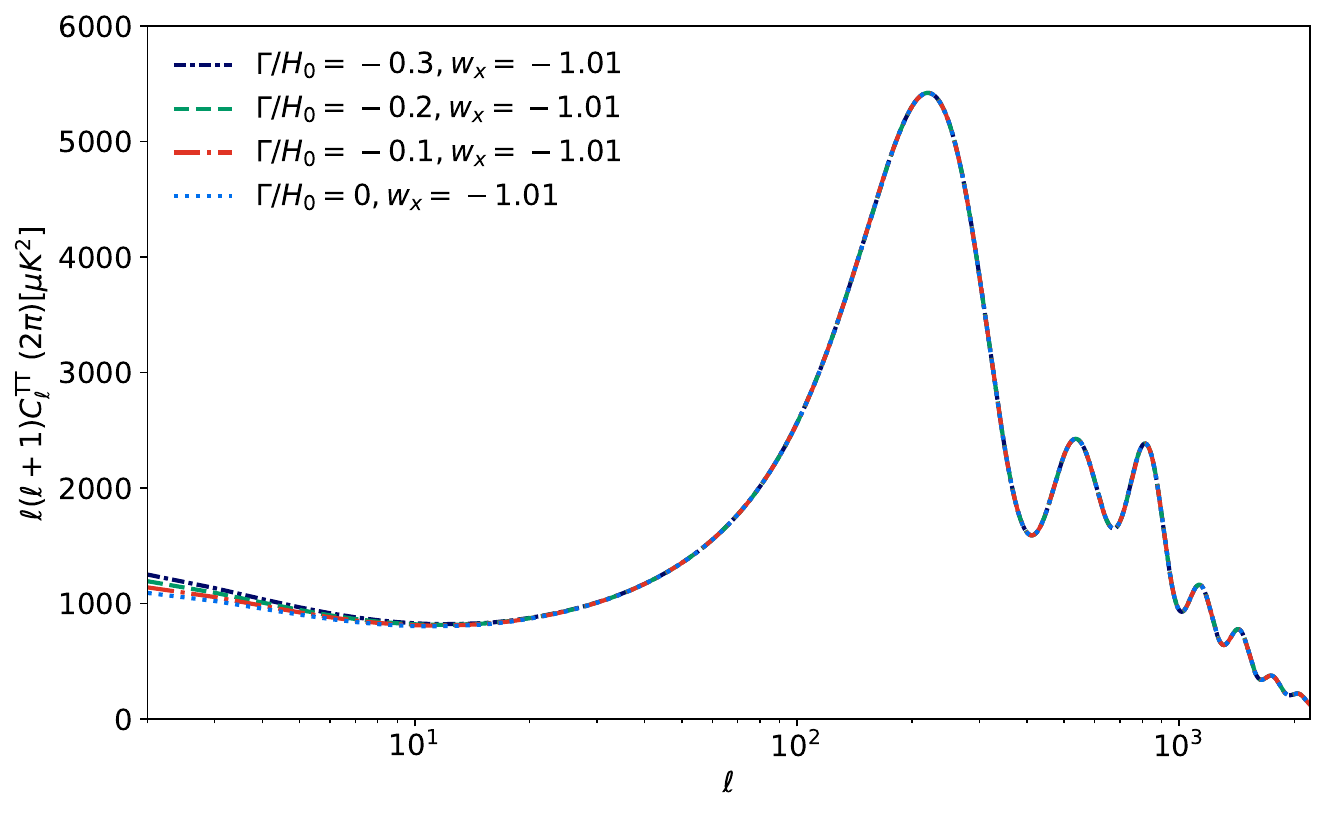}
    \includegraphics[width=0.47\textwidth]{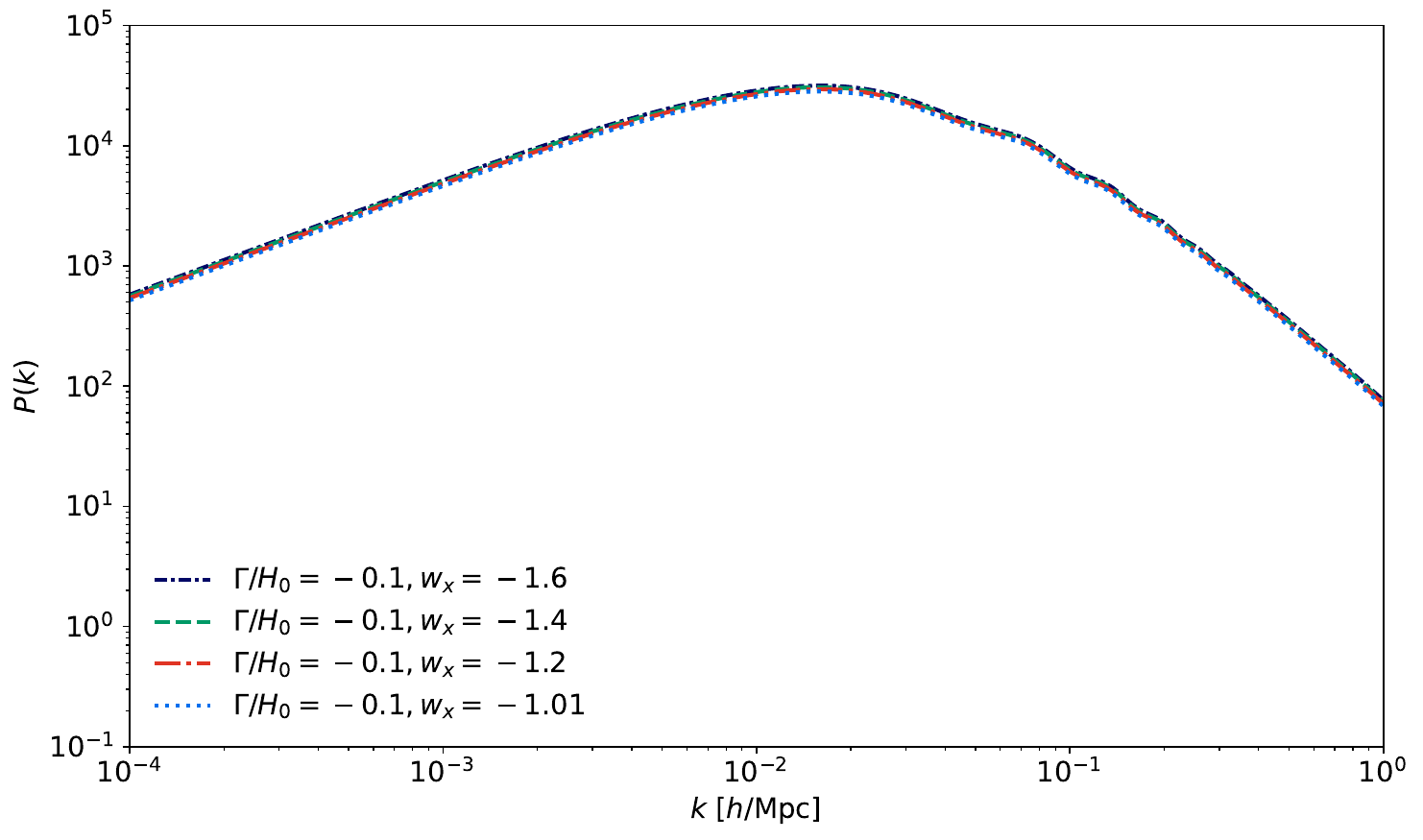}
    \includegraphics[width=0.47\textwidth]{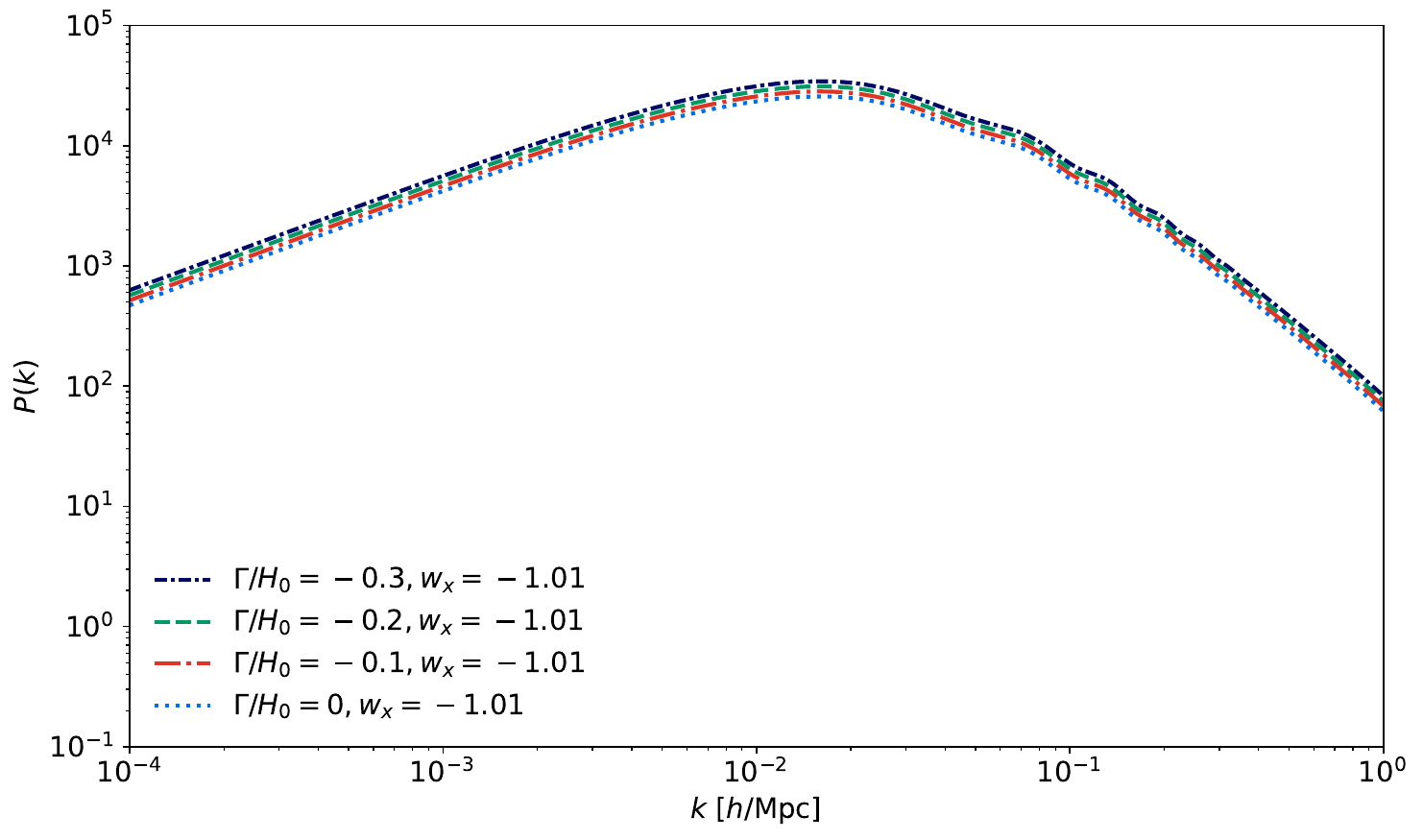}
    \caption{({\bf IDEphan}) The upper panel corresponds to  CMB TT spectrum and the lower panel corresponds to matter power spectrum for different values of $\Gamma/H_0$ and $w_x$ in the IDEphan scenario. In the figures on the l.h.s. $\Gamma/H_0$ is fixed, whereas $w_{x}$ assumes different values. In the figures on the r.h.s. the parameter $w_{x}$ is fixed while $\Gamma/H_0$ assumes different values.  The mean values of the other parameters, e.g. $\Omega_bh^2$, $\Omega_ch^2$ and $H_0$ required to generate the plots are taken from the combined analysis \cmbdesipantheon (Table~\ref{tab:IDEquin}).  }
    \label{fig:idephan-spectra}
\end{figure*}
\begin{figure*}
    \centering
    \includegraphics[width=0.47\textwidth]{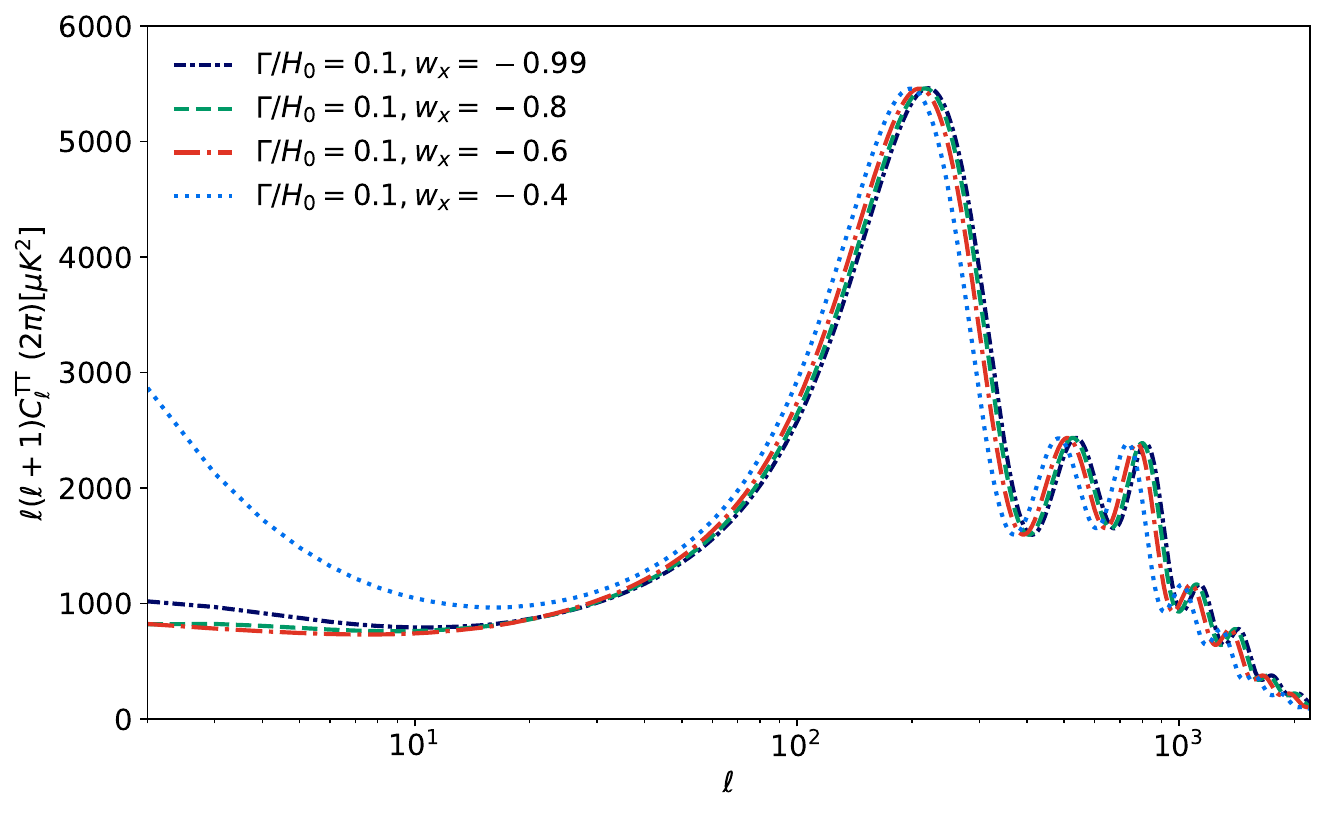}
    \includegraphics[width=0.47\textwidth]{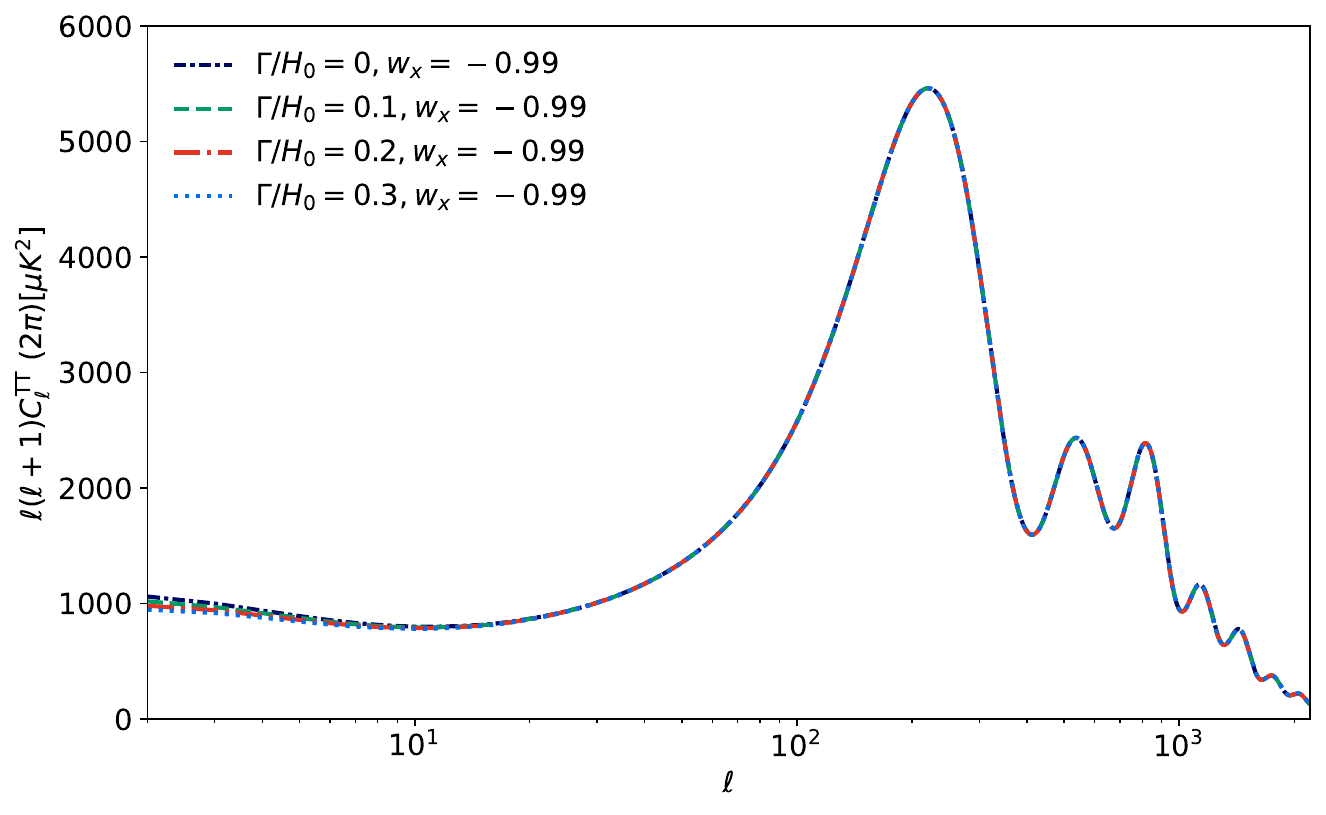}
    \includegraphics[width=0.47\textwidth]{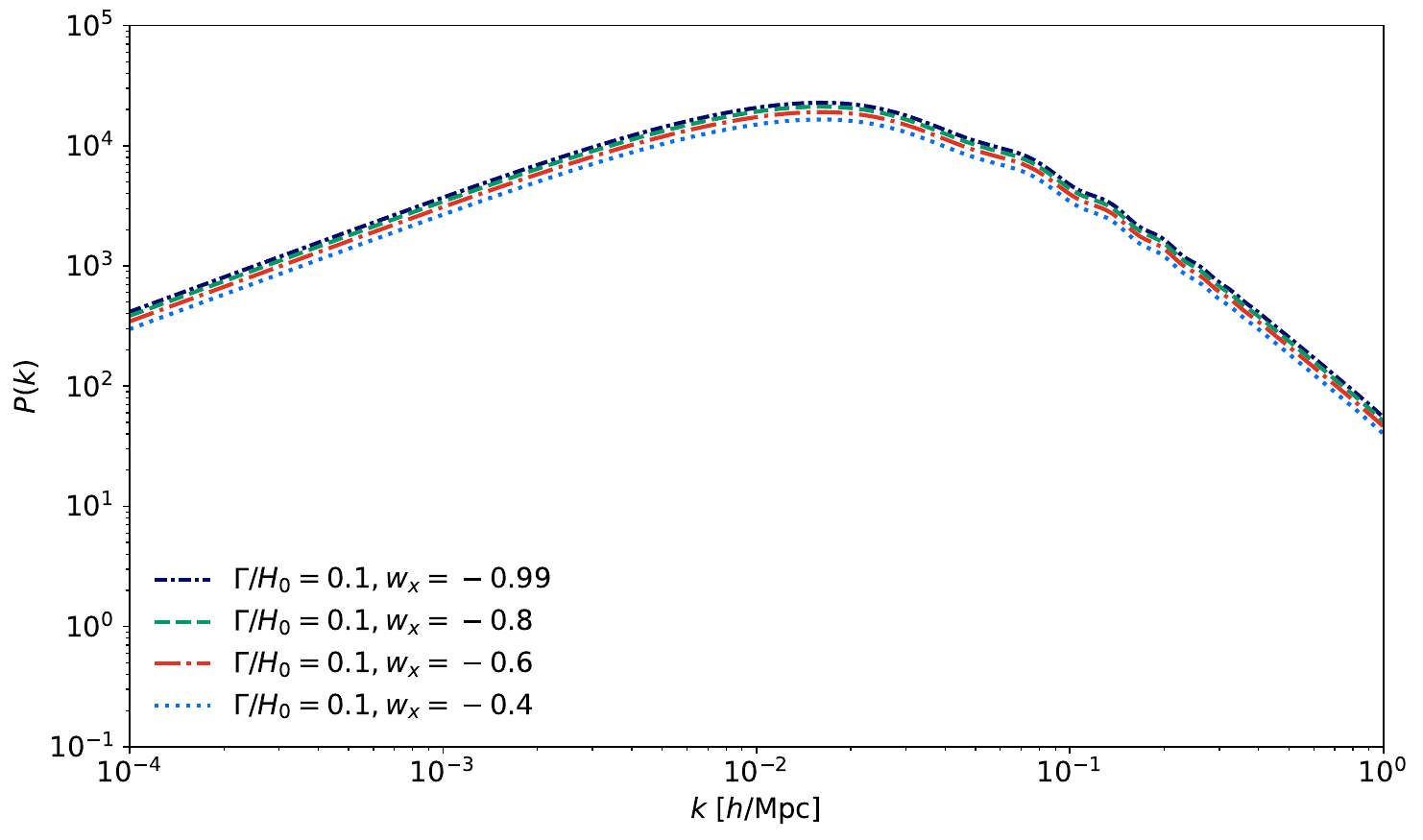}
    \includegraphics[width=0.47\textwidth]{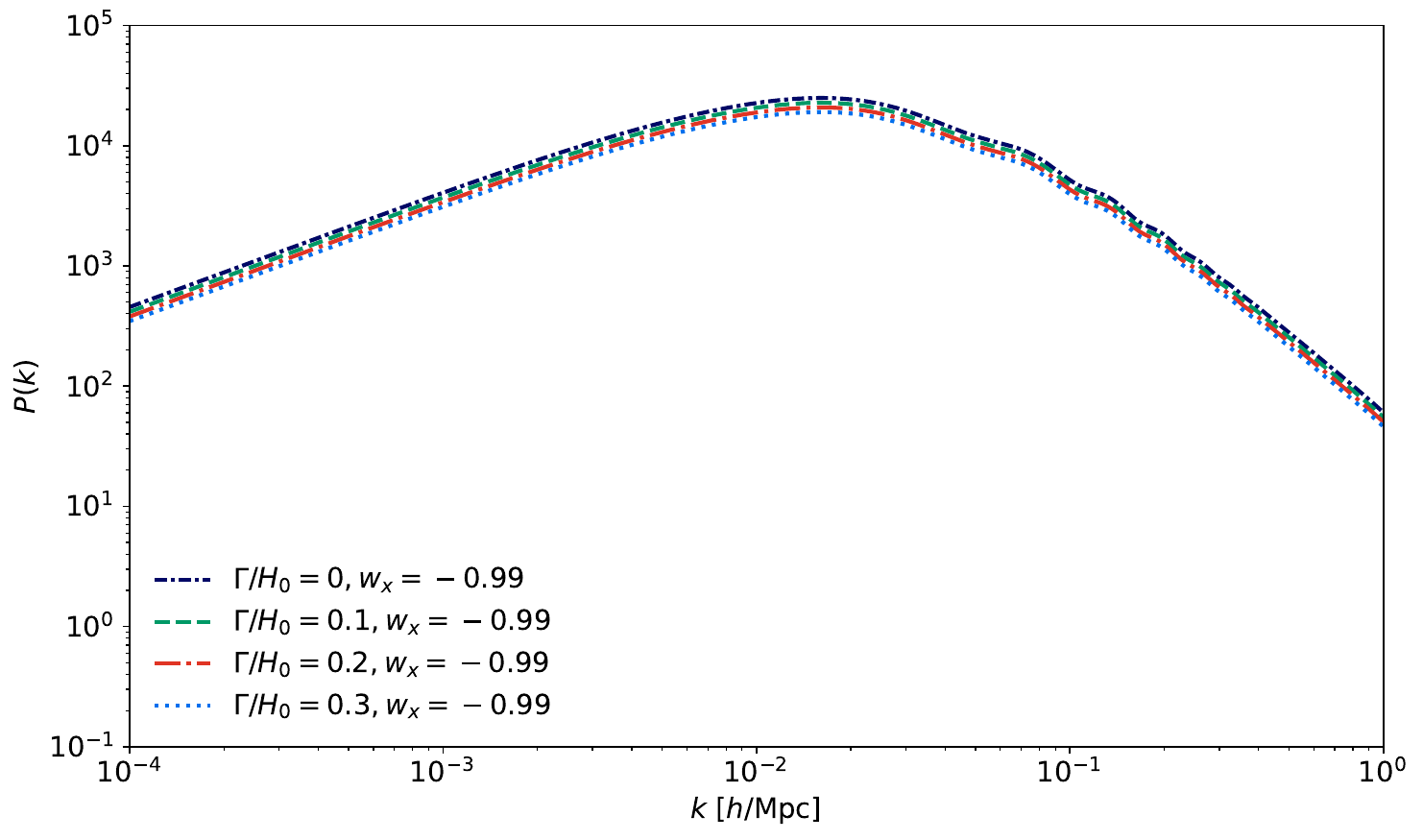}
    \caption{({\bf IDEquin}) The upper panel corresponds to  CMB TT spectrum and the lower panel corresponds to matter power spectrum for different values of $\Gamma/H_0$ and $w_x$ in the IDEquin scenario. In the figures on the l.h.s. $\Gamma/H_0$ is fixed while $w_{x}$ assumes different values. In the figures on the r.h.s. the parameter $w_{x}$ is fixed while $\Gamma/H_0$ assumes different values.  The mean values of the other parameters, e.g. $\Omega_bh^2$, $\Omega_ch^2$ and $H_0$ required to generate the plots are taken from the combined analysis \cmbdesipantheon  (Table~\ref{tab:IDEquin}).  }
    \label{fig:idequin-spectra}
\end{figure*}
We analyze the effect of different positive and negative values of the dimensionless coupling parameter $\Gamma/H_0$  on the CMB TT and matter power spectra. In Fig. ~\ref{fig:ivs-spectra}  we show the results for the {\bf IVS}. Our first impression is that the interaction in the dark sector affects both spectra. In particular, for $\Gamma/H_0 >0$ we observe the suppression of the matter power spectra compared to the non-interacting scenario ($\Gamma/H_0 =0$). This is evident because $\Gamma/H_0 >0$ indicates the flow of energy from DM to DE, and hence, this affects the amplitude of the matter density fluctuations, resulting in a suppressed matter power spectrum. The reverse scenario is observed for $\Gamma/H_0 <0$ because in this case DE decays to DM. Now regarding  the CMB TT spectra, we find that only at large angular scales (i.e., the low multipole region), effects of interaction are observed. These effects correspond to the late-time integrated Sachs-Wolfe effect. Additionally, we notice that such effects are pronounced when $\Gamma/H_0$ increases either in the positive or negative direction, which means that they depend on the strength of the interaction.

In Fig. \ref{fig:idephan-spectra}  we show the CMB TT spectra  and matter power spectra for {\bf IDEphan}, considering two separate cases: fixed $\Gamma/H_0$ but varying $w_{x}$  and fixed $w_x$ but varying $\Gamma/H_0$. In the upper left and lower left graphs of Fig. \ref{fig:idephan-spectra} we show the first case, where $w_x$ varies with a fixed $\Gamma/H_0$. In this case, we do not observe any changes in the matter power spectrum; however, in the CMB TT spectrum, we notice some changes, particularly in the large angular scales. This again corresponds to the late-time integrated Sachs-Wolfe effect.  In the second scenario (upper right and lower right graphs of Fig. \ref{fig:idephan-spectra}) where we fix $w_x$ but vary $\Gamma/H_0$, we can clearly see that the matter power spectrum gets enhanced compared to the non-interacting scenario. This can be related to the increased density of DM obtained from decaying DE, as a result of which the amplitude of the matter density fluctuations increases, and this consequently affects the matter power spectrum.  On the other hand, considering the CMB TT spectrum, we notice that in the low multipole region, mild changes appear when we deviate from the non-interacting phantom scenario ($\Gamma/H_0 = 0$). This is caused by the late-time integrated Sachs-Wolfe effect.  Overall, we see that the coupling parameter is the main ingredient that leaves its imprint on the CMB TT and matter power spectra.

Finally, in Fig. \ref{fig:idequin-spectra} we show the CMB TT spectra and matter power spectra for {\bf IDEquin} considering again two cases, namely a fixed $\Gamma/H_0$ but varying $w_{x}$,  and a fixed $w_x$ but varying $\Gamma/H_0$.  In the former case (see the upper and lower left graphs of Fig. \ref{fig:idequin-spectra}), we see that varying $w_x$ significantly affects the low multipole region of the CMB spectra (late-time integrated Sachs-Wolfe effect) and the matter power spectrum gets mildly suppressed for increasing $w_x$. A similar feature is also observed in the matter power spectrum when we consider a fixed $w_x$ but vary $\Gamma/H_0$ (see the lower right graph of Fig. \ref{fig:idequin-spectra}), however, in the CMB TT spectra, very mild changes appear in the low $\ell$ region.

Overall, we see that varying $w_x$ or  $\Gamma/H_{0}$ significantly affects the low multipole region of the CMB spectra. Since this feature is associated to the  late-time integrated Sachs-Wolfe effect, we expect it to be predominantly sensitive to the combination of these parameters, i.e., $w_{x}^{eff}$. In contrast, the more direct dependence of the matter power spectrum  on $\Gamma/H_{0}$ illustrates one of the avenues that breaks the degeneracy between $w_x$ and $\Gamma/H_{0}$.

In the following section, we introduce the dataset and the numerical methodology used to test the prediction of these models with the most recent observations.

\begin{table*}
    \begin{center}
        \renewcommand{\arraystretch}{1.4}
        \begin{tabular}{|c|c|c|c|}
            \hline
            \textbf{Parameter}               & \textbf{Prior (IVS)}     & \textbf{Prior (IDEphan)} & \textbf{Prior (IDEquin)} \\
            \hline\hline
            $\Omega_{b} h^2$             &$[0.005,0.1]$   &$[0.005,0.1]$   &  $[0.005,0.1]$\\
            $\Omega_{c} h^2$ & $[0.001,0.99]$              & $[0.001,0.99]$   & $[0.001,0.99]$\\
            $\tau$    & $[0.01,0.8]$                   & $[0.01,0.8]$    & $[0.01,0.8]$ \\
            $n_s$  & $[0.8, 1.2]$  & $[0.8, 1.2]$    & $[0.8, 1.2]$ \\
            $\log[10^{10}A_{s}]$         & $[1.61,3.91]$ & $[1.61,3.91]$   & $[1.61,3.91]$ \\
            $100\theta_{MC}$ & $[0.5,10]$  & $[0.5,10]$  & $[0.5,10]$ \\
            $\Gamma/H_0$                 & $[-3, 3]$ & $[-3,0]$        & $[0, 3]$\\
            $w_{x}$ & $-$                     & $[-3,-1]$  & $[-1,0]$ \\
            \hline
        \end{tabular}
    \end{center}
    \caption{Uniform priors imposed on the free parameters of the proposed cosmological scenarios for the statistical analysis. }
    \label{tab:priors}
\end{table*}
\begingroup
\squeezetable
\begin{center}
    \begin{table*}
        \begin{tabular}{cccccc}
            \hline
            Parameters & CMB & \cmbdesi & \cmbdesipantheon & \cmbdesiunion & \cmbdesidovekie \\ \hline
            $\Omega_\mathrm{b} h^2$ & $0.02228_{-0.00015-0.00029}^{+0.00015+0.00029}$ & $0.02250_{-0.00013-0.00026}^{+0.00013+0.00026}$ & $0.02254_{-0.00013-0.00025}^{+0.00013+0.00025}$ & $0.02255_{-0.00013-0.00025}^{+0.00013+0.00026}$ & $0.02255_{-0.00013-0.00026}^{+0.00013+0.00026}$ \\
            $\Omega_\mathrm{c} h^2$ & $0.07840_{-0.03336-0.07516}^{+0.04833+0.05734}$ & $0.10626_{-0.01488-0.03675}^{+0.02096+0.03349}$ & $0.12789_{-0.00641-0.01574}^{+0.00934+0.01389}$ & $0.12954_{-0.00524-0.01757}^{+0.01104+0.01408}$ & $0.13016_{-0.00556-0.01390}^{+0.00845+0.01230}$ \\
            $100\theta_\mathrm{MC}$ & $1.04333_{-0.00340-0.00409}^{+0.00174+0.00527}$ & $1.04167_{-0.00123-0.00196}^{+0.00079+0.00220}$ & $1.04041_{-0.00054-0.00095}^{+0.00047+0.00106}$ & $1.04035_{-0.00061-0.00101}^{+0.00046+0.00112}$ & $1.04033_{-0.00050-0.00088}^{+0.00042+0.00092}$ \\
            $\tau$ & $0.0550_{-0.0075-0.0161}^{+0.0076+0.0159}$ & $0.0590_{-0.0084-0.0148}^{+0.0071+0.0165}$ & $0.0593_{-0.0082-0.0147}^{+0.0072+0.0166}$ & $0.0595_{-0.0084-0.0146}^{+0.0072+0.0165}$ & $0.0589_{-0.0077-0.0157}^{+0.0077+0.0158}$ \\
            $n_\mathrm{s}$ & $0.9716_{-0.0044-0.0085}^{+0.0043+0.0084}$ & $0.9788_{-0.0035-0.0070}^{+0.0036+0.0070}$ & $0.9801_{-0.0034-0.0068}^{+0.0033+0.0065}$ & $0.9803_{-0.0034-0.0067}^{+0.0035+0.0065}$ & $0.9802_{-0.0034-0.0070}^{+0.0034+0.0067}$ \\
            $\ln(10^{10} A_\mathrm{s})$ & $3.057_{-0.015-0.032}^{+0.016+0.033}$ & $3.058_{-0.017-0.031}^{+0.015+0.033}$ & $3.057_{-0.017-0.031}^{+0.015+0.035}$ & $3.057_{-0.018-0.030}^{+0.015+0.034}$ & $3.056_{-0.016-0.032}^{+0.016+0.033}$ \\
            $\Gamma/H_0$ & $0.134_{-0.106-0.211}^{+0.144+0.197}$ & $0.038_{-0.070-0.125}^{+0.062+0.127}$ & $-0.044_{-0.038-0.060}^{+0.030+0.065}$ & $-0.051_{-0.047-0.061}^{+0.025+0.074}$ & $-0.053_{-0.036-0.054}^{+0.026+0.059}$ \\
            $\Omega_\mathrm{m}$ & $0.2162_{-0.1060-0.1610}^{+0.1023+0.1627}$ & $0.2738_{-0.0415-0.0914}^{+0.0513+0.0872}$ & $0.3305_{-0.0183-0.0429}^{+0.0254+0.0385}$ & $0.3352_{-0.0162-0.0485}^{+0.0300+0.0404}$ & $0.3367_{-0.0166-0.0387}^{+0.0228+0.0352}$ \\
            $\sigma_8$ & $0.958_{-0.190-0.201}^{+0.081+0.265}$ & $0.839_{-0.068-0.100}^{+0.038+0.119}$ & $0.775_{-0.027-0.039}^{+0.018+0.046}$ & $0.771_{-0.030-0.041}^{+0.016+0.051}$ & $0.769_{-0.024-0.036}^{+0.016+0.040}$ \\
            $H_0$ [Km/s/Mpc] & $69.39_{-2.59-4.08}^{+2.31+4.21}$ & $68.93_{-1.17-2.01}^{+0.95+2.11}$ & $67.64_{-0.58-0.95}^{+0.46+1.02}$ & $67.54_{-0.65-1.04}^{+0.47+1.16}$ & $67.51_{-0.52-0.89}^{+0.43+1.00}$ \\
            $S_8$ & $0.766_{-0.028-0.194}^{+0.094+0.117}$ & $0.793_{-0.014-0.042}^{+0.022+0.038}$ & $0.812_{-0.011-0.021}^{+0.011+0.020}$ & $0.813_{-0.010-0.022}^{+0.011+0.020}$ & $0.813_{-0.010-0.019}^{+0.010+0.020}$ \\
            $r_{\rm{drag}}$ [Mpc] & $147.06_{-0.29-0.58}^{+0.29+0.57}$ & $147.63_{-0.21-0.43}^{+0.22+0.42}$ & $147.74_{-0.21-0.41}^{+0.20+0.41}$ & $147.74_{-0.21-0.42}^{+0.21+0.40}$ & $147.76_{-0.21-0.41}^{+0.21+0.41}$ \\
            \hline
            $\rm{ln}\mathcal{B}_{ij}$ & $-2.7$ & $-4.8$ & $-5.2$ & $-4.6$ & $-4.4$ \\
            \hline
        \end{tabular}
            \caption{{\bf (IVS)}  68\% and 95\% CL constraints on cosmological parameters for CMB, \cmbdesi, \cmbdesipantheon, \cmbdesiunion and \cmbdesidovekie data combinations.}
        \label{table:IVS}
    \end{table*}
\end{center}
\endgroup
\begin{figure*}
    \includegraphics[width=0.8\textwidth]{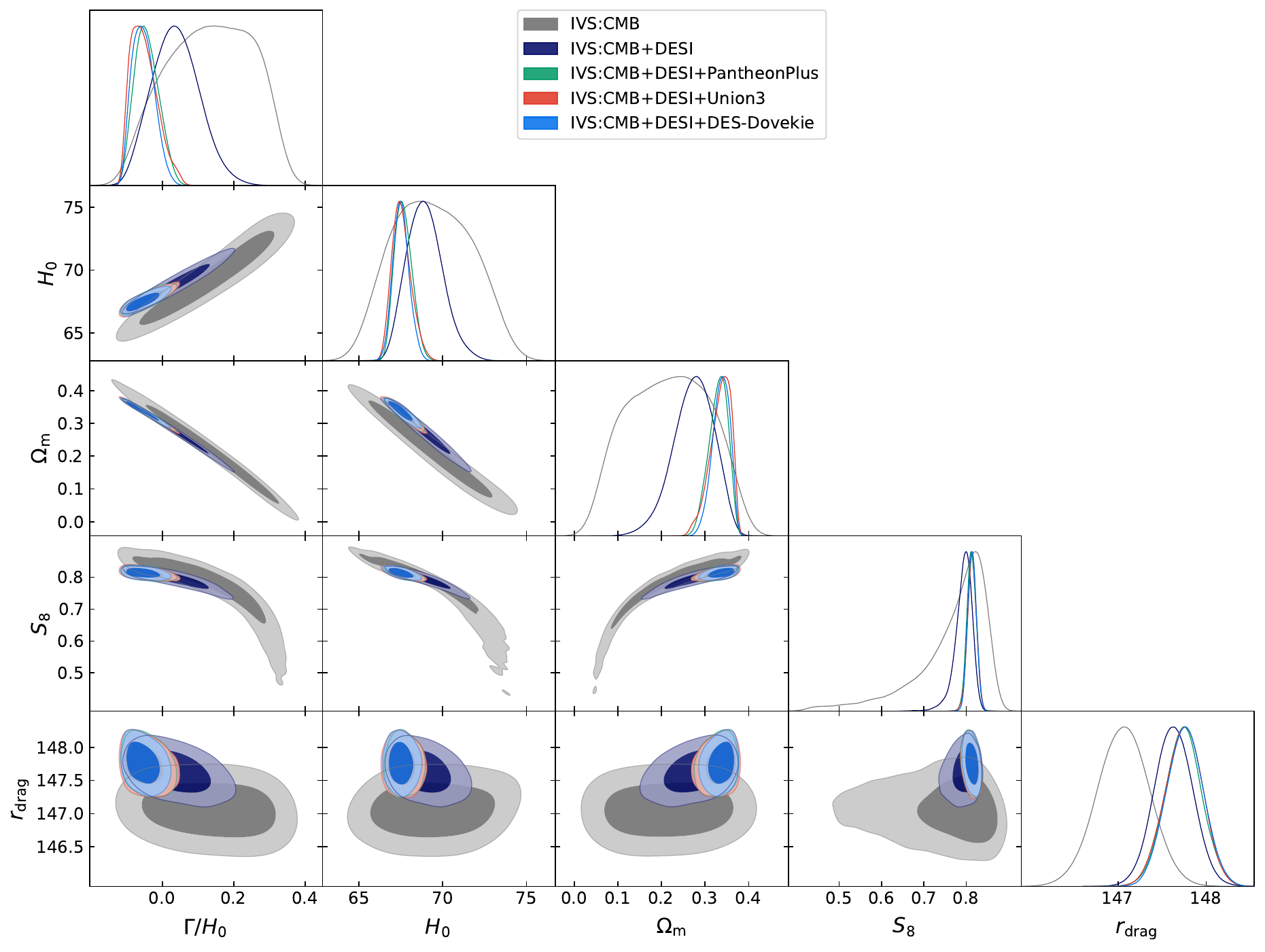}
        \caption{{\bf (IVS)}  One dimensional posterior distributions and two dimensional joint contours for the most relevant parameters using CMB from Planck 2018 and its combination with several cosmological datasets.}
    \label{fig:IVS-1}
\end{figure*}
\begin{figure}
    \includegraphics[width=0.5\textwidth]{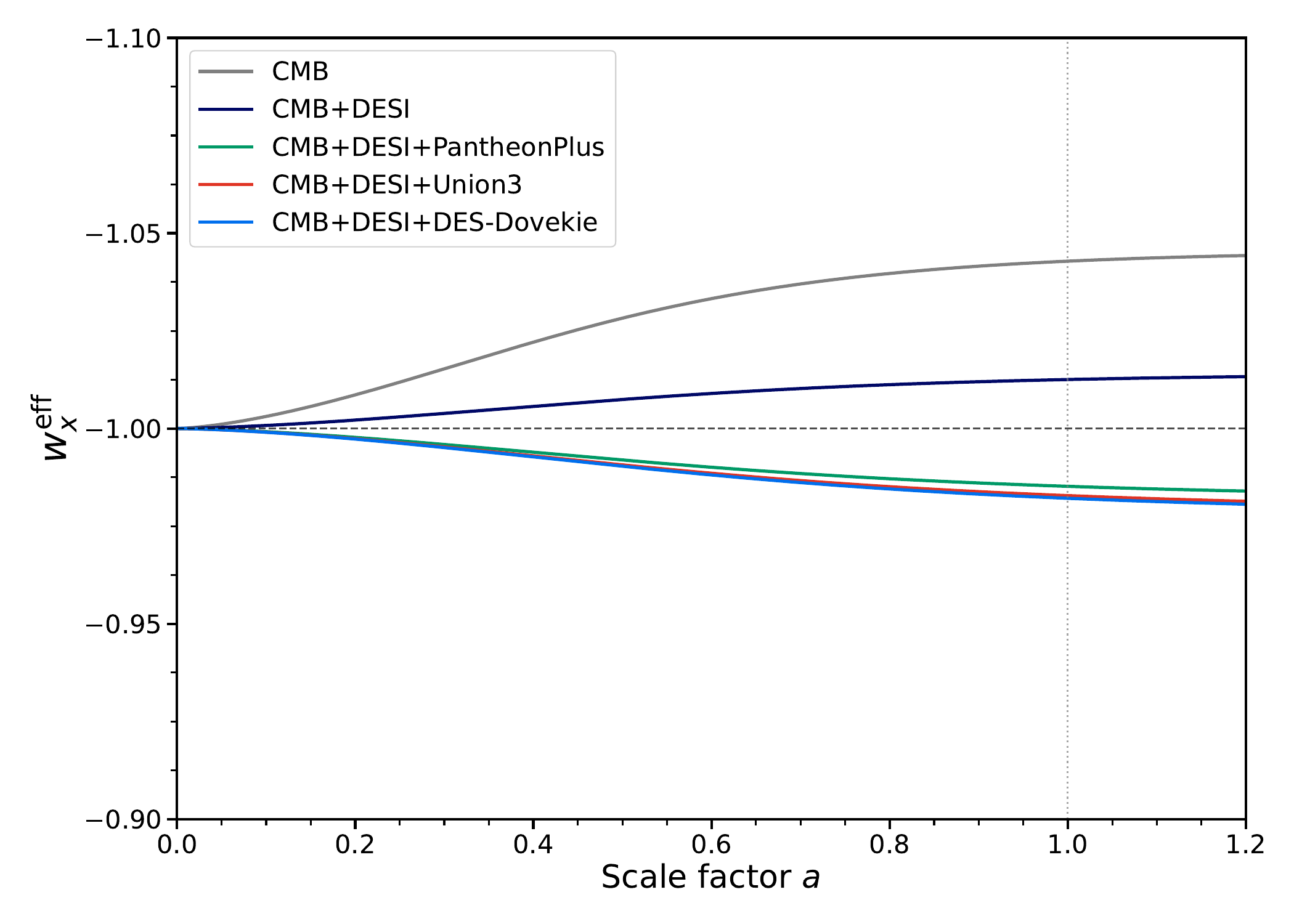}
    \includegraphics[width=0.5\textwidth]{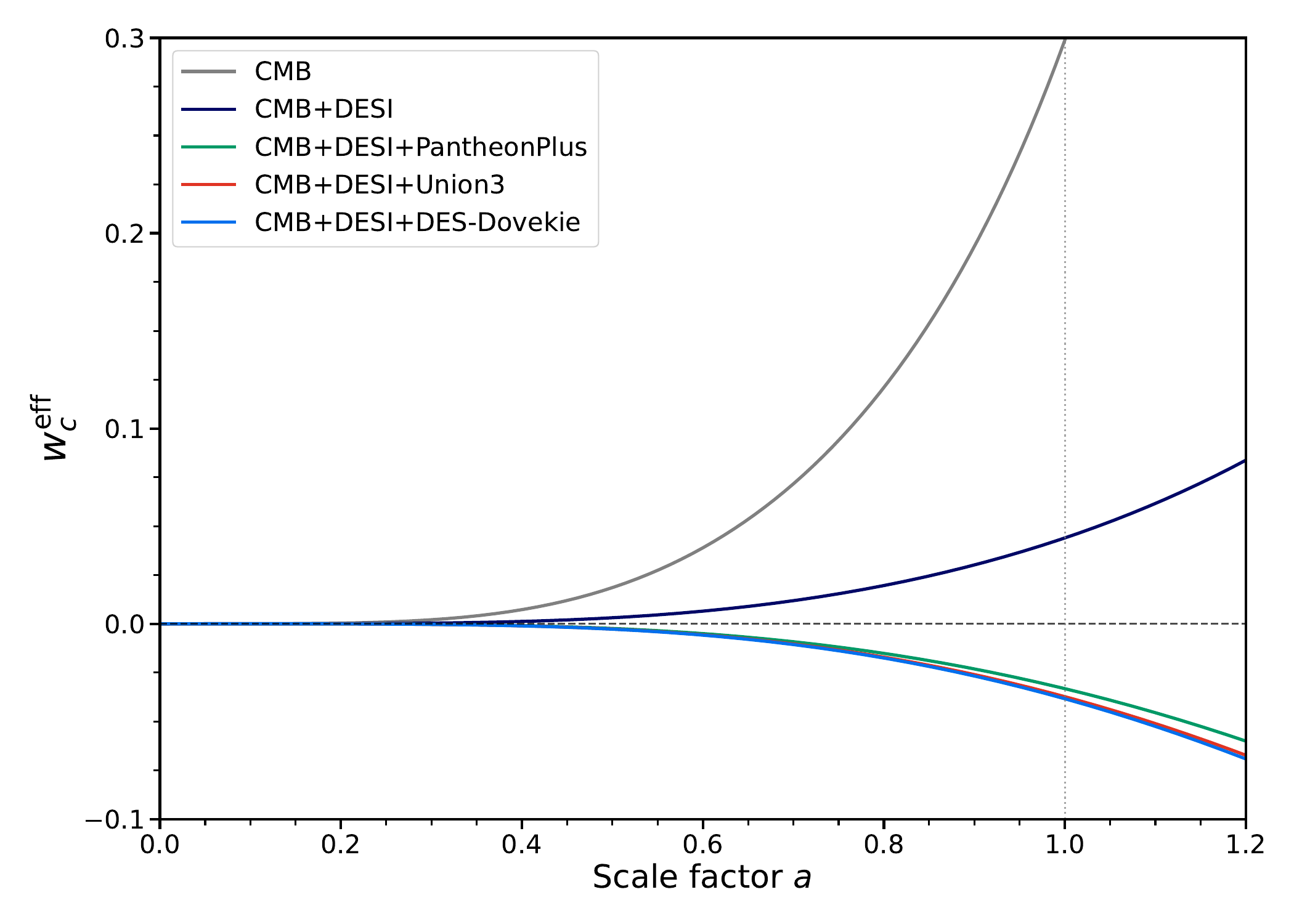}
    \caption{{\bf (IVS)}  Evolution of $w_{x}^{\rm eff}$ (upper graph) and $w_{c}^{\rm eff}$ (lower graph) in terms of their mean curves considering several observational datasets.}
    \label{fig:eff-eos-ivs}
\end{figure}
\begingroup
\begin{center}
    \begin{table*}[t]
        \scalebox{0.85}{
            \begin{tabular}{ccccccccc}
                \hline
                Parameters & CMB & \cmbdesi & \cmbdesipantheon & \cmbdesiunion & \cmbdesidovekie\\ \hline
                $\Omega_\mathrm{b} h^2$ & $0.02233_{-0.00015-0.00029}^{+0.00015+0.00030}$ & $0.02249_{-0.00014-0.00025}^{+0.00013+0.00027}$ & $0.02252_{-0.00013-0.00025}^{+0.00013+0.00026}$ & $0.02252_{-0.00013-0.00025}^{+0.00013+0.00026}$ & $0.02253_{-0.00013-0.00025}^{+0.00013+0.00025}$ \\
                $\Omega_\mathrm{c} h^2$ & $0.13200_{-0.01009-0.01188}^{+0.00587+0.01333}$ & $0.13219_{-0.00596-0.01320}^{+0.01087+0.01225}$ & $0.13552_{-0.00324-0.01273}^{+0.00822+0.00927}$ & $0.13530_{-0.00334-0.01320}^{+0.00843+0.00946}$ & $0.13597_{-0.00290-0.01196}^{+0.00754+0.00870}$ \\
                $100\theta_\mathrm{MC}$ & $1.03998_{-0.00052-0.00092}^{+0.00052+0.00093}$ & $1.04016_{-0.00055-0.00091}^{+0.00050+0.00096}$ & $1.04000_{-0.00049-0.00076}^{+0.00036+0.00091}$ & $1.04003_{-0.00051-0.00079}^{+0.00037+0.00089}$ & $1.03999_{-0.00046-0.00073}^{+0.00036+0.00087}$ \\
                $\tau$ & $0.0541_{-0.0075-0.0150}^{+0.0075+0.0155}$ & $0.0577_{-0.0078-0.0154}^{+0.0079+0.0164}$ & $0.0585_{-0.0082-0.0150}^{+0.0073+0.0164}$ & $0.0585_{-0.0081-0.0153}^{+0.0074+0.0164}$ & $0.0583_{-0.0079-0.0161}^{+0.0079+0.0170}$ \\
                $n_\mathrm{s}$ & $0.9731_{-0.0042-0.0085}^{+0.0042+0.0085}$ & $0.9781_{-0.0035-0.0068}^{+0.0035+0.0068}$ & $0.9794_{-0.0035-0.0066}^{+0.0035+0.0069}$ & $0.9794_{-0.0035-0.0068}^{+0.0034+0.0066}$ & $0.9794_{-0.0034-0.0068}^{+0.0034+0.0067}$ \\
                $\ln(10^{10} A_\mathrm{s})$ & $3.054_{-0.015-0.030}^{+0.015+0.031}$ & $3.056_{-0.016-0.032}^{+0.016+0.034}$ & $3.056_{-0.016-0.032}^{+0.016+0.033}$ & $3.056_{-0.016-0.032}^{+0.016+0.033}$ & $3.056_{-0.017-0.033}^{+0.016+0.035}$ \\
                $\Gamma/H_0$ & $>-0.061>-0.098$ & $>-0.076>-0.104$ & $-0.073_{-0.036-0.038}^{+0.013+0.053}$ & $-0.072_{-0.037-0.039}^{+0.013+0.055}$ & $-0.075_{-0.033-0.036}^{+0.012+0.050}$ \\
                $w_x$ & $>-1.22>-1.43$ & $-1.067_{-0.030}^{+0.045}>-1.132$ & $>-1.042>-1.073$ & $>-1.040>-1.076$ & $>-1.036>-1.062$ \\
                $\Omega_\mathrm{m}$ & $0.3112_{-0.0329-0.0751}^{+0.0421+0.0698}$ & $0.3261_{-0.0205-0.0357}^{+0.0226+0.0351}$ & $0.3438_{-0.0101-0.0328}^{+0.0201+0.0255}$ & $0.3435_{-0.0111-0.0351}^{+0.0216+0.0275}$ & $0.3464_{-0.0093-0.0311}^{+0.0185+0.0240}$ \\
                $\sigma_8$ & $0.812_{-0.036-0.057}^{+0.024+0.064}$ & $0.782_{-0.023-0.036}^{+0.020+0.039}$ & $0.765_{-0.019-0.028}^{+0.012+0.034}$ & $0.765_{-0.021-0.030}^{+0.013+0.035}$ & $0.762_{-0.018-0.028}^{+0.012+0.033}$ \\
                $H_0$ [Km/s/Mpc] & $70.91_{-4.83-5.87}^{+1.76+8.30}$ & $69.04_{-0.98-1.72}^{+0.83+1.77}$ & $67.95_{-0.55-0.99}^{+0.49+1.06}$ & $67.93_{-0.65-1.13}^{+0.55+1.19}$ & $67.79_{-0.47-0.88}^{+0.47+0.96}$ \\
                $S_8$ & $0.824_{-0.023-0.053}^{+0.029+0.049}$ & $0.815_{-0.011-0.022}^{+0.011+0.022}$ & $0.818_{-0.010-0.019}^{+0.010+0.020}$ & $0.818_{-0.011-0.021}^{+0.011+0.021}$ & $0.818_{-0.010-0.020}^{+0.010+0.020}$ \\
                $r_{\rm{drag}}$ [Mpc] & $147.13_{-0.29-0.58}^{+0.30+0.57}$ & $147.56_{-0.22-0.44}^{+0.22+0.45}$ & $147.66_{-0.20-0.40}^{+0.21+0.40}$ & $147.66_{-0.21-0.42}^{+0.22+0.42}$ & $147.67_{-0.20-0.40}^{+0.21+0.39}$ \\
                \hline
                $\rm{ln}\mathcal{B}_{ij}$ & $-6.3$ & $-7.5$ & $-8.0$ & $-7.4$ & $-7.5$ \\
                \hline
        \end{tabular}}
        \caption{{\bf (IDEphan)}  68\% and 95\% CL constraints are presented for CMB, \cmbdesi, \cmbdesipantheon, \cmbdesiunion and \cmbdesidovekie datasets. }
        \label{tab:IDEphan}
    \end{table*}
\end{center}
\endgroup
\begin{figure*}
\includegraphics[width=0.8\textwidth]{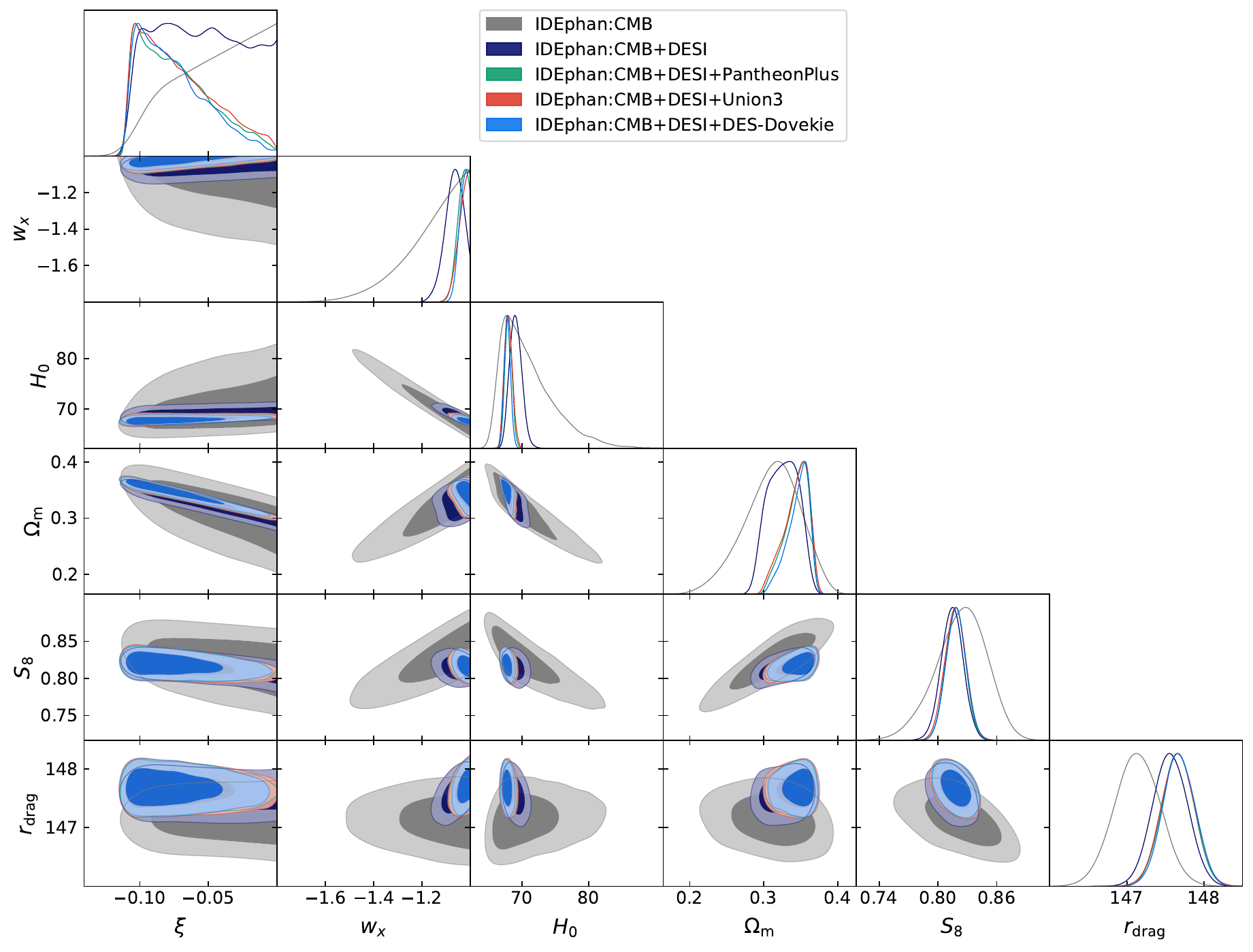}
        \caption{{\bf (IDEphan)}  One dimensional posterior distributions and two dimensional joint contours for the most relevant parameters using different combinations of cosmological measurements. }
    \label{fig:IDEphan-2}
\end{figure*}
\begin{figure}
    \includegraphics[width=0.5\textwidth]{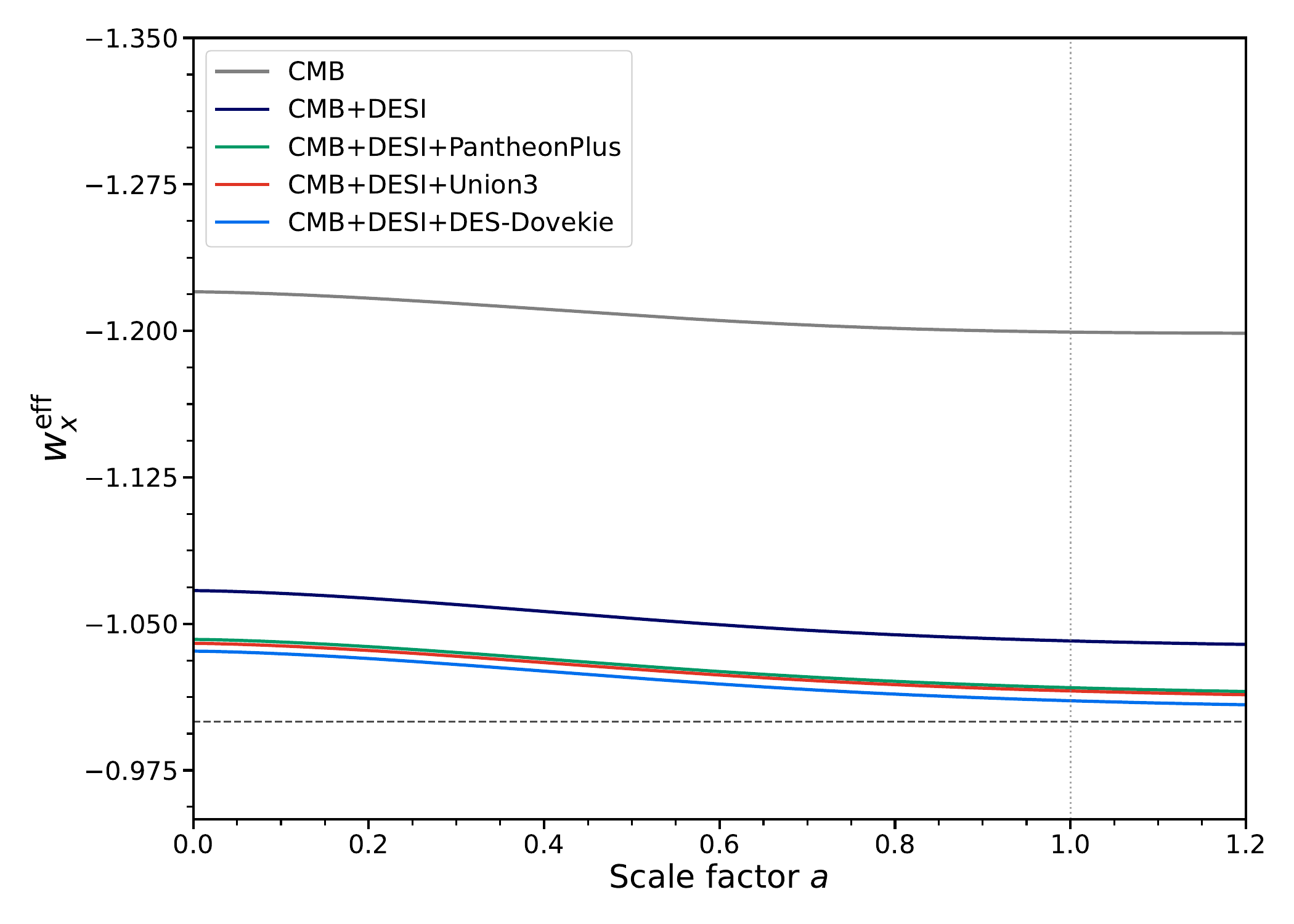}
    \includegraphics[width=0.5\textwidth]{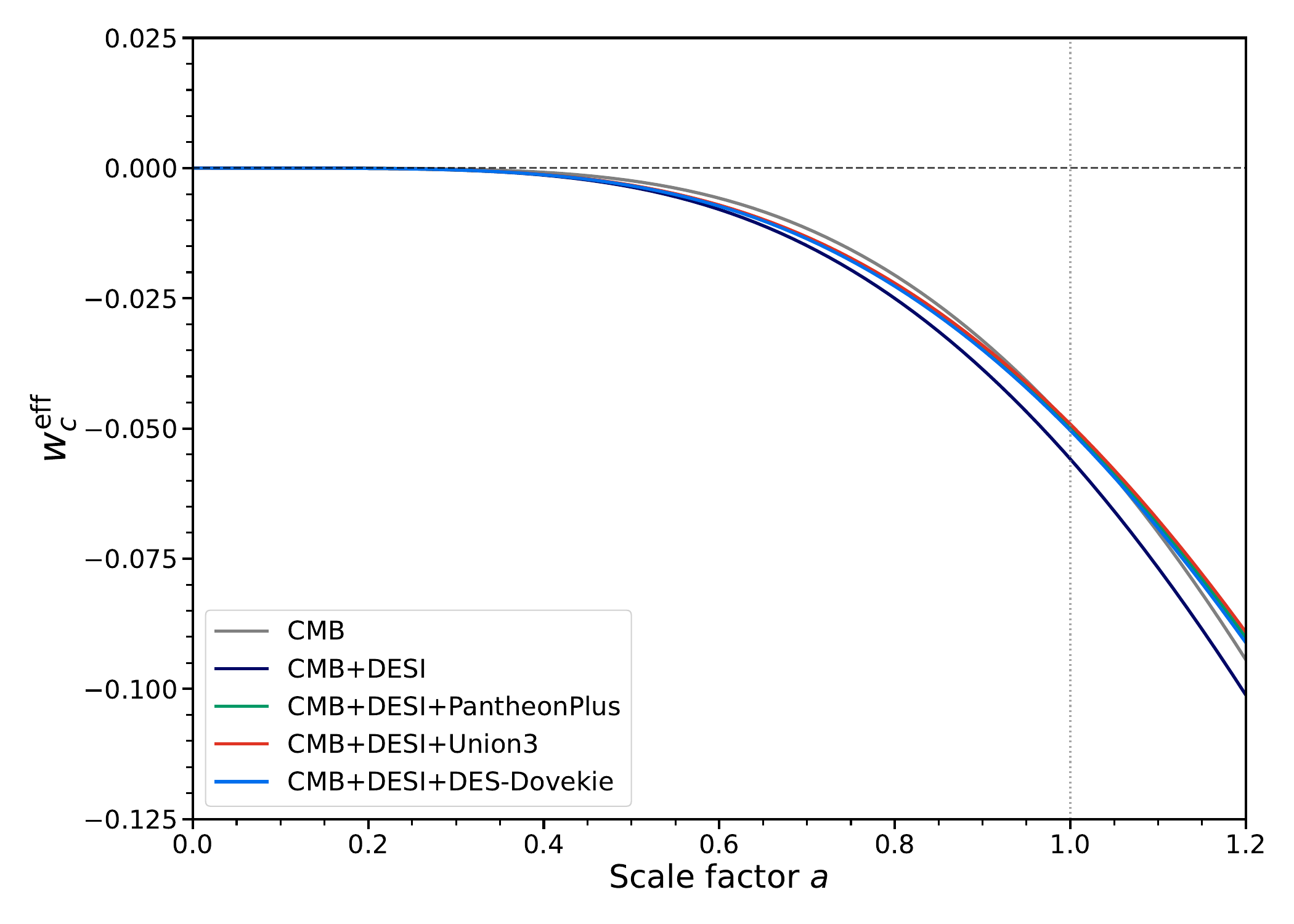}
    \caption{{\bf (IDEphan)}  Evolution of  $w_{x}^{\rm eff}$ (upper graph) and $w_{c}^{\rm eff}$ (lower graph) in terms of their mean curves considering several observational datasets. }
    \label{fig:eff-eos-idephan}
\end{figure}
\begin{table*}
    \scalebox{0.85}{
            \begin{tabular}{ccccccccc}
                \hline
                Parameters & CMB & \cmbdesi & \cmbdesipantheon & \cmbdesiunion & \cmbdesidovekie\\ \hline
                $\Omega_\mathrm{b} h^2$ & $0.02226_{-0.00015-0.00029}^{+0.00015+0.00029}$ & $0.02249_{-0.00014-0.00027}^{+0.00013+0.00026}$ & $0.02255_{-0.00013-0.00026}^{+0.00013+0.00026}$ & $0.02254_{-0.00013-0.00025}^{+0.00013+0.00026}$ & $0.02255_{-0.00013-0.00025}^{+0.00013+0.00026}$ \\
                $\Omega_\mathrm{c} h^2$ & $0.07286_{-0.01946-0.07186}^{+0.04674+0.04443}$ & $0.08212_{-0.01264-0.04806}^{+0.03022+0.03600}$ & $0.10451_{-0.00324-0.02365}^{+0.01241+0.01417}$ & $0.10307_{-0.00281-0.02767}^{+0.01401+0.01541}$ & $0.10589_{-0.00273-0.02110}^{+0.01110+0.01247}$ \\
                $100\theta_\mathrm{MC}$ & $1.04366_{-0.00314-0.00341}^{+0.00124+0.00502}$ & $1.04315_{-0.00192-0.00245}^{+0.00080+0.00340}$ & $1.04178_{-0.00082-0.00118}^{+0.00037+0.00154}$ & $1.04187_{-0.00091-0.00125}^{+0.00032+0.00180}$ & $1.04170_{-0.00073-0.00108}^{+0.00032+0.00140}$ \\
                $\tau$ & $0.0551_{-0.0082-0.0150}^{+0.0072+0.0162}$ & $0.0599_{-0.0087-0.0151}^{+0.0074+0.0167}$ & $0.0601_{-0.0088-0.0152}^{+0.0074+0.0169}$ & $0.0606_{-0.0088-0.0157}^{+0.0077+0.0174}$ & $0.0605_{-0.0085-0.0157}^{+0.0077+0.0169}$ \\
                $n_\mathrm{s}$ & $0.9705_{-0.0043-0.0086}^{+0.0043+0.0085}$ & $0.9786_{-0.0035-0.0069}^{+0.0035+0.0070}$ & $0.9802_{-0.0034-0.0067}^{+0.0034+0.0069}$ & $0.9799_{-0.0034-0.0070}^{+0.0035+0.0070}$ & $0.9803_{-0.0034-0.0068}^{+0.0035+0.0066}$ \\
                $\ln(10^{10} A_\mathrm{s})$ & $3.058_{-0.017-0.031}^{+0.015+0.033}$ & $3.060_{-0.018-0.032}^{+0.016+0.034}$ & $3.059_{-0.018-0.032}^{+0.016+0.035}$ & $3.060_{-0.018-0.032}^{+0.016+0.035}$ & $3.060_{-0.017-0.033}^{+0.017+0.034}$ \\
                $\Gamma/H_0$ & $<0.231<0.343$ & $0.118_{-0.101}^{+0.044}<0.251$ & $<0.054<0.123$ & $<0.059<0.142$ & $<0.048<0.111$ \\
                $w_x$ & $<-0.813<-0.620$ & $<-0.951<-0.901$ & $-0.953_{-0.036}^{+0.019}<-0.901$ & $-0.951_{-0.040}^{+0.018}<-0.893$ & $-0.951_{-0.032}^{+0.022}<-0.903$ \\
                $\Omega_\mathrm{m}$ & $0.2333_{-0.1037-0.1818}^{+0.1082+0.1714}$ & $0.2209_{-0.0341-0.1077}^{+0.0683+0.0863}$ & $0.2777_{-0.0098-0.0539}^{+0.0290+0.0365}$ & $0.2745_{-0.0096-0.0650}^{+0.0334+0.0406}$ & $0.2816_{-0.0081-0.0485}^{+0.0261+0.0321}$ \\
                $\sigma_8$ & $0.944_{-0.175-0.186}^{+0.077+0.257}$ & $0.909_{-0.102-0.123}^{+0.036+0.186}$ & $0.828_{-0.034-0.045}^{+0.012+0.064}$ & $0.833_{-0.039-0.052}^{+0.011+0.081}$ & $0.823_{-0.030-0.042}^{+0.011+0.059}$ \\
                $H_0$ [Km/s/Mpc] & $65.28_{-3.11-8.36}^{+4.84+7.30}$ & $69.21_{-1.13-1.84}^{+0.86+2.02}$ & $67.82_{-0.56-1.06}^{+0.54+1.04}$ & $67.85_{-0.65-1.26}^{+0.63+1.28}$ & $67.72_{-0.52-1.00}^{+0.52+1.00}$ \\
                $S_8$ & $0.786_{-0.041-0.215}^{+0.101+0.149}$ & $0.764_{-0.013-0.082}^{+0.041+0.055}$ & $0.794_{-0.010-0.031}^{+0.016+0.027}$ & $0.793_{-0.010-0.037}^{+0.018+0.030}$ & $0.796_{-0.011-0.028}^{+0.015+0.027}$ \\
                $r_{\rm{drag}}$ [Mpc] & $147.01_{-0.30-0.60}^{+0.30+0.59}$ & $147.65_{-0.22-0.42}^{+0.22+0.43}$ & $147.75_{-0.21-0.44}^{+0.22+0.40}$ & $147.74_{-0.22-0.43}^{+0.22+0.44}$ & $147.76_{-0.21-0.42}^{+0.22+0.43}$ \\
                \hline
                $\rm{ln}\mathcal{B}_{ij}$ & $-4.0$ & $-7.5$ & $-8.4$ & $-7.6$ & $-8.1$ \\
                \hline
        \end{tabular}}
        \caption{{\bf (IDEquin)}  68\% and 95\% CL constraints on cosmological parameters for CMB, \cmbdesi, \cmbdesipantheon, \cmbdesiunion and \cmbdesidovekie data combinations. }
        \label{tab:IDEquin}
    \end{table*}
\begin{figure*}
    \includegraphics[width=0.8\textwidth]{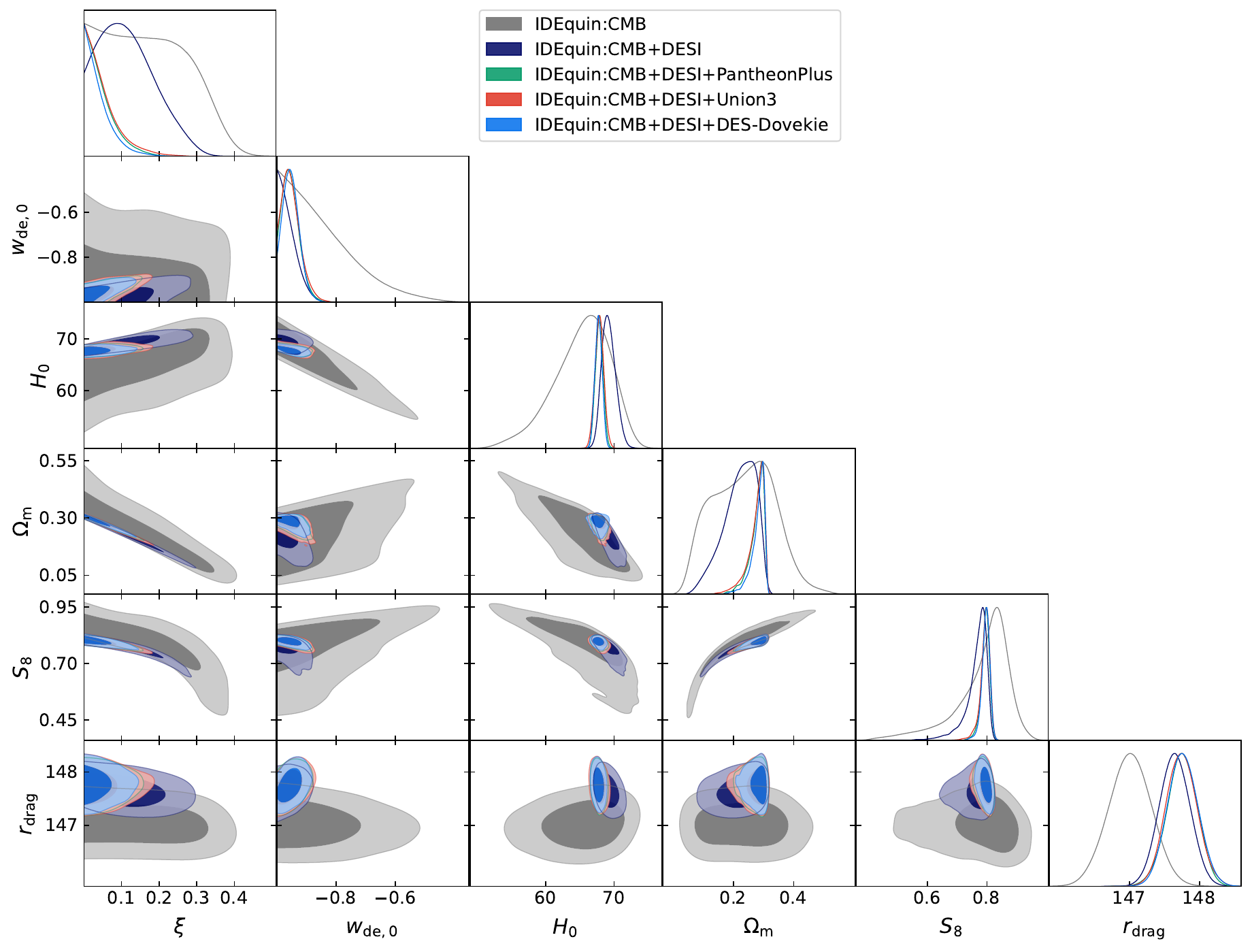}
    \caption{{\bf (IDEquin)} One dimensional posterior distributions and two dimensional joint contours for the most relevant parameters using different combinations of cosmological measurements. }
    \label{fig:IDEquin-2}
\end{figure*}
\begin{figure}
    \includegraphics[width=0.5\textwidth]{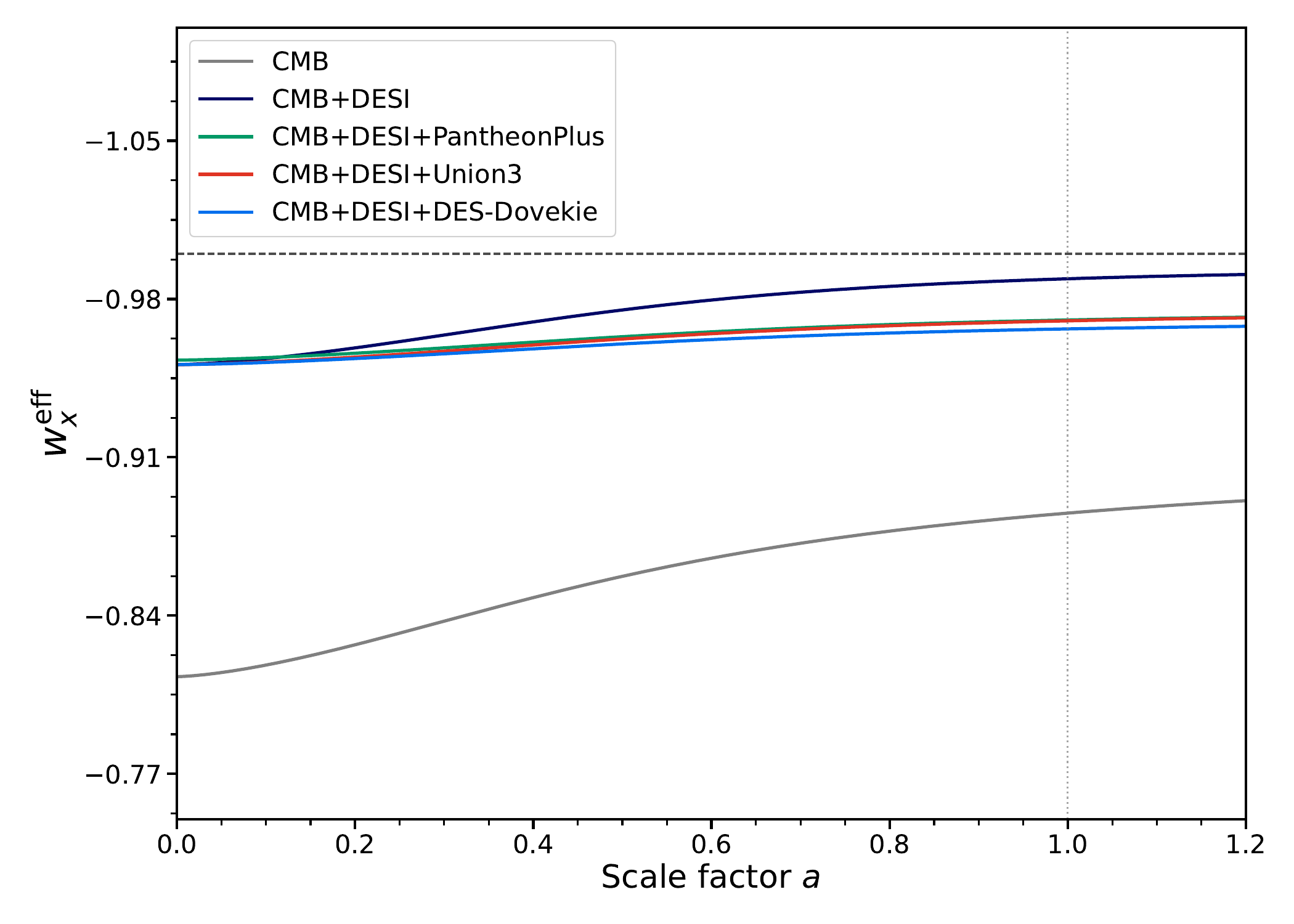}
    \includegraphics[width=0.5\textwidth]{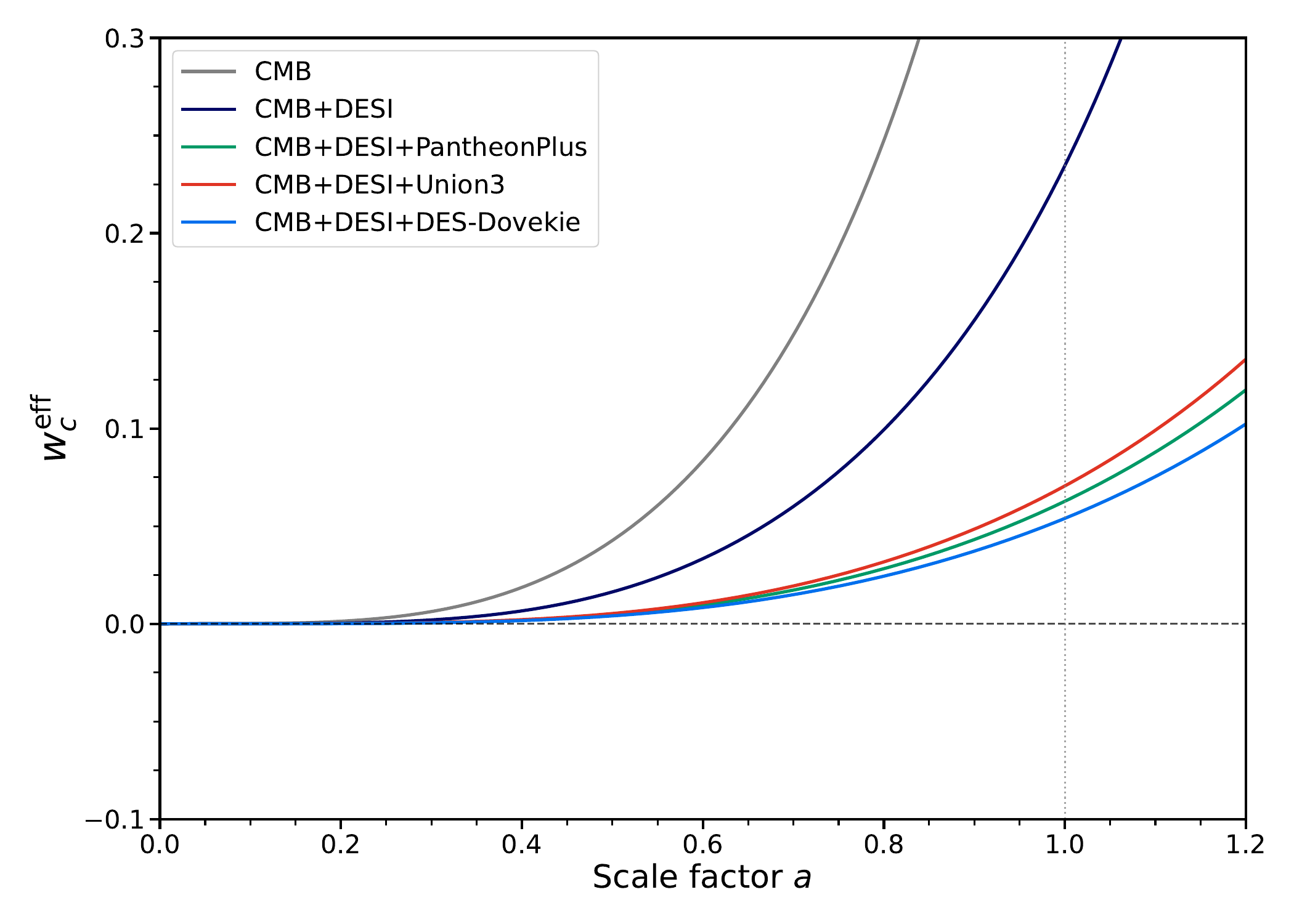}
    \caption{{\bf (IDEquin)}  Evolution of  $w_{x}^{\rm eff}$ (upper graph) and $w_{c}^{\rm eff}$ (lower graph) in terms of their mean curves considering several observational datasets. }
    \label{fig:eff-eos-idequin}
\end{figure}

\section{Datasets and Methodology}
\label{sec-data}

In this section, we describe the datasets used in our analysis, the adopted methodology, and the priors imposed on the free parameters of the model under consideration.

We begin by presenting the dataset used in our analysis, which comprises the following observables:

\begin{enumerate}

\item Cosmic Microwave Background ({\bf CMB}) anisotropies from {\bf Planck 2018}~\cite{Planck:2018vyg,Planck:2019nip}. In this work, we employed the {\it plikTTTEEE+lowl+lowE} likelihood combination, which contains information on the angular power spectra for the temperature and E-mode polarization on low and high multipoles.

\item Baryon Acoustic Oscillations ({\bf BAO}) data obtained from the {\bf DESI DR2}~\cite{DESI:2025zgx} survey. This dataset consists of cosmic distance measurements from various large-scale structure tracers: (i) the bright galaxy sample (BGS); (ii) emission line galaxies (ELG); (iii) luminous red galaxies (LRG) and (iv) quasi-stellar objects (QSO). Lastly, the auto-correlations of Ly-$\alpha$ forest spectra, as well as their cross-correlation with QSO, are also considered in this analysis~\cite{DESI:2025zpo}. As specified in~\cite{DESI:2025zpo}, there is no dependence between catalogs.

\item A set of three wide Type Ia Supernovae ({\bf SNIa}) compilations: (i) {\bf PantheonPlus} sample~\cite{Scolnic:2021amr} with 1701 light curves from 1550 supernovae; (ii) the {\bf Union3} catalog~\cite{Rubin:2023ovl}, incorporating 2087 SNIa from 24 datasets; and (iii) a new re-analysis of 5-year supernova survey data of the {\bf Dark Energy Survey} (labeled as {\bf DES-Dovekie})~\cite{DES:2024jxu}.

\end{enumerate}
In order to conduct our statistical analysis, we utilized a modified version of the cosmology code \texttt{CAMB}~\cite{Lewis:2002} in combination with the sampler \texttt{Cobaya}~\cite{torrado:2021cobaya}. The latter was used to run the Markov Chain Monte Carlo (MCMC) chains, with the Gelman-Rubin statistic as the convergence diagnosis. Lastly, \texttt{GetDist}~\cite{getdist} was used to analyze the chains. Table~\ref{tab:priors} describes the uniform priors employed for the free parameters of each cosmological model considered.

\section{Results}
\label{sec-results}

In this section, we present the results of our statistical analysis, that is, the constraints on the interacting scenarios, and examine whether they  are preferred with respect to the standard $\Lambda$CDM cosmological model using various observational surveys. For each interacting scenario, we performed five analyses, e.g. CMB, \cmbdesi, and three distinct combined analyses \cmbdesi+SNIa are characterized by different variants of SNIa  (PantheonPlus, Union3, and DES-Dovekie).

Moreover, to appraise these competing cosmological scenarios, we calculate the logarithm of the Bayesian evidence, $\ln\mathcal{Z}$, with the \texttt{MCEvidence} methodology developed in \cite{Heavens:2017afc}. This calculation was executed using the {\tt Cobaya} interface bundled in the \texttt{wgcosmo} repository~\cite{Giare:2025}.

Bayes’ rule relates the posterior density for the parameter vector $\Theta$ of a given model $\mathcal{M}_i$ to the data $D$ via
\begin{equation}
P(\Theta|D, \mathcal{M}_i) = \frac{\mathcal{L}(D|\Theta, \mathcal{M}_i) \pi(\Theta|\mathcal{M}_i)}{\mathcal{Z}_i},
\label{eq:bayes_theorem}
\end{equation}
where $\mathcal{L}$ is the maximum likelihood function, $\pi$ the prior probability density, and the evidence $\mathcal{Z}_i$ reads
\begin{equation}
\mathcal{B}_i = \int \mathcal{L}(D|\Theta, \mathcal{M}_i) \pi(\Theta|\mathcal{M}_i) {\rm d}\Theta.
\label{eq:bayesian_evidence}
\end{equation}

The relative performance of two models $i$ and $j$ is quantified by the Bayes factor

\begin{equation}
    \mathcal{B}_{ij} = \frac{\mathcal{B}_i}{\mathcal{B}_j}
    \label{eq:bayes_factor}
\end{equation}
or, equivalently, by the difference in their log–Bayes factors,
\begin{equation}
\ln \mathcal{B}_{ij} \equiv \ln \mathcal{B}_i - \ln \mathcal{B}_j.
\label{eq:relative_log_bayesian_evidence}
\end{equation}
We adopt $i$ for the interacting model ({\bf IVS, IDEphan, IDEquin}) and let $j$ denote the baseline $\Lambda$CDM model; therefore $\ln\mathcal{B}_{ij}>0$ favors the interaction and $\ln\mathcal{B}_{ij} <0$ indicates the reverse (i.e. $\Lambda$CDM is preferred)

To interpret the magnitude of  $\ln\mathcal{B}_{ij}$, we follow the revised Jeffreys' scale~\cite{Kass:1995loi}:
$[0,1]$ (\textit{inconclusive}),
$[1,2.5]$ (\textit{weak}),
$[2.5,5]$ (\textit{moderate}),
$[5,10]$ (\textit{strong}),
and $>10$ (\textit{very strong}) support for the model with the higher evidence.

In Tables~\ref{tab:IDEphan} and \ref{tab:IDEquin} we present the observational constraints on the free and derived parameters of the interacting scenarios, and in Figs. \ref{fig:IVS-1}, \ref{fig:IDEphan-2}, \ref{fig:IDEquin-2} we show the one dimensional posterior distributions of some of the model parameters together with their joint contours at 68\% and 95\% CL. In what follows, we summarize the main results. We split our analysis into three parts corresponding to {\bf IVS}, {\bf IDEphan}, and {\bf IDEquin}.

\subsection{IVS}

Table~\ref{table:IVS} and Fig. \ref{fig:IVS-1} summarize the constraints on {\bf IVS} considering various observational datasets. This is the simplest scenario in this series, where the EoS of DE is $-1$.
Beginning with the CMB alone case, one can notice that the dimensionless coupling parameter includes its null value within 68\% CL ($\Gamma/H_0 = 0.134_{-0.106}^{+0.144}$ at 68\% CL), which indicates that no evidence of interaction in the dark sector is preferred by CMB.
We find a higher Hubble constant, $H_0 = 69.39_{-2.59}^{+2.31}$, in comparison to $\Lambda$CDM from Planck~\cite{Planck:2018vyg}. However, the estimated value of $\Omega_\mathrm{c} h^2$ within this interacting scenario, $\Omega_\mathrm{c} h^2 = 0.07840_{-0.03336}^{+0.04833}$ at 68\% CL, is significantly lower than $\Lambda$CDM, while $\sigma_8$ ($\sigma_8 = 0.958_{-0.190}^{+0.081}$) is higher. As a result, $\Omega_m$ and $S_8$ also lead to lower values than those reported in ~\cite{Planck:2018vyg}.

When DESI is combined with CMB, the coupling parameter shifts to lower values, while maintaining a null interaction at $68\%$ CL ($\Gamma/H_0 = 0.038_{-0.070}^{+0.062}$). Moreover, as $\Gamma/H_0$ remains positive, this suggests a flow of energy from CDM to vacuum, which results in a mildly lower value of $\Omega_{m} = 0.2738_{-0.0415}^{+0.0513}$ at 68\% CL (Planck yields $\Omega_{m} =  0.3153 \pm 0.0073$ at 68\% CL for Planck TT,TE,EE+lowE+lensing~\cite{Planck:2018vyg}). This results in a mild increase in the Hubble constant ($H_0 = 68.93_{-1.17}^{+0.95}$ km/s/Mpc at 68\% CL). In addition to that, $S_8$ takes significantly higher values in comparison with the previous case, alleviating the $S_8$ tension.

For the next three analyses, we notice the preference of a negative coupling ($\Gamma/H_0 < 0$); hence, a flow of energy from DE to CDM is suggested at slightly more than 68\% CL. Therefore, $\Omega_m$ is mildly increased, which in turn slightly decreases the estimated value of the Hubble constant. Furthermore, $S_8$ takes higher values in all three cases, indicating consistency with recent measurements by  Kilo-Degree Survey (KiDS)~ \cite{Wright:2025xka,Stolzner:2025htz}, Dark Energy Survey Year 3 (DES-Y3)~\cite{DES:2025xii} and Planck~\cite{Planck:2018vyg}.

Fig.~\ref{fig:eff-eos-ivs} displays the evolution of  $w_{x}^{\rm eff}$ and $w_{c}^{\rm eff}$ for the IVS. Since the mean value of $\Gamma/H_{0}$ is positive for CMB and \cmbdesi, (see Table~\ref{table:IVS}), deviation from  $w_{x}^{\rm eff}=-1$ in the phantom direction is found where at the present epoch, $w_{x}^{\rm eff}$ assumes $-1.05$ (for CMB-only) and  $-1.01$  for \cmbdesi. For the other datasets, the mean values of $\Gamma/H_{0}$ are negative, and therefore the deviation of $w_{x}^{\rm eff}$ in the the quintessence direction is suggested,  yielding $w_{x}^{\rm eff} \sim -0.982$ at the present epoch for all three combined datasets.

For CDM, we find that ($w_{c}^{\rm eff}$) increases rapidly with the expansion scale factor for both CMB and \cmbdesi analyses where at the present epoch, maximum value of $w_{c}^{\rm eff}$ is attained for the CMB case ($w_{c}^{\rm eff}\sim 0.3$). This implies that, due to the interaction, CDM component acquires a positive effective equation of state ($w_{c}^{\rm eff} > 0$). From the continuity equation of CDM, $d\ln \rho_c/d \ln a  = - 3 (1+w_{c}^{\rm eff})$, one can understand that
a positive $w_{c}^{\rm eff}$, henceforth, $d\ln \rho_c/d \ln a < 0$, causes the DM energy density to decrease more rapidly than the standard $a^{-3}$ evolution (i.e. without the presence of the interaction).
This behavior can be explained by analyzing the direction of the energy transfer. Since $\Gamma/H_0>0$ for CMB and \cmbdesi, the transfer of energy occurs from CDM to DE, thereby leading to a continuous dilution of the CDM density. Now since the interaction function is proportional to the DE density, while the effective equation of state satisfies $w_c^{\rm eff}= \frac{\Gamma}{3H}\frac{\rho_x}{\rho_c}$, therefore,
with the decrease of $\rho_c$, the ratio $\rho_x/\rho_c$ increases. Consequently, a rapid growth of $w_c^{\rm eff}$ is observed. In the asymptotic future, as $ \rho_c\rightarrow 0$, the ratio $\rho_x/\rho_c$ diverges, leading to the divergence of $w_c^{\rm eff}$.

In contrast, for the last three combined analyses, namely, \cmbdesipantheon, \cmbdesiunion and \cmbdesidovekie, the preferred values of the interaction parameter are negative. In this case, the energy transfer is reversed, occurring from DE to CDM. Consequently, $w_c^{\rm eff}$ becomes negative, implying that the DM component acquires a small effective negative pressure. Although the continuous transfer of energy from DE to DM tends to reduce the ratio $\rho_x/\rho_c$, the expansion of the Universe eventually drives this ratio upward again, resulting in a gradual increase of $w_c^{\rm eff}$ in the far future while remaining finite over the observationally relevant epoch.

Finally, in terms of the Bayesian evidence analysis, we notice that, irrespective of the datasets, $\Lambda$CDM is favored over this interacting scenario. However, according to the revised Jeffrey's, this evidence is moderate, being slightly strong ($-5.2$) for the \cmbdesipantheon combination.
In summary, considering the observational constraints and the model comparison analysis, this interacting model can be considered an alternative scenario to the $\Lambda$CDM model for the description of our universe.

For the next three analyses, namely, \cmbdesipantheon, \cmbdesiunion and \cmbdesidovekie, we find that evidence of interaction at slightly more than 68\% CL is suggested by all of them. However, within 95\% CL, one can recover the non-interacting $\Lambda$CDM model.

\subsection{Phantom IDE (IDEphan)}

Table~\ref{tab:IDEphan} and Fig. \ref{fig:IDEphan-2} summarize the constraints on this interacting scenario considering various observational datasets.  The large-scale stability within this model scenario requires $w_x < -1$ and $\Gamma/H_0 < 0$.

We start with the constraints from CMB alone and then gradually move on to the next set of analyses, combining CMB with other datasets. For CMB alone, $\Gamma/H_0$ attains a lower limit; thus, no evidence of interaction is reported in this case. However, this scenario predicts a slightly high value of $H_0$ ($H_0 = 70.91^{+1.76}_{-4.83}$ km/s/Mpc at 68\% CL) compared to the $\Lambda$CDM model~\cite{Planck:2018vyg}. While $\Omega_m$ and $S_8$ estimates are similar to the $\Lambda$CDM predicted values~\cite{Planck:2018vyg}.  In light of the Bayesian evidence analysis, moderate evidence for this model ($\Delta\ln\mathcal{B}_{ij} =-6.3$) was noted, while $\Lambda$CDM remained favored.

When DESI is combined with CMB, a mild phantom nature of $w_x$ is found ($w_x = -1.067^{+0.045}_{-0.030}$ at 68\% CL for \cmbdesi) but $\Gamma/H_0$ again attains an upper limit. Thus, similar to the CMB alone case, \cmbdesi does not predict any evidence of a non-zero coupling between DE and DM. The mean value of $H_0$ is slightly reduced ($\sim 1.87$ km/s/Mpc) but its uncertainties are significantly reduced, leading to $H_0 = 69.04^{+0.83}_{-0.98}$ km/s/Mpc at 68\% CL. On the other hand, as $w_x$ now comes close to $-1$, the expansion rate is slightly decreased compared to the CMB alone case, and as a result, $H_0$ decreases (see Fig. \ref{fig:IDEphan-2} showing an anti-correlation between $w_x$ and $H_0$). The matter density parameter therefore takes a slightly higher value ($\Omega_m =0.3261^{+0.0226}_{-0.0205}$ at 68\% CL), however, $S_8$ parameter takes similar value to that of Planck-$\Lambda$CDM~\cite{Planck:2018vyg}. However, according to the Bayesian evidence analysis, this dataset shows very strong evidence  ($\Delta\ln\mathcal{B}_{ij} =-7.5$) against this interacting scenario with respect to the $\Lambda$CDM model.

We now discuss the constraints for the remaining three combined analyses, namely, \cmbdesipantheon, \cmbdesiunion and \cmbdesidovekie. According to the results, an evidence for a non-vanishing $\Gamma/H_0$ is found at more than $2\sigma$ across all three combined datasets. The constraints on $H_0$ and $S_8$ are similar to the Planck-$\Lambda$CDM model~\cite{Planck:2018vyg}, however, all three combined datasets report a high value of $\Omega_m \sim 0.34$. This is driven by the strong anti-correlation between $\Omega_m$ and $\Gamma/H_0$ as shown in Fig.~\ref{fig:IDEphan-2}. A Bayesian evidence analysis does not support this interacting scenario over the reference model. According to the results, $\ln\mathcal{B}_{ij}=-8.0$ for \cmbdesipantheon, $\ln\mathcal{B}_{ij}=-7.4$ for \cmbdesiunion and $\Delta\ln\mathcal{B}_{ij}=-7.5$ for \cmbdesidovekie.

Figure~\ref{fig:eff-eos-idephan} describes the evolution of  $w_x^{\rm eff}$ and $w_c^{\rm eff}$, for the {\bf IDEphan} model. For all dataset combinations, $w_x^{\rm eff}$ remains in the phantom regime, exhibiting only a mild evolution with the scale factor. In the CMB-only analysis, $w_x^{\rm eff}$ is more deeply confined within the phantom region than other cases, leading to a more pronounced effective phantom behavior. On the other hand, $w_c^{\rm eff}$ evolves monotonically toward negative values with the expansion of the universe. For the CMB-only analysis, it reaches $w_c^{\rm eff}\simeq -0.05$ at the present epoch, while the combined datasets predict a slightly less negative value $w_c^{\rm eff}\simeq-0.045$. Thus, the interaction induces a small but non-zero effective negative pressure for the DM component. Overall, we find that for the present interaction function, phantom $w_x$ does not lead to a competitive scenario for describing the present universe.

\subsection{Quintessential IDE (IDEquin)}

Table~\ref{tab:IDEquin} and Fig. \ref{fig:IDEquin-2} summarize the constraints on this model scenario using the same datasets. The allowed region for this scenario in terms of the large-scale stability of the interacting model is $w_x > -1$ and $\Gamma/H_0 > 0$.

As before, we first discuss the constraints from the CMB alone and then examine how they are modified by the inclusion of low-redshift observations. Using only the CMB data, neither the interaction strength nor the dark-energy equation of state is well constrained, yielding the 95\% CL upper limits ($\Gamma/H_0<0.343$) and ($w_x<-0.620$), respectively. The model predicts relatively low values of the Hubble constant ($H_0=65.28^{+4.84}{-3.11}$ km/s/Mpc), the matter density parameter, ($\Omega_m=0.2333^{+0.1082}{-0.1037}$), and the clustering parameter, ($S_8=0.786^{+0.101}{-0.041}$), all quoted at the 68\% CL. The preference for a lower matter density is qualitatively consistent with the positive coupling, which corresponds to a transfer of energy from the DM sector to the DE sector. However, the relatively low values of ($H_0$) and ($\Omega_m$) inferred from the CMB alone should be interpreted with caution, as they are influenced by the well-known geometrical degeneracies among ($H_0$), ($\Omega_m$), ($w_x$), and the coupling parameter.
However, according to the Bayesian evidence analysis,  $\Lambda$CDM model is still favored over this model ($\ln\mathcal{B}_{ij} =-4.0$).

When DESI BAO is combined with CMB, we find that $\Gamma/H_0 = 0.118^{+0.044}_{-0.101}$ at 68\% CL, thereby indicating a mild evidence of interaction. However, $w_x$ attains an upper bound ($w_x< -0.901$ at 95\% CL) but $H_0$ assumes a higher value ($H_0 = 69.21^{+0.86}_{-1.13}$ Km/s/Mpc at 68\% CL). Interestingly, similar to the CMB alone case,  $S_8$ assumes a low value ($S_8 = 0.764^{+0.041}_{-0.013}$ at 68\% CL) compared to Planck-based $\Lambda$CDM paradigm~\cite{Planck:2018vyg}. Moreover, this combined dataset also leads to a lower value of $\Omega_m$ ($= 0.2209^{+0.0683}_{-0.0341}$ at 68\% CL) compared to the non-interacting scenario, where DM density evolves as $a^{-3}$.
However, Bayesian evidence analysis showed strong evidence ($\ln\mathcal{B}_{ij} =-7.5$) for this model when considering this dataset.

For the last three combined datasets with CMB, DESI, and SNIa (PantheonPlus, Union3, and DES-Dovekie), we see that $\Gamma/H_0$ attains an upper limit across all datasets.
On the other hand, 68\% CL constraints on $w_x$ are available and they are close to $-1$, but within 95\% CL, $w_x$ attains its upper limit.  The constraints on $H_0$ are almost similar to those of the non-interacting $\Lambda$CDM model by Planck~\cite{Planck:2018vyg}, however, $S_8$ measurements are still slightly low:
$S_8 \sim 0.79$ for all three combined datasets.

Figure~\ref{fig:eff-eos-idequin} shows the evolution of  $w_{x}^{\rm eff}$ and $w_{c}^{\rm eff}$ for the {\bf IDEquin} model.
We notice that for the CMB-only analysis, $w_{x}^{\rm eff}$ attains a higher value ($w_{x}^{\rm eff}\sim -0.9$) at the present epoch, whereas the combined datasets favor values in the range $w_{x}^{\rm eff}\sim -0.96$  -- $-0.98$. In all cases $w_{x}^{\rm eff}$ exhibits only a very mild evolution toward the cosmological-constant boundary.

For the DM component, $w_{c}^{\rm eff}$ increases rapidly with the scale factor, reaching very large values already in the recent past.
This behavior is qualitatively similar to that found for the {\bf IVS} model shown in Fig.~\ref{fig:eff-eos-ivs}. Since the interaction transfers energy from CDM to DE (as $\Gamma/H_0>0$), the DM density is continuously decreases. As a consequence, the ratio $\rho_x/\rho_c$ grows rapidly, and according to $w_c^{\rm eff}= \frac{\Gamma}{3H}\frac{\rho_x}{\rho_c}$, the effective EoS of DM increases accordingly. This establishes a self-reinforcing mechanism: the reduction of DM density increases the ratio $\rho_x/\rho_c$, which results in higher values of $w_c^{\rm eff}$. In the asymptotic future, as $\rho_c\rightarrow 0$, the ratio $\rho_x/\rho_c$ diverges, and so does $w_c^{\rm eff}$.

In light of the Bayesian evidence analysis, $\Lambda$CDM remains in the preferred place compared to {\bf IDEquin}: $\ln\mathcal{B}_{ij}=-8.4$, $-7.6$, $-8.1$ for \cmbdesipantheon, \cmbdesiunion and \cmbdesidovekie, respectively.

\section{Conclusions and Prospects}
\label{sec-summary}

Cosmological probes from a variety of astronomical surveys have consistently argued that our universe is mainly dominated by two dark fluids, namely  DM and DE, which occupy nearly 96\% of the total energy budget of the universe. Many theoretical models have been developed to reveal their nature, but they remain mysterious so far. Existing models focusing on DE and DM can be divided into two distinct categories: models in which DE and DM evolve independently and models in which DM and DM interact. This later class of models represent a general construction of cosmological scenarios and they are promising in the cosmological domain because, according to the existing records, interacting cosmological models can alleviate several longstanding cosmic puzzles, such as the cosmic coincidence problem and cosmological tensions. Interestingly, issues related to the cosmological constant problem do not appear in this framework.

In the present article, we considered a coupled DE-DM scenario (DM is pressureless and DE has a constant EoS) where the coupling function $Q$ is proportional to the energy density of DE as, $Q = \Gamma \rho_{x}$, where $\Gamma$ is a constant  coupling parameter and its dimension is equal to the dimension of the Hubble rate (therefore, in the statistical simulations we have considered the dimensionless parameter $\Gamma/H_0$).
Because this interaction function does not involve external parameters such as the expansion rate and scale factor, the mechanism of interaction depends on the intrinsic nature of DE. This model is usually referred to as the local interaction rate because of the absence of such external parameters.  One of the important observations in the interacting cosmologies is that any interacting scenario between DE and DM is equivalent to a non-interacting prescription between DE and DM, in which the EoS parameters of these dark fluids are time-dependent. This allows us to extract more qualities of the interacting scenarios on theoretical and observational grounds. As the nature of DE plays a crucial role in the large-scale structure of the universe and the DE EoS could influence the stability of the underlying interacting model, we considered three separate regions of $w_x$ where the interacting model is stable, and we label these scenarios as {\bf IVS}, {\bf IDEphan}, and {\bf IDEquin}. The observational constraints are summarized in various tables and figures: Table~\ref{table:IVS} and Fig. \ref{fig:IVS-1} (for {\bf IVS}); Table~\ref{tab:IDEphan} and Fig. \ref{fig:IDEphan-2} (for {\bf IDEphan}); Table~\ref{tab:IDEquin} and Fig.~\ref{fig:IDEquin-2} (for {\bf IDEquin}).
Our main results are the following:

\begin{enumerate}[label=(\roman*)]

\item The interaction between DE-DM could offer a reasonable justification for the dynamical DM. The mechanism of energy transfer between the dark sectors allows us to realize a phantom crossing without invoking any phantom scalar field. Thus, phantom $\longrightarrow$ quintessence or quintessence $\longrightarrow$ phantom regimes are just two different outcomes enabled by the direction of energy transfer between the dark sectors.
It has also been observed that for $w_x = \Gamma/3H$, $w_{x}^{\rm eff}$ vanishes, and as a result of which DE evolves as $a^{-3}$ and DM evolves with an effective dynamical EoS $w_{c}^{\rm eff} = (\Gamma/3H) \times \rho_x \rho_{c}^{-1}$.

\item Coupling parameter ($\Gamma/H_0$): In {\bf IVS}, except for CMB alone and \cmbdesi, all other datasets indicate an evidence of interaction  at slightly more than 68\% CL;
{\bf IDEphan} shows evidence of interaction at more than 95\% CL for \cmbdesipantheon, \cmbdesiunion and \cmbdesidovekie;
{\bf IDEquin} did not show any evidence of interaction. These findings suggest that the indication of an interaction in the dark sector depends on many factors, such as the nature of DE, interaction model and the underlying datasets.

\item \textbf{$S_8$:} In {\bf IVS} $S_8$ values for CMB alone ($S_8 \sim 0.766$) and \cmbdesi ($S_8 \sim 0.793$) are low compared to the $\Lambda$CDM-based Planck~\cite{Planck:2018vyg} but for the remaining three combined datasets in presence of SNIa, $S_8$ values are almost identical with Planck ~\cite{Planck:2018vyg}
and other measurements~\cite{Wright:2025xka,Stolzner:2025htz,DES:2025xii}; in {\bf IDEphan}, $S_8$ values are similar to Planck ~\cite{Planck:2018vyg}
and other measurements~\cite{Wright:2025xka,Stolzner:2025htz,DES:2025xii} across all datasets.
Only {\bf IDEquin} scenario leads to mildly smaller values of $S_8$ for all datasets. The estimated values lie in the range $S_8 \sim 0.764 - 0.796$. This is an interesting outcome of this scenario.

\item $H_0$ (km/s/Mpc): For the combined datasets with CMB, DESI and SNIa, all three scenarios yield $H_0 \sim 67$ km/s/Mpc, however, for CMB alone and \cmbdesi, {\bf IVS} and {\bf IDEphan} report slightly higher values. Maximum value of $H_0$ is attained in the {\bf IDEphan} scenario leading to $H_0 =70.91_{-4.83}^{+1.76}$ (CMB alone). For \cmbdesi, all three scenarios yield $H_0 \sim 69$ km/s/Mpc.

\item $\ln\mathcal{B}_{ij}$: According to the Bayesian evidence analysis, $\Lambda$CDM remains preferred over all the interacting scenarios across all the datasets. Specifically, moderate evidence is observed for {\bf IVS} except for the dataset \cmbdesipantheon; {\bf IDEphan} shows {\bf strong  evidence}; and considering {\bf IDEquin}, we see the transition from moderate (CMB alone) to strong (for remaining datasets) evidence.

\end{enumerate}

In summary, in this article we showed that the present interacting model has appealing features in  different regimes of the DE EoS. Although we observed that for this local interaction rate, alleviation of the $H_0$ tension is not homogeneously valid across all the datasets, and the EoS of DE also plays a crucial role in this direction, and additionally, estimated values of $S_8$ are different in different interacting scenarios,  however, we strongly argue here that these results could be influenced by the choice of the interaction model. Furthermore, relaxing the constancy of the coupling, and a different choice of the coupling function other than $Q \propto \rho_x$ could also offer interesting outcomes.

\section*{Acknowledgments}
We thank the referee for some useful comments that helped us improve the overall quality of the manuscript. WY has been supported by the National Natural Science Foundation of China under Grant Nos. 12547110 and 12175096. L.G is supported by research grants from Conselho Nacional
de Desenvolvimento Cientıfico e Tecnologico (CNPq),
Grant No. 307636/2023-2 and from the Fundacao Carlos
Chagas Filho de Amparo a Pesquisa do Estado do Rio
de Janeiro (FAPERJ), Grant No. E-26/204.598/2024.

\section*{Data Availability}
The observational data that support the findings of this article are publicly available from the corresponding data releases for Planck 2018 [Refs. \onlinecite{Planck:2018vyg,Planck:2019nip}], DESI DR2 [Refs. \onlinecite{DESI:2025zgx,DESI:2025zpo}], PantheonPlus [Ref. \onlinecite{Scolnic:2021amr}], Union3 [Ref. \onlinecite{Rubin:2023ovl}], and DES-Dovekie [Ref. \onlinecite{DES:2024jxu}]. The software packages used for the cosmological analysis are publicly available. The custom model implementation and analysis scripts developed for this work are available from the authors upon reasonable request.

\appendix

\section{Additional plots}
\label{sec-appendix}

In this section, we show the DE density contrast fluctuations $\delta_x$ at $k = 0.1/\mathrm{Mpc}$ and $f\sigma_8$ curves for the {\bf IDEquin} scenario for different values of $c_{s,x}^2$ (Fig. \ref{fig:idequin_deltax_fs8_cs2}). The $f\sigma_8$ curves for different values of $\Gamma/H_0$ are also shown for {\bf IVS} (Fig. \ref{fig:ivs_fs8_xi}). For {\bf IDEquin}, variations with $w_x$ and $\Gamma/H_0$ are shown in Figs. \ref{fig:idequin_deltax_fs8_w0} and \ref{fig:idequin_deltax_fs8_xi}, respectively. We see that the nature of the quantities as depicted in Figs. \ref{fig:idequin_deltax_fs8_cs2}, \ref{fig:idequin_deltax_fs8_w0}, and \ref{fig:idequin_deltax_fs8_xi} are almost identical with {\bf IDEphan}. Finally, considering {\bf IDEquin}, in Fig. \ref{fig:idequin-cs2}, we show the CMB TT and matter power spectra for different values of $c_{s,x}^2$ assuming $\Gamma/H_0 =0.1$ and $w_x =-0.99$. Fig. \ref{fig:idequin-cs2} clearly shows that the CMB and matter power spectra remain insensitive to $c_{s,x}^2$. This is what we have observed in the case with {\bf IDEphan}. This motivated us to constrain the present interacting cosmological scenarios fixing $c_{s,x}^2 =1$.

\FloatBarrier

\begin{table}
    \begin{center}
        \renewcommand{\arraystretch}{1.4}
        \begin{tabular}{ l c c c}
        \hline
        \hline
        Survey & $z$ & $f\sigma_8$ & Reference \\ \hline
        ALFALFA & $0.013$ & $0.46 \pm 0.06$ & \cite{avila2021} \\ \hline
        $6$dFGS+SDSS & $0.035$ & $0.338 \pm 0.027$ & \cite{Said2020} \\ \hline
        \multirow{2}{4em}{GAMA} & $0.18$ & $0.29 \pm 0.10$ & \cite{simpson2016} \\
        & $0.38$ & $0.44 \pm 0.06$ & \cite{Blake2013} \\ \hline
        \multirow{4}{4em}{WiggleZ} & $0.22$ & $0.42 \pm 0.07$ & \multirow{4}{2em}{\cite{Blake2011}} \\
        & $0.41$ & $0.45 \pm 0.04$ & \\
        & $0.60$ & $0.43 \pm 0.04$ & \\
        & $0.78$ & $0.38 \pm 0.04$ & \\ \hline
        \multirow{2}{7em}{DR$12$ BOSS} & $0.32$ & $0.427 \pm 0.056$ & \multirow{2}{2em}{\cite{gil2016clustering}}\\
        & $0.57$ & $0.426 \pm 0.029$ & \\ \hline
        \multirow{2}{4em}{VIPERS} & $0.60$ & $0.49 \pm 0.12$ & \multirow{2}{2em}{\cite{mohammad2018vimos}}\\
        & $0.86$ & $0.46 \pm 0.09$ & \\ \hline
        VVDS & $0.77$ & $0.49 \pm 0.18$ & \cite{guzzo2008test, song2009reconstructing}\\ \hline
        FastSound & $1.36$ & $0.482 \pm 0.116$ & \cite{okumura2016subaru}\\ \hline
        eBOSS Quasar & $1.48$ & $0.462 \pm 0.045$ & \cite{hou2021completed}\\
        \hline
        \hline
            \end{tabular}
        \end{center}
    \caption{Summary of $f\sigma_8$ values at different redshifts taken from Ref. \cite{Toda:2024fgv}. }
    \label{tab:fs8_values}
\end{table}

\renewcommand{\thefigure}{A\arabic{figure}}
\begin{figure}[tbp]
    \centering
    \includegraphics[width=0.47\textwidth]{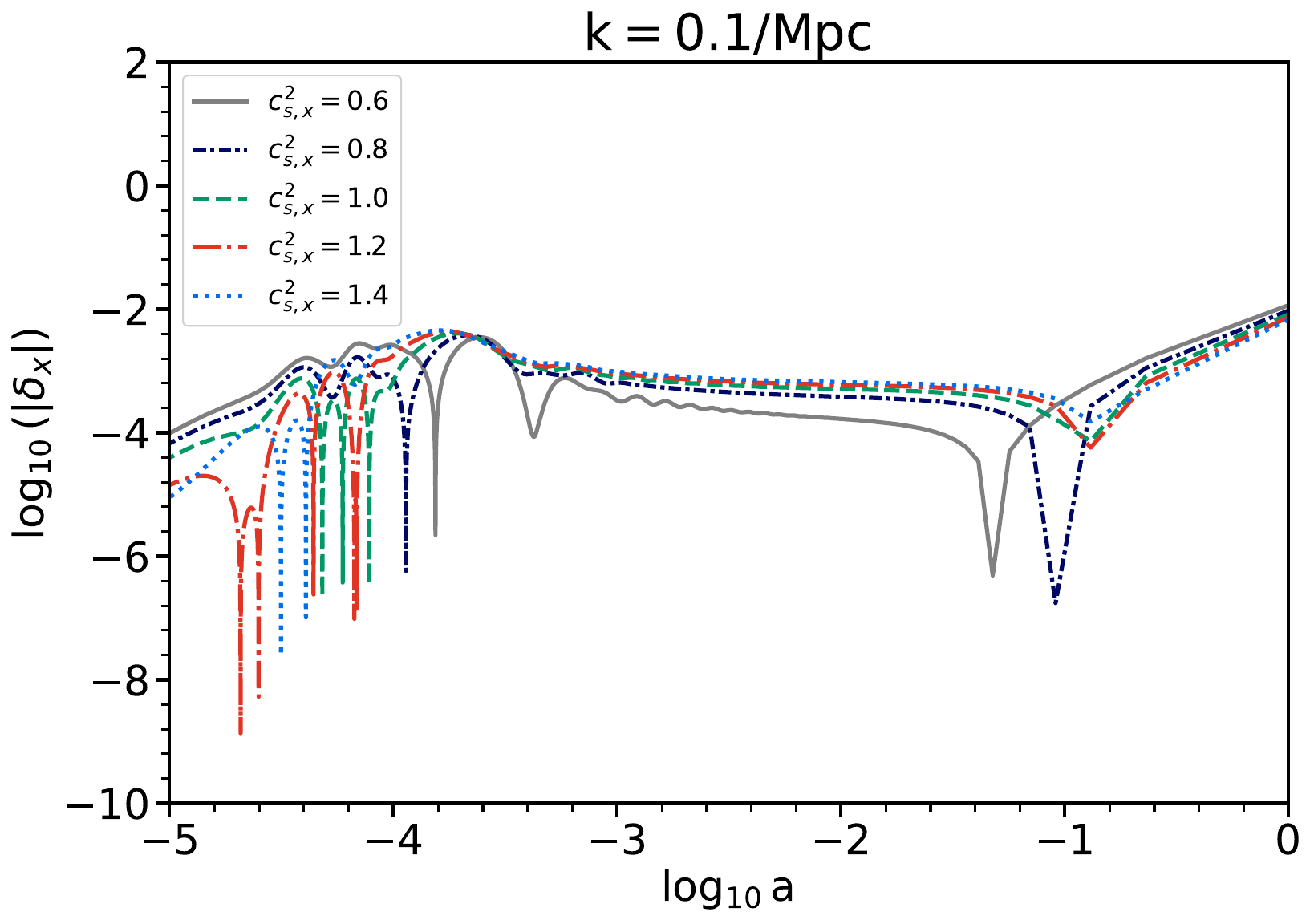}
    \includegraphics[width=0.47\textwidth]{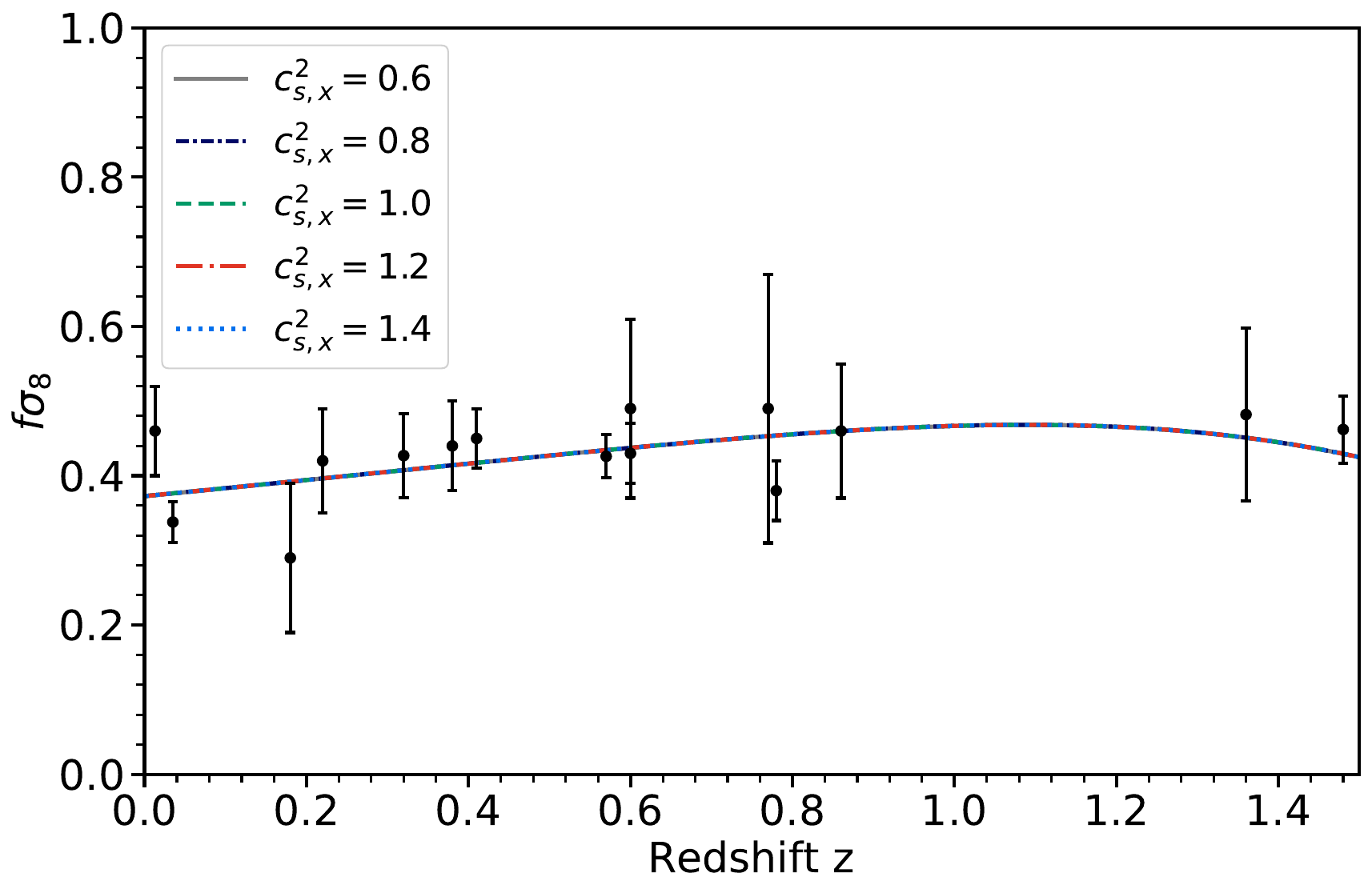}
    \caption{{\bf (IDEquin)} DE density contrast fluctuations $\delta_x$ at $k = 0.1/\mathrm{Mpc}$ (left) and $f\sigma_8$ curves (right) for variations of DE sound speed $c_{s,x}^2$. The $f\sigma_8$ measurements (black) can be found in Table \ref{tab:fs8_values}. For plotting $\delta_x$, we set $w_x = -0.99$, $\Gamma/H_0 = 0.1$, $H_0=67.5$ km/s/Mpc, $\Omega_b h^2=0.022$, $\Omega_c h^2=0.122$, and for the $f\sigma_8$ plot we fix $n_s =0.965$. }
    \label{fig:idequin_deltax_fs8_cs2}
\end{figure}
\begin{figure}[tbp]
    \centering
    \includegraphics[width=0.47\textwidth]{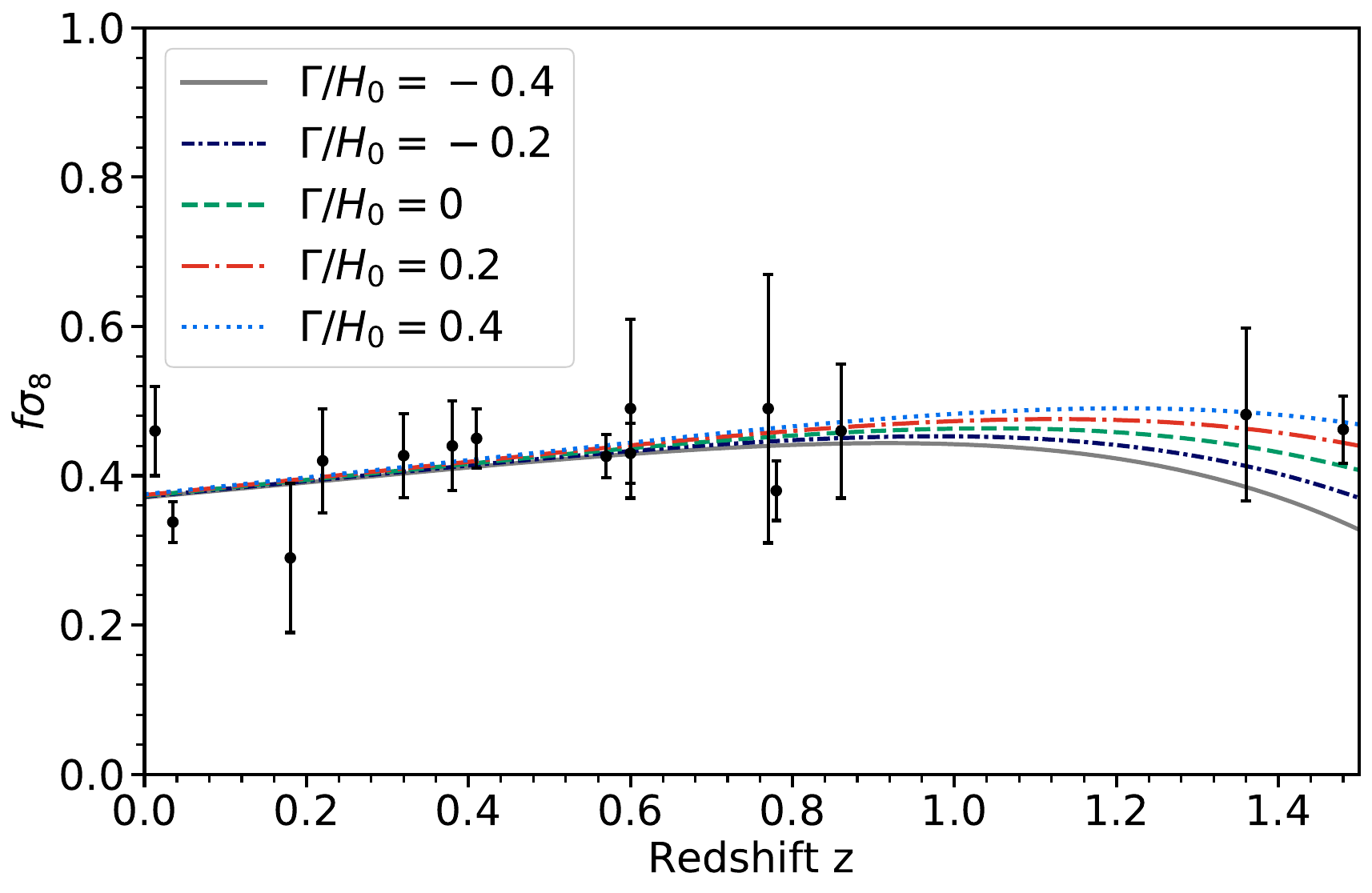}
    \caption{{\bf (IVS)} $f\sigma_8$ for different values of $\Gamma/H_0$. The $f\sigma_8$ measurements (black) can be found in Table \ref{tab:fs8_values}. Here we have fixed $n_s =0.965$.  }
    \label{fig:ivs_fs8_xi}
\end{figure}
\begin{figure*}
    \centering
    \includegraphics[width=0.47\textwidth]{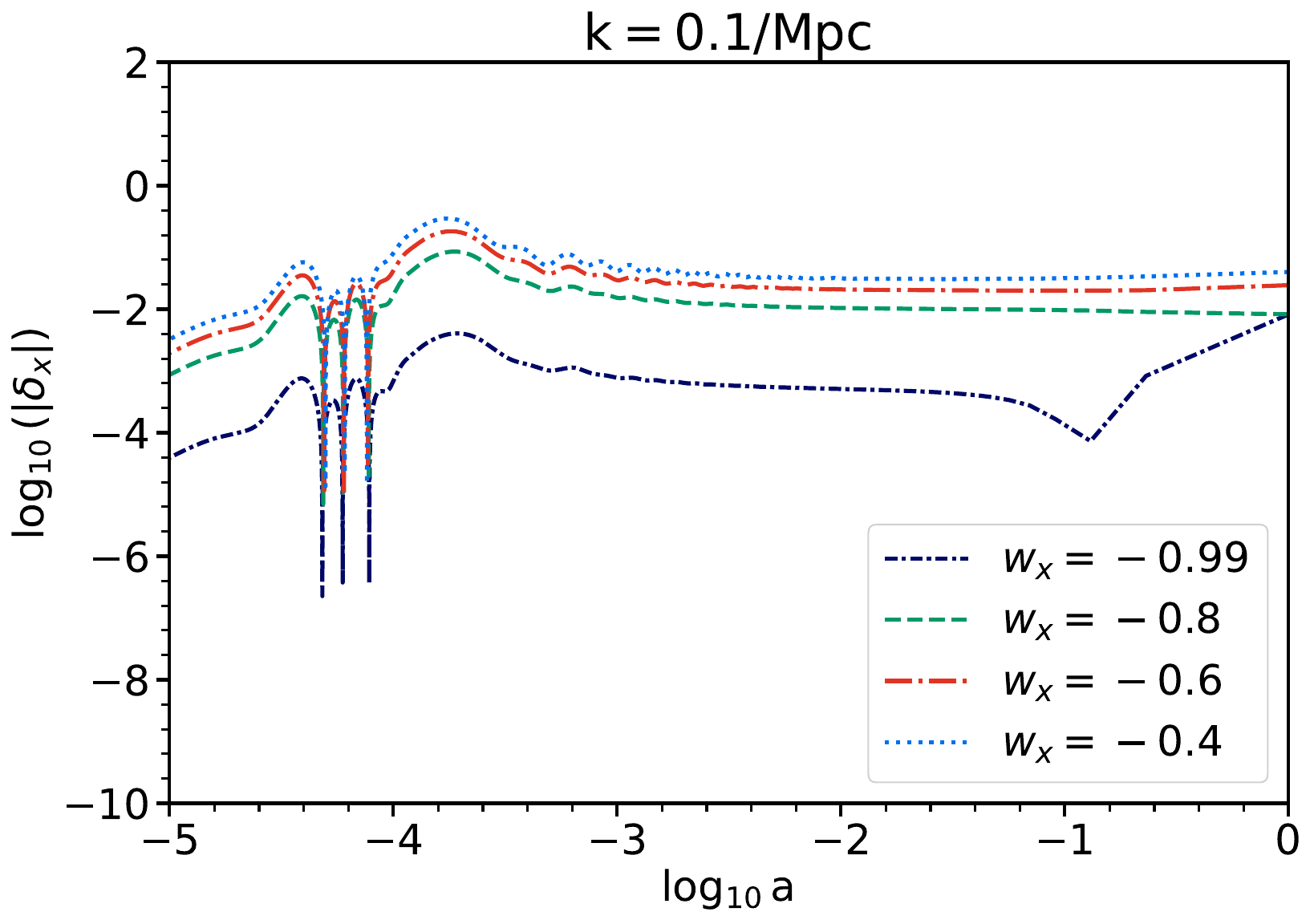}
    \includegraphics[width=0.47\textwidth]{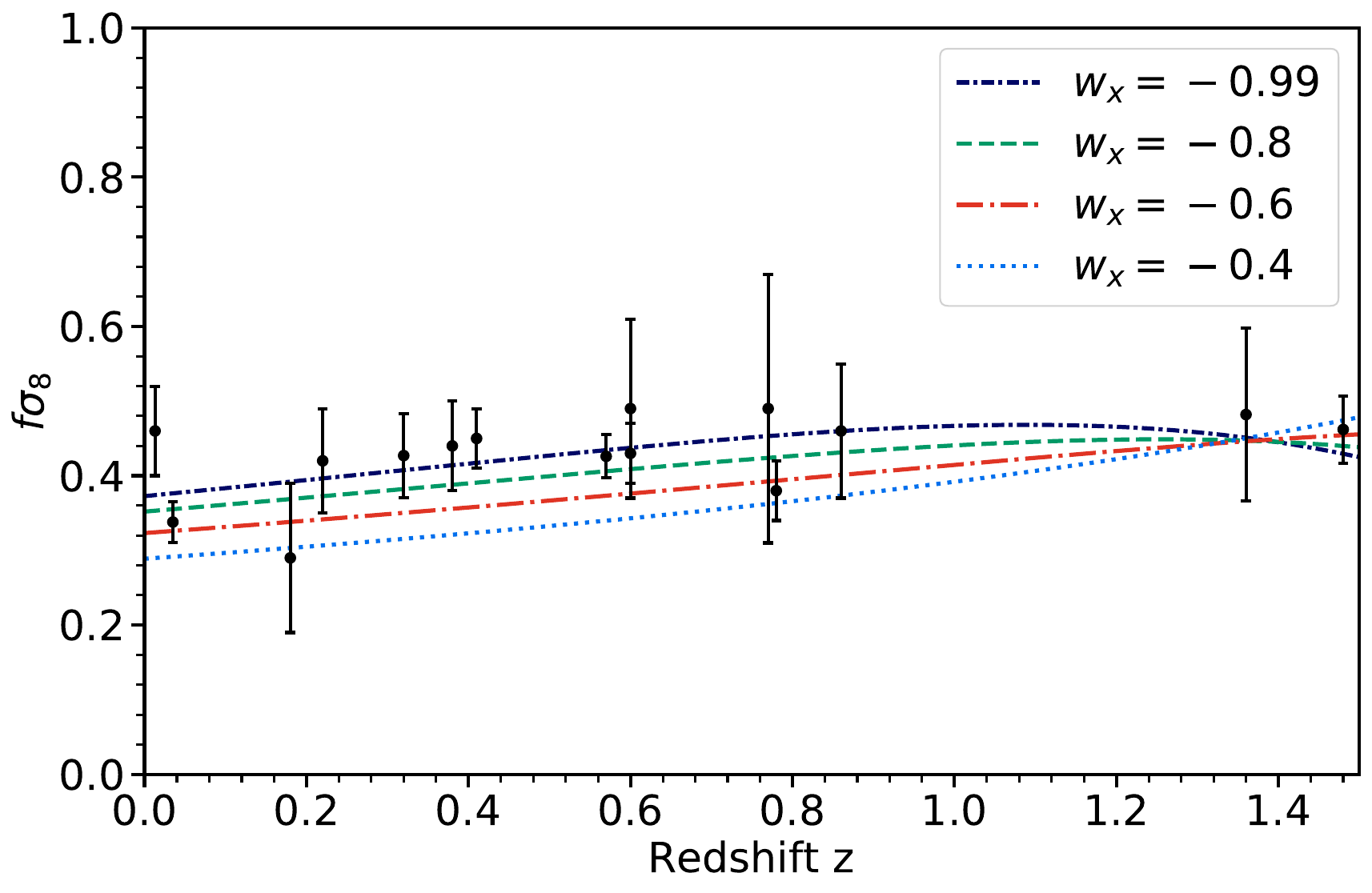}
    \caption{{\bf (IDEquin)} DE density contrast fluctuations $\delta_x$ at $k = 0.1/\mathrm{Mpc}$ (left) and $f\sigma_8$ curves (right) for variations of EoS parameter $w_{x}$. The $f\sigma_8$ measurements (black) can be found in Table \ref{tab:fs8_values}. For plotting $\delta_x$, we set $\Gamma/H_0 = 0.1, c_{s,x}^2 = 1$, $H_0=67.5$ km/s/Mpc, $\Omega_b h^2=0.022$,  $\Omega_c h^2=0.122$, and for the $f\sigma_8$ plot we fix $n_s =0.965$.  }
    \label{fig:idequin_deltax_fs8_w0}
\end{figure*}
\begin{figure*}
    \centering
    \includegraphics[width=0.47\textwidth]{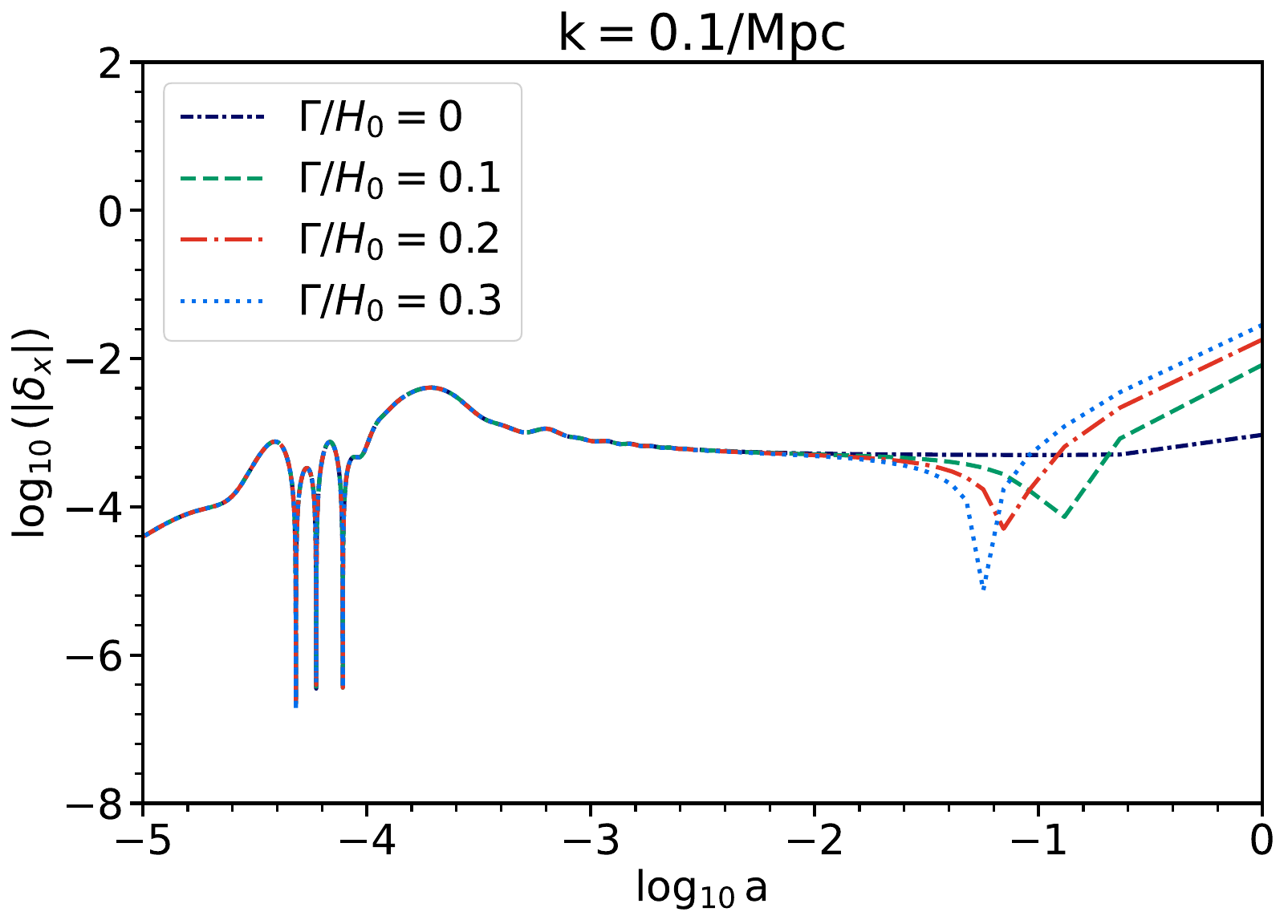}
    \includegraphics[width=0.47\textwidth]{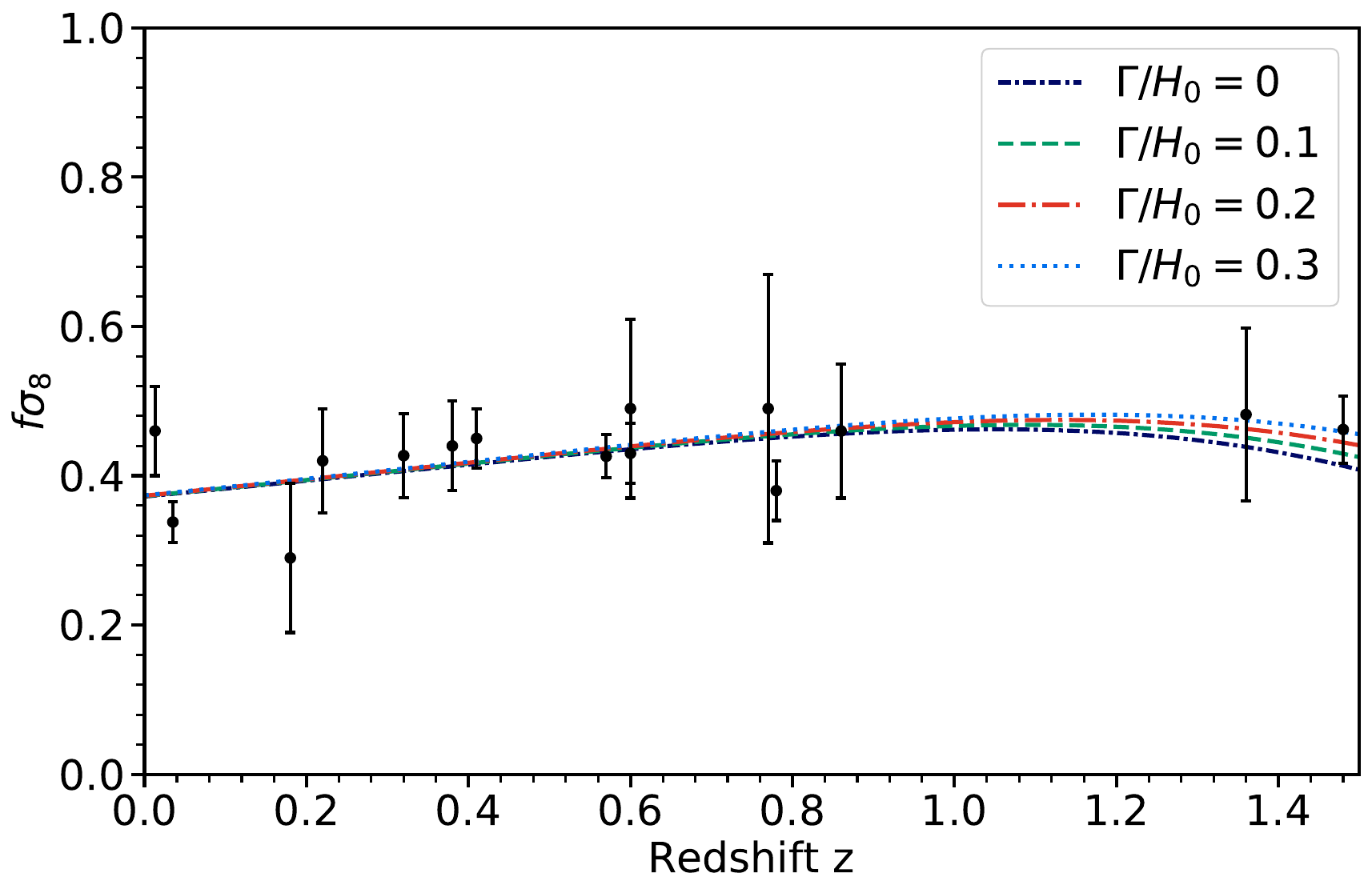}
    \caption{{\bf (IDEquin)} DE density contrast fluctuations $\delta_x$ at $k = 0.1/\mathrm{Mpc}$ (left) and $f\sigma_8$ curves (right) for variations of $\Gamma/H_0$. The $f\sigma_8$ measurements (black) can be found in Table \ref{tab:fs8_values}. For plotting $\delta_x$, we set $w_x = -0.99, c_{s,x}^2 = 1$, $H_0=67.5$ km/s/Mpc, $\Omega_b h^2=0.022$,  $\Omega_c h^2=0.122$, and for the $f\sigma_8$ plot we fix $n_s =0.965$. }
    \label{fig:idequin_deltax_fs8_xi}
\end{figure*}
\begin{figure*}
    \centering
    \includegraphics[width=0.47\textwidth]{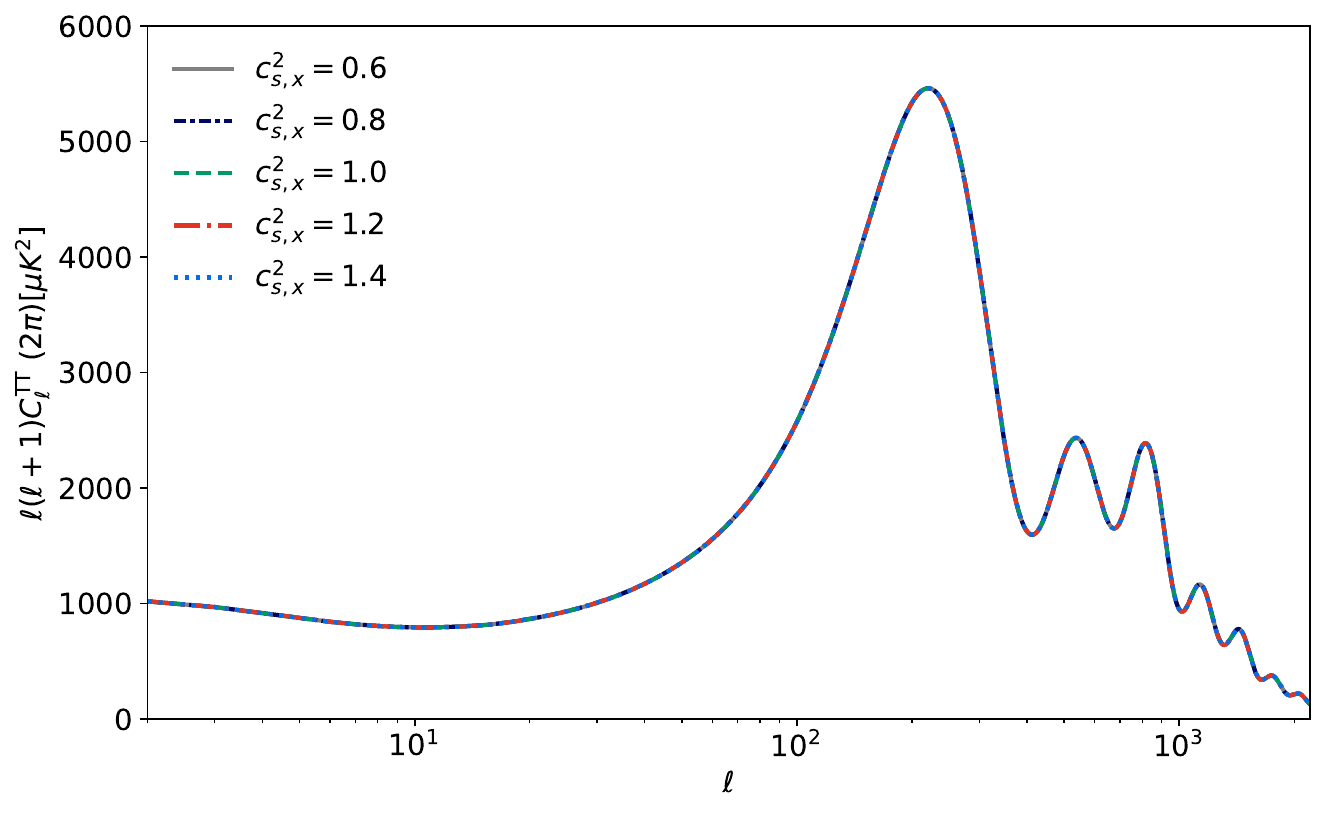}
    \includegraphics[width=0.47\textwidth]{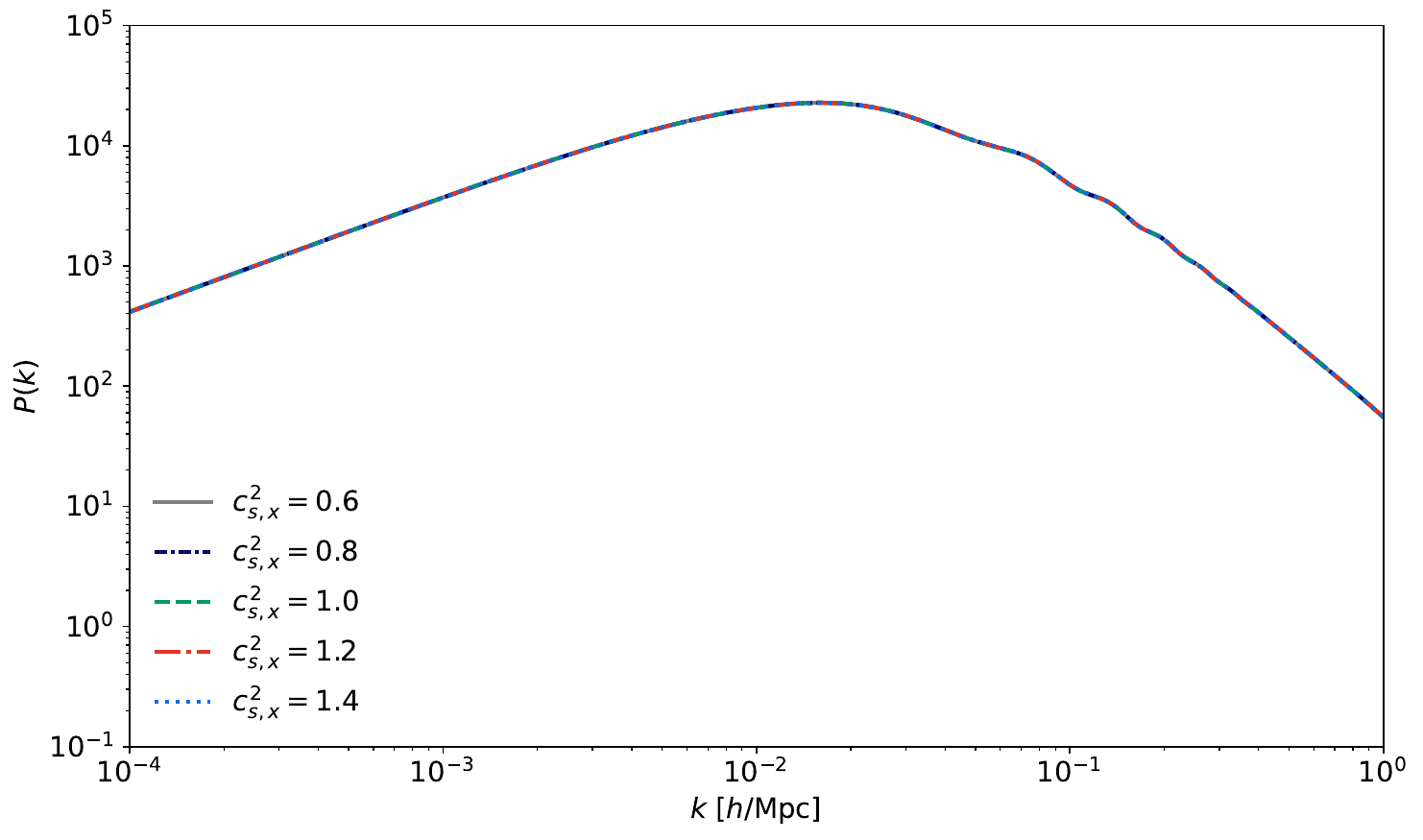}
    \caption{{\bf (IDEquin)} CMB TT spectrum (left) and matter power spectrum (right) for different values of $c_{s,x}^2$ referring to {\bf IDEquin}. We have fixed $\Gamma/H_0 =0.1$ and $w_{x} = -0.99$. The mean values of the other parameters, e.g. $\Omega_bh^2$, $\Omega_ch^2$ and $H_0$ required to generate the plots are taken from the combined analysis \cmbdesipantheon
 (Table~\ref{tab:IDEquin}).  }
    \label{fig:idequin-cs2}
\end{figure*}

\FloatBarrier

%

\end{document}